\documentclass[journal=jctcce,manuscript=article,usetitle=true]{achemso}

\usepackage[version=3]{mhchem} 
\usepackage[T1]{fontenc}       
\usepackage{graphics}
\usepackage{graphicx}
\usepackage{epsfig}
\usepackage{dcolumn}
\usepackage{amssymb}
\usepackage{multirow}
\usepackage{longtable}
\usepackage{amsmath}
\usepackage{braket}
\usepackage{mathtools}
\usepackage{gensymb}
\usepackage{times}
\usepackage{changes}
\author{Zhou Lin}
\author{Troy Van Voorhis}
\email{tvan@mit.edu}
\affiliation{Department of Chemistry, Massachusetts Institute of Technology, Cambridge, MA 02139}

\title{Triplet-Tuning: A Novel Family of Non-Empirical Exchange--Correlation Functionals} 


\begin{document}

\maketitle


\begin{abstract}
	
	In the framework of density functional theory (DFT), the lowest triplet excited state, T$_1$, can be evaluated using multiple formulations, the most straightforward of which are unrestricted DFT (UDFT) and time-dependent DFT (TDDFT). 
	Assuming the exact exchange--correlation (XC) functional is applied, UDFT and TDDFT provide identical energies for T$_1$ ($E_{\rm T}$), which is also a constraint that we require our XC functionals to obey. However, this condition is not satisfied by most of the popular XC functionals, leading to inaccurate predictions of low-lying, spectroscopically and photochemically important excited states, such as T$_1$ and the lowest singlet excited state (S$_1$). 
	Inspired by the optimal tuning strategy for frontier orbital energies {[T. Stein, L. Kronik, and R. Baer, {\it J. Am. Chem. Soc.} {\bf 131}, 2818 (2009)]}, we proposed a novel and {\it non-empirical} prescription of constructing an XC functional in which the agreement between UDFT and TDDFT in $E_{\rm T}$ is strictly enforced. Referred to as ``triplet tuning'',  our procedure allows us to formulate the XC functional on a case-by-case basis using the molecular structure as the exclusive input, without fitting to any experimental data.
	The first triplet tuned XC functional, TT-$\omega$PBEh, is formulated as a long-range-corrected (LRC) hybrid of Perdew--Burke--Ernzerhof (PBE) and Hartree--Fock (HF) functionals {[M. A. Rohrdanz, K. M. Martins, and J. M. Herbert, {\it J. Chem. Phys.} {\bf 130}, 054112 (2009)]} and tested on four sets of large organic molecules. Compared to existing functionals, TT-$\omega$PBEh manages to provide more accurate predictions for key spectroscopic and photochemical observables, including but not limited to $E_{\rm T}$, the optical band gap ($E_{\rm S}$), the singlet--triplet gap ($\Delta E_{\rm ST}$), and the vertical ionization potential ($I_\perp$), as it adjusts the effective electron-hole interactions to arrive at the correct excitation energies. 
	This promising triplet tuning scheme can be applied to a broad range of systems that were notorious in DFT for being {extremely challenging}.
\end{abstract}

\begin{tocentry}
	\centering
	\includegraphics[width = 7.75cm]{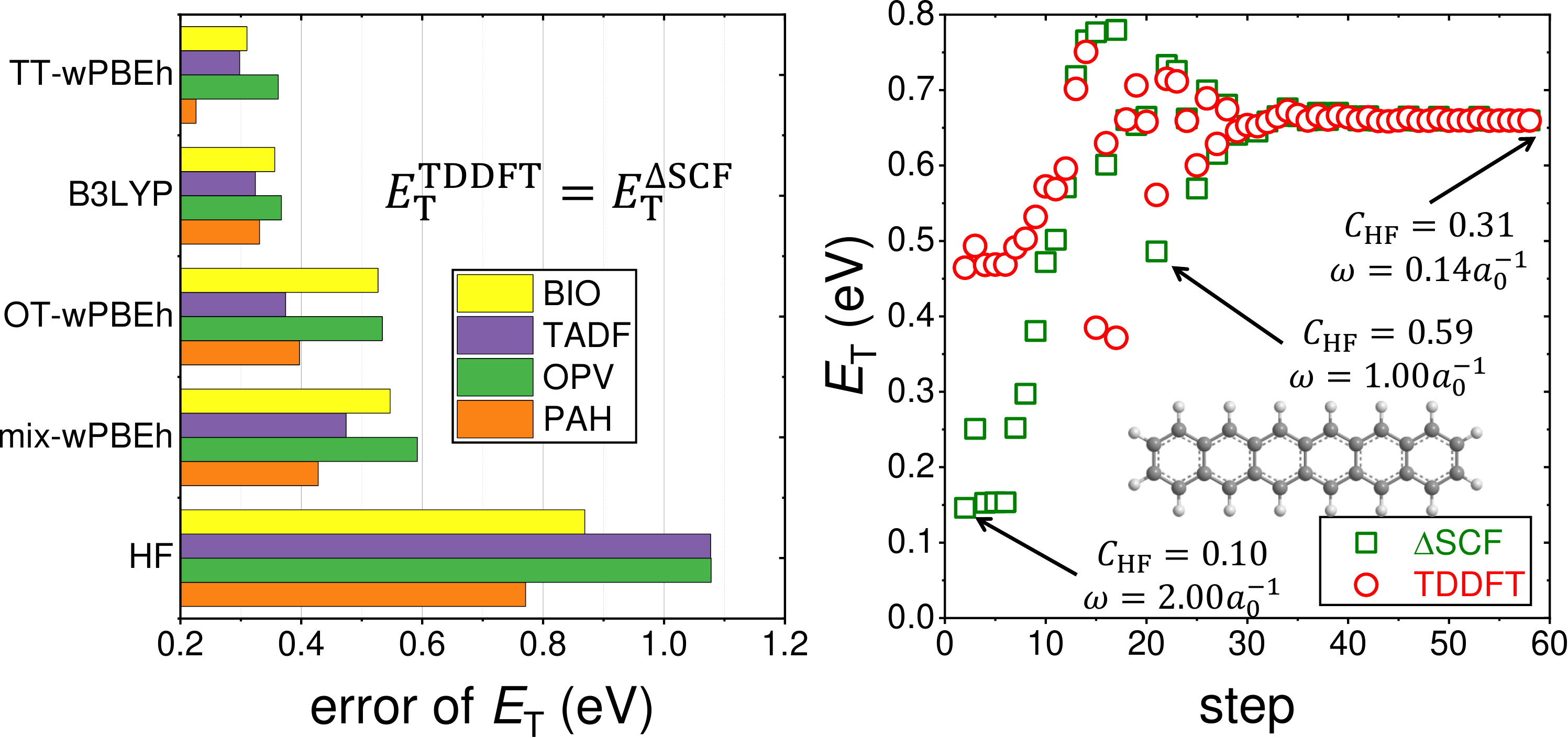}
	\label{fig:toc}
\end{tocentry}

\newpage
\section{Introduction}
\label{sec:intro}

Due to its affordable computational cost, density functional theory (DFT) has become the {main theoretical }workhorse for large molecules {where} most wave function based approaches {are} infeasible.\cite{PhysRev.136.B864, Kohn1965,parr1994density,1.1390175} 
DFT was originally established as a ground-state approach, while various excited-state extensions {were also} formulated to predict experimental observables. Representative excited-state methods include time-dependent DFT (TDDFT)\cite{PhysRevLett.52.997}, spin-flip DFT (SFDFT)\cite{SPDFT}, $\Delta$ self-consistent field ($\Delta$SCF)\cite{PhysRevB.78.075441}, and restricted open-shell Kohn--Sham (ROKS).\cite{kowalczyk2013jcp}
Despite being formally exact, {DFT has several} fundamental issues {which place} a glass ceiling over its predictive power for excited states, especially for organic molecules with large $\pi$-conjugated structures or charge transfer characters.
For example, the accuracy of a (semi-)empirical density functional heavily relies on the selection of parameters in its exchange--correlation (XC) component ($E_{\rm XC}$). Optimal parameters are sensitive to the {configurational }features of {species included} in the fitting database. 
Also, the self-interaction error (SIE) and the locality problem can lead to unphysical density distributions in both short and long ranges for {those difficult} molecules.\cite{PhysRevB.23.5048,dreuw2004jacs,dreuw2005cr,vydrov2005,Medvedev49,Brorsen2017jpcl,Hait2018jctc,hait2018jcp,Bao2018}

{Within the recent decade}, many methodological efforts have been reported to resolve {above-mentioned} issues. 
{For instance, }new DFT variants have been {developed}, such as  self-interaction corrected DFT (SIC-DFT)\cite{Polo2002}, average density self-interaction correction (ADSIC)\cite{CIOFINI200567}, constrained DFT (CDFT),\cite{wu2006jpca} and constrained variational DFT (CV($\infty$)-DFT)\cite{Park2016}. 
Without modifying the {conventional DFT-based approaches}, new XC functionals have also been {constructed} to {approximate} the exact {density}. 
Typical examples include $meta$-generalized gradient approximation ($meta$-GGA) that involves the second gradient of the density (TPSS\cite{PhysRevLett.91.146401}, SCAN\cite{PhysRevLett.115.036402}, M06-L\cite{zhao2006jcp}, and M06-2X\cite{Zhao2008}, {\it etc.}), and range-separation treatment that partially or fully reproduces the asymptotic density decay by implementing the Hartree--Fock (HF) exchange in the long range {using an empirically determined range-separation parameter, $\omega$} (CAM-B3LYP\cite{YANAI200451}, CAM-QTP\cite{jin2016jcp}, $\omega$B97X-V\cite{C3CP54374A}, $\omega$B97M-V,\cite{Mardirossian2016jcp} LRC-$\omega$PBE\cite{vydrov2006jcp,Rohrdanz2008jcp}, and LRC-$\omega$PBEh\cite{vydrov2006jcp2,Rohrdanz2009},  {\it etc.}). 

More recently, several optimally tuned {(OT) versions} of range-separated XC functionals have been proposed and applied {to large molecules}, including OT-BNL\cite{B617919C}, OT-$\omega$B97XD\cite{ct2003447}, and OT-$\omega$PBEh,\cite{PhysRevB.86.205110} {\it etc.}
These functionals allow the users to optimize $\omega$ in a {\it non-empirical}, system-dependent fashion, and can be considered as a ``black box'' in which the molecular structure serves as the exclusive input.
{As an outstanding example}, Kronik, Baer, and co-workers developed the most popular {optimal tuning} scheme\cite{B617919C,ja8087482,baer2010tuned} based on Iikura's idea of range separation\cite{1.1383587} and Koopmans' theorem.\cite{KOOPMANS1934104}  
They enforce an agreement between the vertical ionization potential ($I_{\perp}$) and the negative eigenvalue for the highest occupied molecular orbital ($-\varepsilon^{(N)}_{\rm HOMO}$), which is satisfied by the exact XC functional.
These optimally tuned functionals have successfully predicted {diverse} (pseudo-)one-electron properties such as the fundamental band gap, the photoelectron spectrum, and the charge transfer excitation energy. 
Also, they behave reasonably well for spectroscopic and photochemical properties like the optical band gap ($E_{\rm S} = E_{{\rm S}_1} -  E_{{\rm S}_0}$), the phosphorescence energy ($E_{\rm T} = E_{{\rm T}_1} -  E_{{\rm S}_0}$), {and the singlet--triplet gap ($\Delta E_{\rm ST} = E_{{\rm S}_1} -  E_{{\rm T}_1}$)}.\cite{jp1057572,1.3656734,PhysRevB.84.075144,ct2002804,ct2009363,ar500021t,ct5000617,acs.jctc.7b01058}
However, as these functionals are not explicitly tuned for low-lying {energetics}, their {for accuracy these quantities is} not expected to be comparable with {(pseudo-)}one-particle properties. 

Motivated by reaching more accurate predictions of spectroscopically and photochemically important excited states, we proposed the novel triplet tuning scheme based on the lowest triplet excited state (T$_1$). {Our} idea {borrows a page from} the above-mentioned optimal tuning {prescription},\cite{B617919C,ja8087482,baer2010tuned} in which we optimized the values of one or two selected parameters in a {\it non-empirical} manner by matching a particular energy level evaluated using two different approaches. Instead of equaling $I_\perp$ to $-\varepsilon^{(N)}_{\rm HOMO}$, we enforced the agreement in the lowest triplet excitation energy ($E_{\rm T}$) between two excited-state DFT {formulations}, namely $\Delta$SCF\cite{PhysRevB.78.075441} and TDDFT.\cite{PhysRevLett.52.997}

In the present study, we developed the first triplet tuned {(TT)} functional in this series, TT-$\omega$PBEh, by imitating the formula of LRC-$\omega$PBEh\cite{vydrov2006jcp,Rohrdanz2008jcp} and leaving two {adjustable }parameters: $\omega$ and the percentage of the HF exchange in the short range, $C_{\rm HF}$. 
The {methodological} details will be provided in Sec. \ref{sec:theory}. 
In Sec. \ref{sec:result}, the accuracy and stability of TT-$\omega$PBEh will be analyzed using four groups of large organic molecules which possess rich spectroscopic and photochemical information and are notoriously difficult for theoretical investigations, even for low-lying excited states.
As will be illustrated by {Sec. \ref{sec:result}}, TT-$\omega$PBEh provides the most excellent agreement with experiments regarding $E_{\rm T}$, $E_{\rm S}$, and {$\Delta E_{\rm ST}$}, due to its accurate reproduction of the screening of the electron-hole interaction. Concerning $I_{\perp}$, we will show that the {incorporation} of triplet tuning will {enhance} the predictive power of the conventional optimal tuning prescription. The conclusion and the future work will be discussed in Sec. \ref{sec:conc}.

\section{Theory}
\label{sec:theory}

\subsection{Triplet Excitation Energy} 
\label{sec:t1}

Constructed by Kohn and Sham (KS-DFT), the most popular version of DFT {allows us to} evaluate the energy and density based on molecular orbitals.\cite{Kohn1965,parr1994density} 
In this framework, there exists a universal, variational functional that provides the correct ground-state energy from the exact ground-state density. 
It is {also} straightforward to generalize the KS functional to one that gives the lowest {state} of a system with {\it any given symmetry} when constraining the density to the corresponding symmetry\cite{PhysRevA.47.2783}.
As a result, ground-state DFT is able to access {an} excited state that {is} the lowest state within a particular symmetry.
The most common case of this has to do with the spin configuration. 
When the true ground state for a closed-shell molecule {(like most organic molecules)} is a {spin-restricted} singlet (S$_0$), the lowest state with $M_S=\pm1$ must be a triplet (T$_1$). 
{These closed-shell species} can achieve T$_1$ by promoting one electron from HOMO to the lowest unoccupied molecular orbital (LUMO) and flipping its spin accordingly. {S$_0$ and T$_1$} can be most easily resolved by doing a restricted DFT (RDFT) calculation with $M_S = 0$ in the former case and an UDFT calculation with $M_S = \pm1$ in the latter. \cite{pople1954jcp}
Therefore, the triplet excitation energy, $E_{\rm T}$, which corresponds to an experimental measure of $E_{{\rm T}_1}$, can be evaluated following the $\Delta$SCF scheme, 
\begin{equation}
E_{\rm T}^{\Delta{\rm SCF}} = E_{{\rm T}_1}^{\rm UDFT} -  E_{{\rm S}_0}^{\rm RDFT}.\label{eq:dscf}
\end{equation}
{Under} the exact functional, {Eq. (\ref{eq:dscf})} gives the true $E_{\rm T}$.
In an alternative {strategy}, T$_1$ can be treated as an excited state using linear-response TDDFT.\cite{PhysRevLett.52.997,mcweeny1992methods}
With the exact XC kernel TDDFT yields {exact} energies for \emph{all} excited states including T$_1$. \cite{Marques2004}
As such, $E_{\rm T}$ can also be obtained using
\begin{equation}
E_{\rm T}^{\rm TDDFT}  = E_{{\rm T}_1}^{\rm TDDFT} - E_{{\rm S}_0}^{\rm RDFT}.\label{eq:tddft}
\end{equation}

\subsection{Triplet Tuning Protocol}
\label{sec:tuning}

Since {RDFT, UDFT, and TDDFT are all formally exact approaches}, 
Eqs. (\ref{eq:dscf}) and (\ref{eq:tddft}) would provide identical values of $E_{\rm T}$ assuming the exact XC functional and kernel are used,
\begin{equation}
E_{\rm T}^{\Delta{\rm SCF}} \equiv E_{\rm T}^{\rm TDDFT}. \label{eq:cond} 
\end{equation}
However, we note that for an approximate functional, Eq. (\ref{eq:cond}) does not necessarily hold, and the difference between $E_{\rm T}^{\Delta{\rm SCF}}$ and $E_{\rm T}^{\rm TDDFT}$ {strictly }measures one aspect of the inaccuracy of the underlying functional.
Eq. (\ref{eq:cond}) is analogous to the {Koopmans'} theorem that the frontier orbital energies, $-\varepsilon^{(N)}_{\rm HOMO}$ and $-\varepsilon^{(N)}_{\rm LUMO}$, should match $I_\perp$ and the vertical electron affinity ($A_\perp$) {respectively} under the exact XC functional.\cite{parr1994density,jensen2017introduction,KOOPMANS1934104,B617919C,ja8087482,baer2010tuned}
{However, our case differs from the conventional optimal tuning by using an excited state as a target of interest rather than an ionized ground state.}

Thus, {learning} from the optimal tuning approach,\cite{B617919C,ja8087482,baer2010tuned} our key {concept} is to tune the underlying functional on a molecule-by-molecule basis so that Eq. (\ref{eq:cond}) is fulfilled as {rigorously} as possible.
In practice, we minimized the following objective function,
\begin{equation}
J^2_{\rm TT} = \left(E_{\rm T}^{\Delta{\rm SCF}} - E_{\rm T}^{\rm TDDFT}\right)^2, \label{eq:comp}
\end{equation}
in order to obtain optimal parameters.
We refer to this approach as ``triplet tuning'', which can be utilized to {improve the performance of} any XC functional.
We should note that TT {\it does not introduce any empiricism} to the underlying functional, as the calculated {$E_{\rm T}^{\Delta{\rm SCF}}$ and $E_{\rm T}^{\rm TDDFT}$ are} not matched to {any} experimental value -- instead the parameters are {optimized} to achieve an internal consistency {of $E_{\rm T}^{\Delta{\rm SCF}}$ and $E_{\rm T}^{\rm TDDFT}$} that the exact functional is known to possess.

{Similar to} optimal tuning, our {triplet tuning} protocol can be partly justified based on the adiabatic connection\cite{PhysRevLett.94.043002}:
the exact {$E_{\rm XC}$} is related to the coupling-constant-average of the electron-electron
interaction energy, for which the value at the low-coupling limit is just the exchange energy ($E_{\rm X}$) {and the stronger coupling is embodied in the correlation energy ($E_{\rm C}$)}. 
The precise weight of weakly and strongly interacting systems in the integral depends 
on the system and {the} range of the interaction and must typically be determined semi-empirically 
\cite{PhysRevA.70.062505,vydrov2006jcp,vydrov2006jcp2}.
The triplet tuning scheme offers the attractive possibility of performing {the} adiabatic connection
in a \emph{non-empirical} fashion, that is, minimizing $J_{\rm TT}^2$ in Eq. (\ref{eq:comp}) to define the optimal form of {$E_{\rm XC}$}. 
{In the present study}, we focused on the tuning of $E_{\rm X}$ as its weight is {significant }heavier than $E_{\rm C}$ for most weakly interacting molecules{ of interest}.

Bearing this in mind, we performed triple tuning on the basis of the range-separated hybrid formula of PBE and HF,\cite{vydrov2006jcp,vydrov2006jcp2,chai2008jcp,Rohrdanz2008jcp,Rohrdanz2009} and named it as TT-$\omega$PBEh. TT-$\omega$PBEh {was} designed to predict the accurate density in both short and long ranges and to reproduce the correct electron-hole interactions in molecules. Its formula separates $E_{\rm X}$ into the short-range (SR) and long-range (LR) contributions,\cite{1.1383587}
\begin{equation}
E_{\rm XC} = E^{\rm SR}_{\rm X} + E^{\rm LR}_{\rm X} + E_{\rm C}. \label{eq:rsh}
\end{equation}
{This can be accomplished} by re-expressing the Coulomb operator, $1/r_{12}$, {into}\cite{1.1383587,szabo2012modern}
\begin{equation}
\frac{1}{r_{12}} = \underbrace{\frac{1-\rm{erf}(\omega r_{12})}{r_{12}}}_\text{SR} + \underbrace{\frac{\rm{erf}(\omega r_{12})}{r_{12}}}_\text{LR}, \label{eq:omega}
\end{equation}
in which $r_{12} = |\vec{r}_1-\vec{r}_2|$ is the interelectron distance, and ``erf'' represents the Gauss error function.\cite{abramowitz1964handbook}
Eq. (\ref{eq:omega}) introduces $\omega$ as the range-separation parameter, which is the reciprocal of the distance at which $E^{\rm SR}_{\rm X}$ transitions to $E^{\rm LR}_{\rm X}$\cite{1.1383587}. 
As the non-local HF exchange exhibits the correct asymptotic behavior,\cite{szabo2012modern} we selected 
\begin{equation}
E^{\rm LR}_{\rm X} \equiv E^{\rm HF}_{\rm X}.
\end{equation} 
Meanwhile, $E^{\rm SR}_{\rm X}$ takes a hybrid form of HF and a selected (semi-)local DFT exchange functional in order to balance the localization error from HF and delocalization error from this DFT functional:\cite{PhysRevLett.100.146401} 
\begin{equation}
E^{\rm SR}_{\rm X} = C_{\rm HF} E^{\rm SR}_{\rm X,HF} + C_{\rm DFT} E^{\rm SR}_{\rm X,DFT}. \label{eq:sr}
\end{equation}
$C_{\rm HF}$ and $C_{\rm DFT}$ represent the fractions of the HF and DFT components in $E^{\rm SR}_{\rm X}$, and they are related to each other through the uniform electron gas (UEG) constraint,\cite{PhysRev.136.B864,Kohn1965,parr1994density,1.1390175} 
\begin{equation}
C_{\rm HF} + C_{\rm DFT} = 1. \label{eq:chf}
\end{equation}
Eqs. (\ref{eq:sr}) and (\ref{eq:chf}) suggest the second to-be-determined parameter, $C_{\rm HF}$.
In {existing }range-separated hybrid formulations, $E_{\rm C}$ and $E^{\rm SR}_{\rm X,DFT}$ can be selected {among} LDA\cite{Kohn1965}, GGA,\cite{PhysRevA.38.3098,PhysRevB.28.1809,PhysRevB.46.6671} or $meta$-GGA\cite{PhysRevLett.91.146401,PhysRevLett.115.036402,zhao2006jcp,Zhao2008} {expressions} based on the {requirement} of the system. In {TT-$\omega$PBEh} one utilizes the Perdew--Burke--Ernzerhof (PBE) formulas for $E^{\rm SR}_{\rm X,DFT}$ and $E_{\rm C}$.\cite{PhysRevLett.77.3865} 

In the present study, we {examined} three variants of TT-$\omega$PBEh, one with both parameters tuned ({the original} TT-$\omega$PBEh), and two with {a single} tunable parameter, either $\omega$ (TT-$\omega$PBEh$\omega$, similar to LRC-$\omega$PBEh\cite{Rohrdanz2009}) or $C_{\rm HF}$ (TT-$\omega$PBEh$C$). 
{Such a comparison} enables us 
to validate the {insufficiency of the single-parameters versions and the }necessity of tuning both $\omega$ and $C_{\rm HF}$. 

\subsection{One-Electron Property}
\label{sec:ot}

As was discussed in Sec. \ref{sec:intro}, our triplet tuning scheme {reproduces}  correct electron-hole interactions while the optimal tuning prescription of Kronik and Baer provides accurate one-electron properties.\cite{B617919C,ja8087482,baer2010tuned} Therefore a performance comparison between these two approaches {is} necessary. 
To {realize this} we also {re-}constructed the range-separated hybrid OT-$\omega$PBEh functional 
as described in Eqs. (\ref{eq:rsh})--(\ref{eq:chf}). The objective function was established following Koopmans' theorem,\cite{KOOPMANS1934104,B617919C,ja8087482,baer2010tuned} 
\begin{equation}
J^2_{\rm OT} = \left[I_\perp+\varepsilon_{\rm HOMO}^{(N)}\right]^2+\left[A_\perp+\varepsilon_{\rm HOMO}^{(N+1)}\right]^2, \label{eq:jia}
\end{equation}
where $\varepsilon_{\rm HOMO}^{(N+1)}$ stands for the HOMO energy of the anionic species {with an identical molecular structure to the neutral species} and is a substitute of $\varepsilon_{\rm LUMO}^{(N)}$ due to the incorrect physical interpretation of virtual orbitals in the KS {framework}.\cite{parr1994density,jensen2017introduction} 
$I_\perp$ and $A_\perp$ are evaluated using $\Delta$SCF,
\begin{eqnarray}
I_\perp = E_{{\rm C}^+}^{\rm UDFT} - E_{{\rm S}_0}^{\rm RDFT}, \label{eq:ip} \\
A_\perp = E_{{\rm S}_0}^{\rm RDFT} - E_{{\rm A}^-}^{\rm UDFT}, \label{eq:ea}
\end{eqnarray}
where C$^+$ and A$^-$ represent cationic and anionic species {respectively}. 

In an attempt to take into account both one-electron properties and electron-hole interactions using the very same functional, we also combined the recipes of {OT-$\omega$PBEh} and TT-$\omega$PBEh to formulate another functional {termed as} mix-$\omega$PBEh. 
Its objective function is consequently
\begin{equation}
J_{\rm mix}^2 = \frac{J_{\rm TT}^2+J_{\rm OT}^2}{2}. \label{eq:jmix}
\end{equation}

\subsection{{Molecular} Test Sets}

\begin{figure}[t]
	\centering
	\includegraphics[width=16cm]{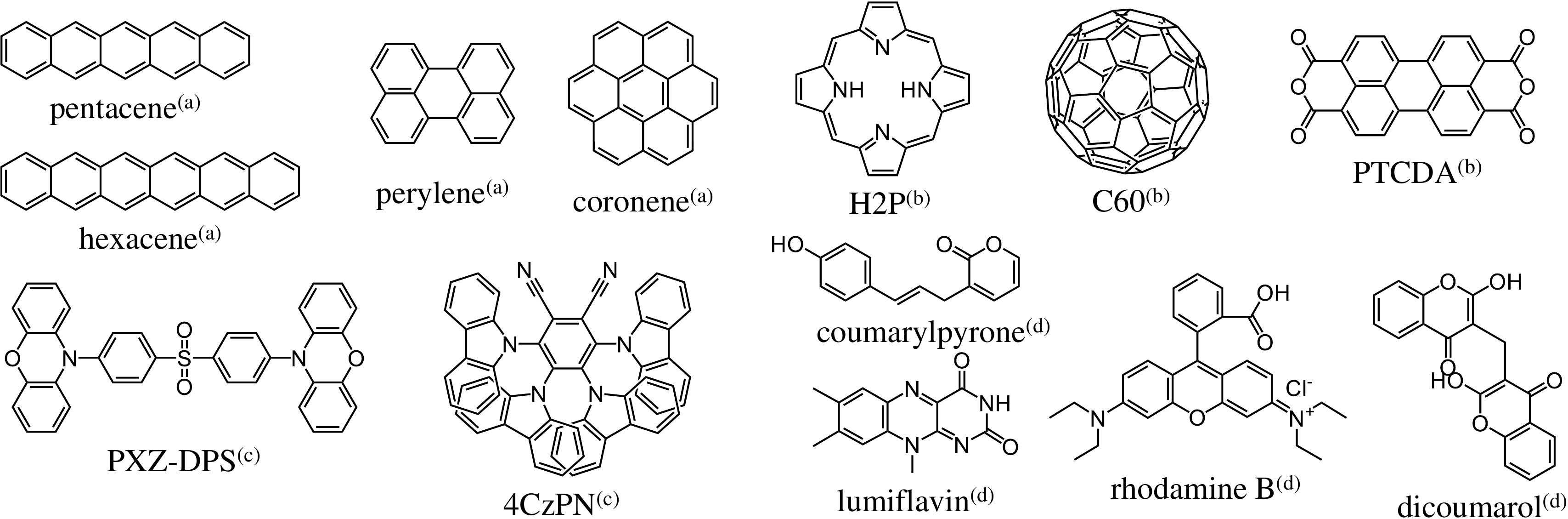}
	\caption{Representative molecules from the test sets of (a) PAH, (b) OPV, (c) TADF, and (d) BIO.}
	\label{fig:all_mole}
\end{figure}

In order to evaluate the performance and applicability of TT-$\omega$PBEh and mix-$\omega$PBEh and compare their behaviors to OT-$\omega$PBEh and existing functionals, 
we selected a total of 110 organic molecules{, most of which posses semiconducting structures or charge transfer characters} and reliable experimental measurements of $E_{\rm T}$, $E_{\rm S}$, $\Delta E_{\rm ST}$, and $I_\perp$. 
In spite of being closed-shell, {these species} are notorious {as challenging} cases for theoretical investigations. {M}olecules were divided into four test sets based on their {structural features}, excited-state properties, and real-life applications, including polycyclic aromatic hydrocarbons (PAH),\cite{murov1993handbook,doering1969jcp,BOLOVINOS1984240,HELLNER197671,CLAR1950116,ALLAN1989219,mcclure1951jcp,mcclure1949jcp,PHP:PHP31,moodie1954jcp,mcglynn1964jcp,perkampus1992uv,birks1970photophysics,ANGLIKER1982208,CLARKE1969309,paris1961jcp,duncan1981jcp,QUA:QUA21237,Lyapustina2000jcp,SCHAFER197591,ando2007jcp,ELAND1972214,becker1966jcp,schmidt1977jcp,hagen1988ac,masaaki2007jpca,STAHL1984613,crocker1993jacs,SCHIEDT199718,SHCHUKA198987,boschi1974jcp,clar1981jacs,hajgato2008jcp,obenland1975jacs} organic photovoltaics (OPV) materials,\cite{hung1991jpc,arbogast1991jpc,PHP:PHP31,F29787401870,GOUTERMAN197488,vincett1971jcp,petruska1961jcp,palummo2009jcp,dvorak2012jcp,becker1996jpc,seixas1999jcp,shizuka1982jpc,marchetti1967jacs,murov1993handbook,kearns1966jacs,ghoshal1981effects,kuboyama1967phosphorescence,borkman1967jcp,KANDA19611,berlman1971handbook,lim1970jcp,bulliard1999jpca,adams1973jacs,grimme1999jcp,carsey1979jacs,A806072J,wasserberg2005jpcb,mcclure1951jcp,dyck1962jcp,SALTIEL1980233,beljonne1996jacs,PHP:PHP655,BURKE1973574,Herkstroeter1975jacs,perkampus1992uv,tway1982jpc,zander1985z,goodman1959jcp,FT9949000411,JR9590002753,najbar1980japp,suga1982static,anderson1994jacs,F29736901155,ELAND1969471,muigg1996jpb,huang2014jcp,Dewar431,WOJNAROVITS1981511,F29736901808,KHANDELWAL1975355,butler1980jacs,edward1998jpcrd,F29817701621,PhysRevB.84.075144,chen1991jacs,PASINSZKI201085,berkowitz1979jcp,C004332J,PhysRevB.73.195208,OMS:OMS1210160409,wentworth1975jpc,paul1989jacs,polevoi1987formation,OMS:OMS1210141011,MAEYAMA200818,OMS:OMS1210080121,kobayashi1983conformational,BBPC:BBPC19740780503,debies1974ic,potapov1971photoionization,lipert1990jcp,hudson1976jacs,C0CP02712J,lu1992jpc,shiedt2000jcp,JOCHIMS1992159,SCHAFER197591,QUA:QUA560180739,VANDENHAM1972447,dillow1989cjc,HLCA:HLCA19720550131} thermally activated delayed fluorescence (TADF) emitters,\cite{huang2013jctc,C3TC30699B,C3TC31936A,zhang2014efficient,hait2016} and $\pi$-conjugated bioorganic (BIO) molecules.\cite{longworth1966jcp,Daniels675,lin1980jacs,aflatooni1998jpca,gueron1967jcp,nguyen2004jpca,li2006jcp,SCHIEDT1998511,DOUGHERTY1978379,dougherty1976jacs,hendricks1996jcp,yu1981jpc,kearns1971jacs,dvornikov1979phosphorescence,C39800000533,F19848001151,becker1971jacs,thomson1969jcp,C39920001175,PHP:PHP17,marsh1970jacs,hansen1975mass,case1970jcp,P29860001217,F29868200135,mantulin1973jacs,PHP:PHP31,usacheva1984rjpc,HLCA:HLCA19750580619,shunsuke2004jcp,PHP:PHP215,0036-021X-52-1-R03,berlman1971handbook,sikorska2004jpca,PALMER1980243,timoshenko1981ionization} 
{M}olecules in the PAH, OPV, and BIO sets possess locally excited {T$_1$ and S$_1$} states with different extents of delocalized $\pi$-bonds, while TADF emitters have charge transfer {T$_1$ and S$_1$} states.
The structures of all molecules are listed in Figs. S1--S9 in the Supporting Information, and representative species are provided in Fig. \ref{fig:all_mole}.

\subsection{Computational Details}
\label{sec:details}

\begin{figure}[h]
	\centering
	\includegraphics[width=8cm]{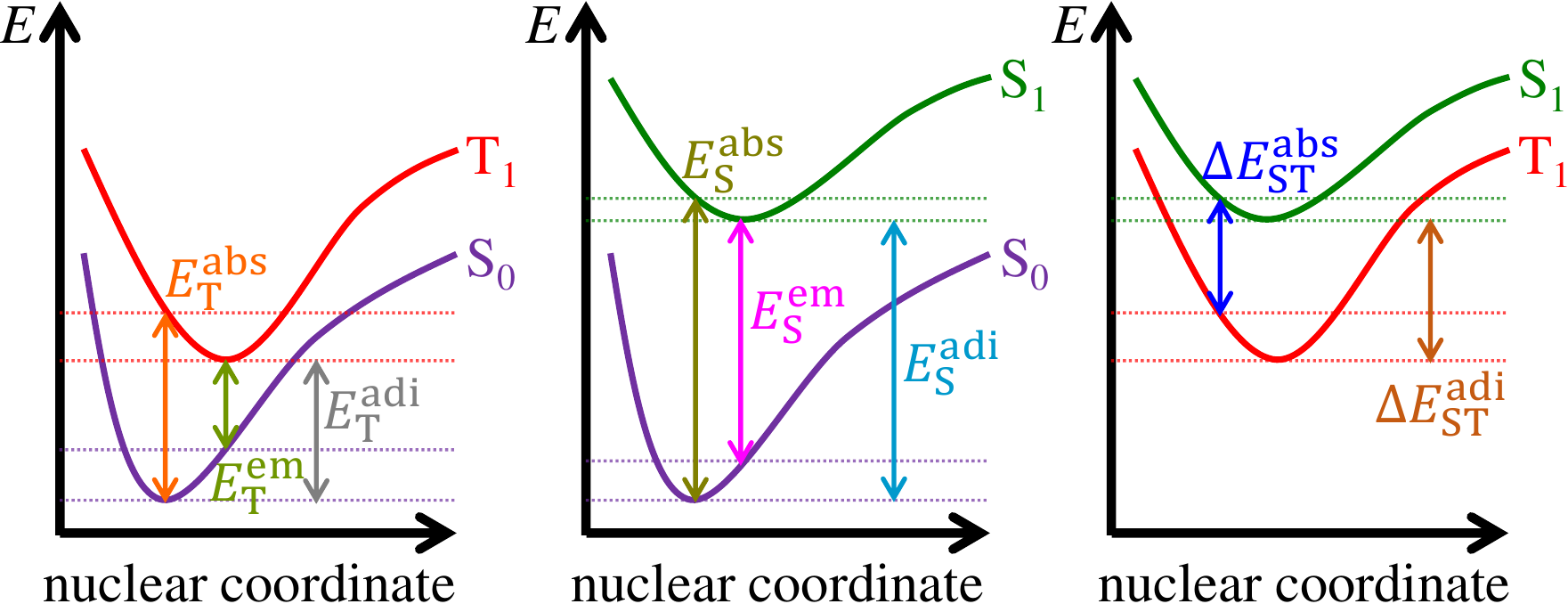}
	\caption{Geometric variants of $E_{\rm T}$, $E_{\rm S}$, and $\Delta E_{\rm ST}$ applied in the present study. ``Abs'', ``em'' and ``adi'' represent the absorption, emission, and adiabatic energy gaps.}
	\label{fig:geometry}
\end{figure}

With nuclear relaxation in consideration, we evaluated $E_{\rm T}$, $E_{\rm S}$, and $\Delta E_{\rm ST}$ under three geometric variants, ``absorption'' (abs), ``emission'' (em) and ``adiabatic'' (adi) (Fig. \ref{fig:geometry}), and compared each of them to its experimental counterpart. $I_\perp$, $A_\perp$, $-\varepsilon^{(N)}_{\rm HOMO}$, and $-\varepsilon^{(N+1)}_{\rm HOMO}$ were evaluated at the S$_0$ geometry. In all cases, the molecular configurations associated with S$_0$ and T$_1$ were {optimized} using RDFT with $M_S=0$ and UDFT with $M_S=\pm1$ respectively,\cite{PhysRevB.37.785} while those of the first singlet excited states (S$_1$) were {optimized} using ROKS\cite{kowalczyk2013jcp}. All {geometry} optimizations utilized the B3LYP functional\cite{PhysRevA.38.3098,PhysRevB.37.785,becke1993jcp}. 
System-dependent $\omega$ and $C_{\rm HF}$ {were {\it non-empirically} tuned} at the optimized S$_0$ geometries by minimizing Eqs. (\ref{eq:comp}), (\ref{eq:jia}), or (\ref{eq:jmix}). 
For {single}-parameter versions of TT-$\omega$PBEh, we optimized $C_{\rm HF}$ at $\omega = 0.200a_0^{-1}$ (TT-$\omega$PBEh$\omega$, $a_0\equiv$ bohr), or optimized $\omega$ at $C_{\rm HF} = 0.20$ (TT-$\omega$PBEh$C$), {by performing a gold-section search of} the one-dimensional {minimum}.\cite{kiefer1953sequential,press2007numerical} 
For TT-$\omega$PBEh, OT-$\omega$PBEh, and mix-$\omega$PBEh with tunable $\omega$ and $C_{\rm HF}$, two-dimensional minimizations were carried out using the simplex algorithm.\cite{press2007numerical}

{Based on the optimized values of} $\omega$ and $C_{\rm HF}$, $E_{\rm T}$, $E_{\rm S}$, $\Delta E_{\rm ST}$, and $I_\perp$ were evaluated using single-point $\Delta$SCF, TDDFT, and ROKS. {For molecules that suffer from failed diagonalization of the linear-response matrix}, the Tamm--Dancoff approximation (TDA)\cite{HIRATA1999291} was included in the TDDFT calculations. $\varepsilon^{(N)}_{\rm HOMO}$ and $\varepsilon^{(N+1)}_{\rm HOMO}$ were extracted as eigenvalues of KS orbitals produced in the ground-state calculations of neutral and anionic species. All {such single-point} energies were compared with the experimental measurements performed at the appropriate geometric variants shown in Fig. \ref{fig:geometry}{, and the accuracy was described using the absolute error (AE)}:
\begin{equation}
{\rm AE} (X)= \left|E^{\rm calc}_X - E^{\rm expt}_X\right|.
\end{equation}
If {more than one} geometric variants {for any species} are experimentally available, we averaged {these} AEs . The accuracy of any XC functional was calibrated using the mean absolute error (MAE) that was averaged over all molecules within each test set. 

Besides OT-$\omega$PBEh, ten existing functionals were also compared with TT-$\omega$PBEh and mix-$\omega$PBEh, including HF,\cite{szabo2012modern} B3LYP,\cite{beckejcp1993-2} CAM-B3LYP,\cite{YANAI200451} PBE,\cite{PhysRevLett.77.3865} PBE0,\cite{adamo1999jcp} LRC-$\omega$PBE ($\omega = 0.300a_0^{-1}$ and $C_{\rm HF} = 0.00$),\cite{Rohrdanz2008jcp} LRC-$\omega$PBEh ($\omega = 0.200a_0^{-1}$ and $C_{\rm HF} = 0.20$),\cite{Rohrdanz2009} TPSS,\cite{PhysRevLett.91.146401} M06-2X\cite{Zhao2008}, and M06-L.\cite{zhao2006jcp} 
All calculations {reported in the present study} used the cc-pVDZ basis set\cite{dunning1989jcp} and the Q-Chem 4.4 package.\cite{QCHEM4}

\section{Results and Discussions}
\label{sec:result}

\subsection{Optimized Parameters}
\label{sec:para}

\subsubsection{Necessity of {Double} Parameters}
\label{sec:1d}

In the {current sub}section, we will show the necessity to set both $\omega$ and $C_{\rm HF}$ as tunable in TT-$\omega$PBEh by presenting the insufficiency of TT-$\omega$PBEh$\omega$ and TT-$\omega$PBEh$C$ based on two {oligomer }series from our test sets: oligoacenes ($n=1-6$, PAH) and $\alpha$-oligothiophenes ($n=1-7$, OPV). 

\begin{figure}[h]
	\centering
	\includegraphics[width=8cm]{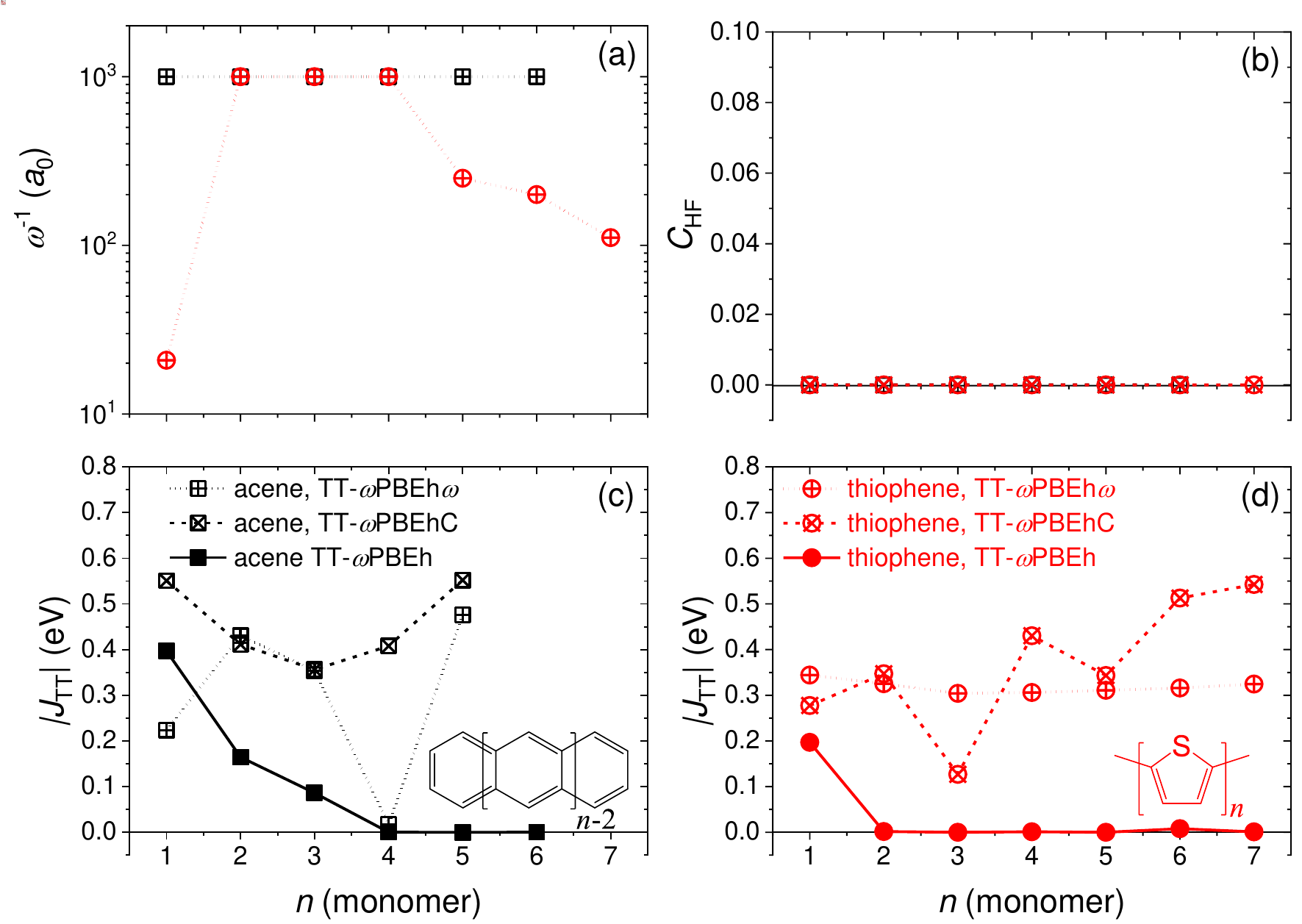}
	\caption{(a) Optimized $\omega^{-1}$ ($a_0^{-1}$) in TT-$\omega$PBEh$\omega$ ($C_{\rm HF}=0.20$, hollow with plus) for oligoacenes (black square) and $\alpha$-oligothiophenes (red circle). (b) Optimized $C_{\rm HF}$ in TT-$\omega$PBEh$C$ ($\omega=0.200a_0^{-1}$, hollow with cross) for oligoacenes and $\alpha$-oligothiophenes. (c) {Minimized} $|J_{\rm TT}|$ {(eV)} for oligoacenes {using TT-$\omega$PBEh$\omega$, TT-$\omega$PBEh$C$,} and TT-$\omega$PBEh{ (solid)}. (d) {Minimized} $|J_{\rm TT}|$ for $\alpha$-oligothiophenes {using TT-$\omega$PBEh$\omega$, TT-$\omega$PBEh$C$,} and TT-$\omega$PBEh.} 
	\label{fig:1d}
\end{figure}

The reciprocal of $\omega$ ($\omega^{-1}$) provides the interelectron distance where the {short-range} PBE--HF hybrid exchange transitions to {the long-range pure} HF{ exchange}. A small $\omega$ represents a small overall {HF} fraction, and {\it vice versa}. 
Fig. \ref{fig:1d}(a) illustrates optimal $\omega^{-1}$ for TT-$\omega$PBEh$\omega$ for {both oligomer series}. Although the trend of the overall HF fraction is not obvious {for either series except for an increase in larger $\alpha$-oligothiophenes ($n\ge5$)}, {a large $\omega^{-1}$ ($\omega^{-1} > 100 a_0$)} in optimized TT-$\omega$PBEh$\omega$ indicate a negligible HF contribution, or more possibly, the non-existence of a minimum on the one-dimensional surface of $J_{\rm TT}^2(\omega)$ at a typical value of $C_{\rm HF} =0.20$. Similarly, TT-$\omega$PBEh$C$ presents a constant of $C_{\rm HF}=0.00$ for both oligomer {series} (Fig. \ref{fig:1d}(b)), also showing the difficulty in finding the optimal $C_{\rm HF}$ at a typical value of $\omega=0.200a_0^{-1}$. 

To evaluate how well the exact {triplet tuning} constraint (Eq. (\ref{eq:cond})) is satisfied {by} TT-$\omega$PBEh$\omega$, TT-$\omega$PBEh$C$, {and TT-$\omega$PBEh under optimized $\omega$ or $C_{\rm HF}$}, we also compare the value of $|J_{\rm TT}|$, the {square root of the }minimized {$J_{\rm TT}^2$ (Eq. (\ref{eq:comp}))}, 
in Fig. \ref{fig:1d}(c) and (d).
TT-$\omega$PBEh$\omega$ and TT-$\omega$PBEh$C$ both exhibit huge $|J_{\rm TT}|$'s for most molecules (as large as $\simeq$0.6 eV), while TT-$\omega$PBEh presents very small $|J_{\rm TT}|$'s ($< 0.1$ eV) except for {the smallest} benzene, naphthalene, and thiophene. This {observation further validates} that a reasonable minimum of  $J^2_{\rm TT}$ {\it does not necessarily exist} unless at least two parameters are used in the triplet tuning procedure. 

\subsubsection{{Independent }Parameters}
\label{sec:ind}

\begin{figure}[h]
	\centering
	\includegraphics[width=8cm]{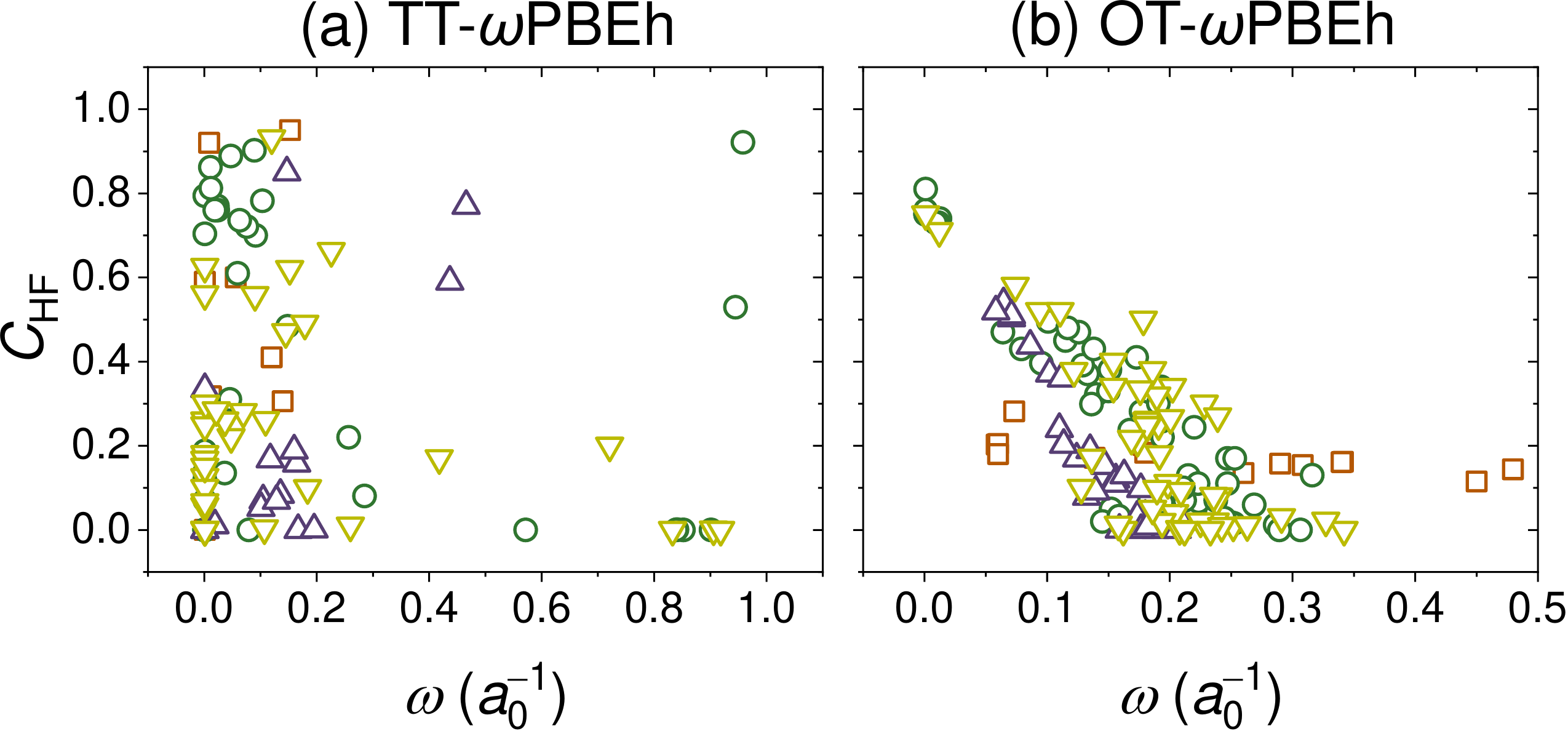}
	\caption{Correlation between optimized $\omega$ ($a_0^{-1}$) and $C_{\rm HF}$ for (a) TT-$\omega$PBEh and (b) OT-$\omega$PBEh.}
	\label{fig:parameters}
\end{figure}

As both $\omega$ and $C_{\rm HF}$ are related to the overall fractions of PBE and HF in our long-range-corrected (LRC) hybrid formula, it is necessary to {eliminate any} possible correlation between $\omega$ and $C_{\rm HF}$ for TT-$\omega$PBEh. In Fig. \ref{fig:parameters} we summarize optimized $\omega$ and $C_{\rm HF}$ for all molecules {in question}. 
For {the majority of species}, $\omega<0.5a_0^{-1}$, indicating that the long-range HF exchange usually takes effect when $r_{12}>2a_0$. In addition, $C_{\rm HF}$ can range from 0 to 1, but {most} molecules need $C_{\rm HF}< 0.4$ to reach the desired accuracy. 

To satisfy Koopmans' theorem, molecules with similar chromophoric sizes are expected to possess similar overall HF exchange fractions. Therefore in OT-$\omega$PBEh a smaller value of $\omega$ is usually compensated by a larger value of $C_{\rm HF}$, and {this expected negative correlation is presented in} Fig. \ref{fig:parameters}(b). 
For TT-$\omega$PBEh, on the other hand, $\omega$ and $C_{\rm HF}$ are statistically independence from each other ({Fig. \ref{fig:parameters}(a)}) and can thus be {\it non-empirically} and {\it mutually} tuned, and molecules with similar-sized chromophores might end up with very different sets of optimized parameters.

\subsection{Triplet Excitation Energy}
\label{sec:st}

As was mentioned in Secs. \ref{sec:intro} and \ref{sec:theory}, our proposed TT-$\omega$PBEh functional {arrives at a} {\it non-empirical} matching of ${E_{\rm T}}$ between $\Delta$SCF and TDDFT (Eq. (\ref{eq:cond})). {Accordingly} the accuracy of $E_{\rm T}$ {becomes} a natural calibration of TT-$\omega$PBEh. Given the well-minimized $J_{\rm TT}^2$, we expected TT-$\omega$PBEh to reach the best approximation of the exact XC functional and thus to provide the most accurate prediction of ${E_{\rm T}}$ among all XC functionals in {comparison}, especially for {badly-behaved} molecules with large-scale $\pi$-conjugations or {significant} charge transfer excitations. 
Because TT-$\omega$PBEh allows us to evaluate $\omega$ and $C_{\rm HF}$ {\it independently} from any experimental {measurements}, it presents {a} minimal fitting artifact and can be safely compared to experiments.

As was described in Sec. \ref{sec:theory}, 
the error of {a functional is} 
characterized using its MAEs. 
We report the MAEs of $E_{\rm T}$'s 
using three single-point DFT variants: $\Delta$SCF, TDDFT, and TDDFT/TDA 
(Tables S1, S5, S9, and S11). 
We also summarize {TDDFT-evaluated MAEs} in the left part of Table \ref{tab:st_summary}.
In these tables{ and those that follow}, the {\bf bold}, \underline{underlined}, and {\it italic} numbers represent the smallest, second smallest, and largest MAEs within each column, respectively.

{In an} ideal situation, all DFT variants provide identical values for $E_{\rm T}$. However, this is difficult to accomplish as the minimized $J_{\rm TT}^2$ 
can hardly reach a rigorous zero.
To quantify the error arising from a non-zero $J_{\rm TT}^2$, as well as to characterize the lower and upper bounds of {M}AEs, for every single molecule we evaluated $E_{\rm T}$ using all three variants and picked the best and worst results (the smallest and largest AEs). The averages of these {\it best} and {\it worst} AEs were defined as the {\it best} and {\it worst} MAEs, respectively, and are included along with their differences for 
all $E_{\rm T}$'s in Tables S1, S5, S9, and S11.
{\it Worst} MAEs are {summarized} in the right part of Table \ref{tab:st_summary}. 

\begin{figure}[t]
	\centering
	\includegraphics[width=16cm]{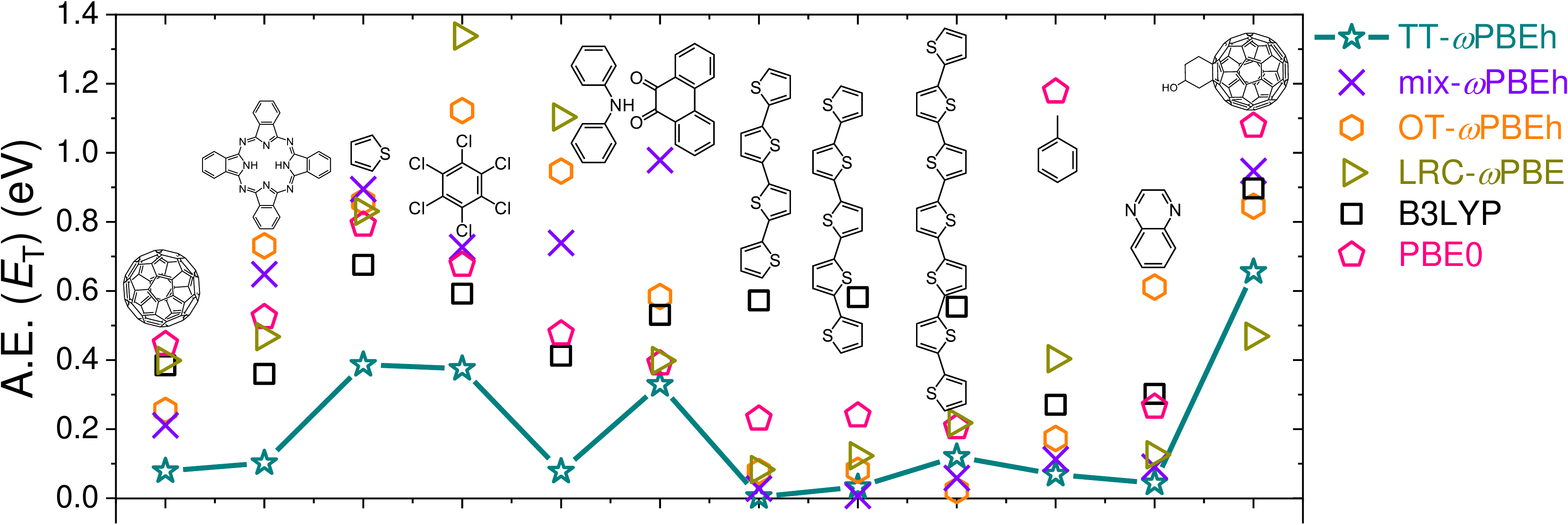}
	\caption{{AEs} of {$E_{\rm T}$'s} (eV) {are} illustrated for selected OPV materials. The results are compared among TT-$\omega$PBEh (dark cyan star), mix-$\omega$PBEh (violet cross), OT-$\omega$PBEh (orange hexagon), LRC-$\omega$PBE (dark yellow right triangle), B3LYP (black square), and PBE0 (pink pentagon).} 
	\label{fig:pah_st}
\end{figure}

In addition, it is worthwhile to {make a systematic comparison between our proposed TT-$\omega$PBEh} and {existing functionals, especially} the very popular {hybrid} B3LYP,\cite{PhysRevA.38.3098,PhysRevB.37.785,becke1993jcp} {PBE,\cite{PhysRevLett.77.3865} and PBE0\cite{adamo1999jcp} }in the last two decades.
Therefore we defined
\begin{equation}
\Delta{\rm AE}_A={\rm AE}_A-{\rm AE}_{\rm{TT-}\omega\rm{PBEh}} \label{eq:dmae}
\end{equation}
and present molecules with significant $\Delta$AE$_{\rm B3LYP}$'s {and $\Delta$AE$_{\rm PBE0}$'s} in Figs. \ref{fig:pah_st} and S11 {evaluated using} 
the {\it best} AE from TT-$\omega$PBEh. 

\begin{table}[h]
	\centering
	\caption{MAEs (eV) of $E_{\rm T}$'s are compared across various functionals for all test sets.$^a$}
	\label{tab:st_summary}
	\begin{tabular}{c|cccc|cccc}
		\hline\hline
		energy & \multicolumn{4}{c}{TDDFT} & \multicolumn{4}{|c}{\it worst}\\
		XC functional & PAH & OPV & TADF & BIO & PAH & OPV & TADF & BIO \\
		\hline
		TT-$\omega$PBEh & {\bf 0.158} & {\bf 0.219} & 0.258 & \underline{0.239} & {\bf 0.226} & {\bf 0.362} & {\bf 0.298} & {\bf 0.310}  \\
		OT-$\omega$PBEh & 0.315 & 0.432 & {\bf 0.212} & 0.365 & 0.397 & 0.534 & 0.374 & 0.527  \\
		mix-$\omega$PBEh & 0.306 & {\it 0.489} & 0.341 & 0.404 & 0.428 & 0.592 & 0.474 & 0.547  \\
		HF & 0.235 & 0.290 & 0.288 & 0.272 & {\it 0.771} & {\it 1.078} & {\it 1.077} & {\it 0.869}  \\
		B3LYP & 0.255 & 0.322 & 0.272 & 0.313 & 0.331 & {\underline{0.367}} & 0.324 & 0.356  \\
		CAM-B3LYP & 0.244 & {0.374} & 0.298 & 0.350 & \underline{0.330} & {0.480} & 0.545 & 0.445  \\
		PBE0 & 0.331 & {0.310} & \underline{0.236} & 0.340 & 0.392 & {0.884} & \underline{0.316} & 0.406  \\
		LRC-$\omega$PBE & {\it 0.367} & {0.390} & 0.299 & {\it 0.428} & 0.455 & {0.516} & 0.700 & 0.514  \\
		M06-2X & \underline{0.206} & {\underline{0.274}} & {\it 0.352} & {\bf 0.192} & 0.405 & {0.444} & 0.557 & \underline{0.322}  \\
		\hline\hline
		\multicolumn{9}{l}{$^a${\bf Bold}, \underline{underlined}, and {\it italic} numbers represent the smallest, second smallest,}\\
		\multicolumn{9}{l}{and largest MAEs within each column.}
	\end{tabular}
\end{table}

\subsubsection{Locally Excited Triplet State}
\label{sec:localized}

In the present subsection, we will show the excellent performance of TT-$\omega$PBEh for the test sets of PAH, OPV, and BIO. Each of these molecules possesses a moderate- or large-scale $\pi$-conjugation {over its chromophore} and a {\it locally excited} T$_1$ state. 
Based on Tables \ref{tab:st_summary} and S1, S5, S9, and S11, as well as Figs. \ref{fig:pah_st} and S11, TT-$\omega$PBEh achieves the most accurate predictions of $E_{\rm T}$'s among all XC functionals in question. 

For {PAH and }OPV {sets}, using TDDFT TT-$\omega$PBEh provides the smallest MAE, outperforming any other functionals including B3LYP {and PBE0}.
Especially, OT-$\omega$PBEh and mix-$\omega$PBEh demonstrate doubled MAEs from TT-$\omega$PBEh, indicating {their expected inferior} performances {(Sec. \ref{sec:ind})} to TT-$\omega$PBEh (Figs. \ref{fig:pah_st} and S11).
A similar trend {was} observed for BIO {molecules} as well {(outperformed by M06-2X only)}.
{{In addition, although TT-$\omega$PBEh does not necessarily provide the smallest} {\it best} MAEs, its {\it worst} counterparts are always superior to every other functional across all molecule sets.} 
{Also, TT-$\omega$PBEh exhibits on average the smallest} 
difference between {\it worst} and {\it best} MAEs, validating the best satisfaction of the {triplet tuning} constraint (Eq. (\ref{eq:cond})). 

{Combining these two results}, we can draw two major conclusions: (1) {TT-$\omega$PBEh reproduces the most accurate electron-hole interactions and accomplishes the best approximation to the exact functional and the best stability across DFT variants when applying to a} locally excited T$_1$ state.
{(2) The consideration of Koopmans' theorem {deteriorates} this advantage and make OT-$\omega$PBEh and mix-$\omega$PBEh worse than non-tuned functionals like B3LYP and PBE0.} 
{Based on these conclusions, we can assert that} 
TT-$\omega$PBEh is advantageous over other functionals {for locally excited T$_1$ states}. 

In a systematic comparison, we found 
a large fraction of molecules {where TT-$\omega$PBEh exhibits} significant advantages over B3LYP{ and PBE0}. For instance, more than one third of OPV materials in question exhibit $\Delta{\rm AE}_{\rm B3LYP}>0.20$ eV and are shown in Fig. \ref{fig:pah_st}. Configurationally, most molecules {illustrated here} possess very extensive $\pi$-conjugations in one (like $\alpha$-oligothiophene) or two dimensions (like fullerene), {for which} the locality of LDA and GGA components in B3LYP and PBE0 are known to {introduce a} serious SIE and thus an inaccurate $E_{\rm T}$.\cite{PhysRevB.23.5048,dreuw2004jacs,dreuw2005cr,vydrov2005,Medvedev49,Brorsen2017jpcl,Hait2018jctc,hait2018jcp,Bao2018}
{However, the LRC and triplet tuned formula of TT-$\omega$PBEh minimizes the SIE in both short and long ranges\cite{vydrov2006jcp,vydrov2006jcp2,Rohrdanz2008jcp,Rohrdanz2009,B617919C,ct2003447,PhysRevB.86.205110,ja8087482,baer2010tuned} and reaches the best accuracy of low-lying T$_1$ states. Also, the {\it non-empiricism} of TT-$\omega$PBEh shows no bias towards any species and makes it robust for newly-{synthesized} molecules that are not {yet} included in any fitting database.} 

\begin{figure}[h]
	\centering
	\includegraphics[width=8cm]{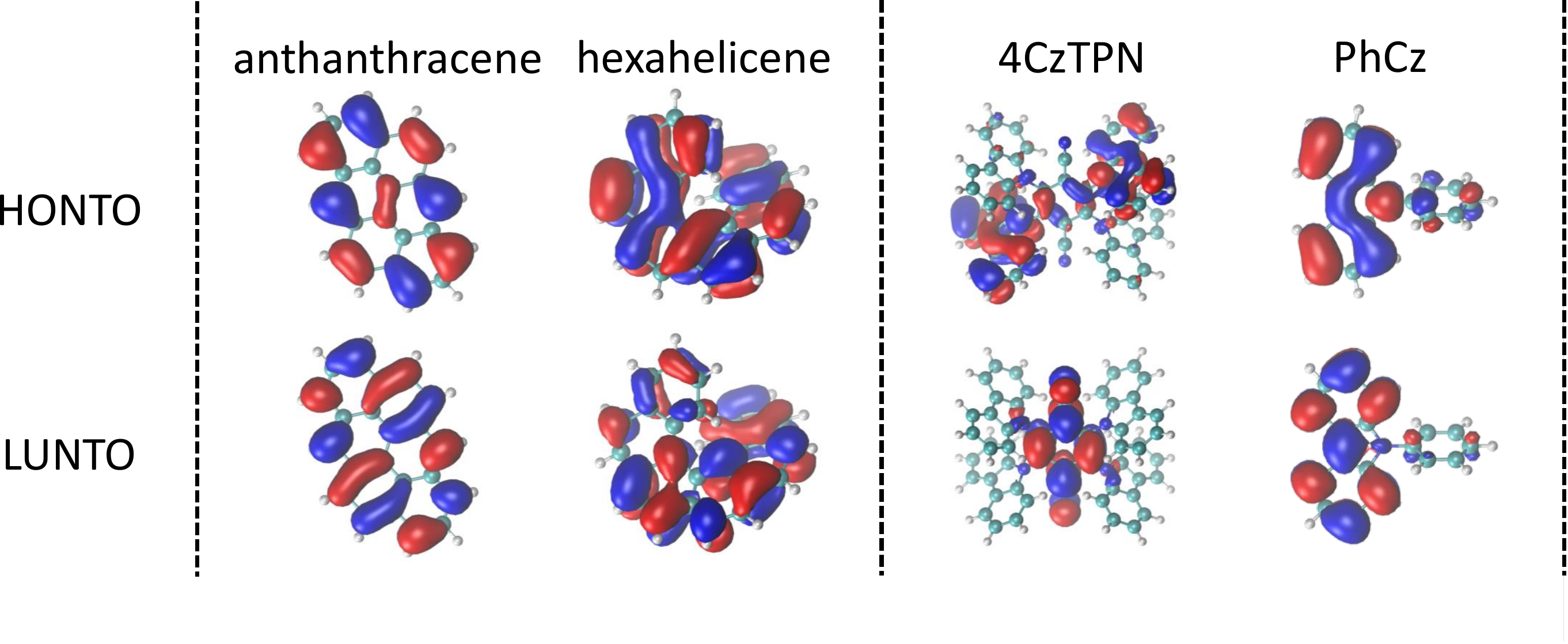}
	\vspace{-0.2 cm}
	\caption{HONTO and LUNTO of the T$_1$ state for anthanthracene and hexahelicene from PAH {(left)}, and 4CzTPN and PhCz from TADF {(right)}.}
	\label{fig:somo_ts}
\end{figure}

{Finally}, to {depict} the locality of T$_1$ states discussed in the present subsection, we illustrate the {perfect spatial overlaps between the highest occupied NTO (HONTO) and the lowest unoccupied NTO (LUNTO)} in the left panel of Fig. \ref{fig:somo_ts} for {two selected} PAH molecules, anthanthracene and hexahelicene. 
For {a similar locally excited T$_1$ state} the accuracy of TT-$\omega$PBEh 
can be guaranteed. 

\subsubsection{Charge Transfer Triplet State}
\label{sec:ct}

In the present subsection, we will show that TT-$\omega$PBEh exhibits a strong predictive power for a {\it charge transfer} T$_1$ state
which is {essentially a pseudo-}one-particle property, {such as that of a TADF emitter}. 

OT-$\omega$PBEh was {constructed} to {largely} reduce the SIE associated with a charge transfer excited state,\cite{B617919C,ja8087482,baer2010tuned,PhysRevB.23.5048,dreuw2004jacs,dreuw2005cr,vydrov2005,Medvedev49,Brorsen2017jpcl,Hait2018jctc,hait2018jcp,Bao2018}
{and }it proves to exhibit the smallest MAE as expected (Tables \ref{tab:st_summary} and S9).
{Interestingly, arriving at {an analogous} long-range HF exchange,\cite{tu2017acsomega,C5CP01782C,zhang2016jpcc} TT-$\omega$PBEh shows a comparable predictive power to OT-$\omega$PBEh and PBE0, and a {smaller} AE than B3LYP for three quarters of TADF emitters.} {Meanwhile, the {\it worst} MAE of TT-$\omega$PBEh is {still} the smallest of all, while OT-$\omega$PBEh is even outperformed by non-tuned B3LYP and PBE0.} 
{Although the {ordinary} performance of mix-$\omega$PBEh indicates a difficulty in {coordinating} the electron-hole interaction and the asymptotic {density} using the same set of parameters}, the {\it unexpected} excellent performance of TT-$\omega$PBEh indicates the gain of complying with Eq. (\ref{eq:cond}) at {only a tiny cost} of {the long-range behavior}. {{In conclusion,} TT-$\omega$PBEh is still the best functional on balance even for a charge transfer {system} for which OT-$\omega$PBEh was originally designed.}

To extend our discussion, we also explored 
{the sensitivity of} TT-$\omega$PBEh on T$_1$ {to} 
the charge separation {extent in the space}.\cite{PhysRevB.23.5048,dreuw2004jacs,dreuw2005cr,vydrov2005,Medvedev49,Isborn2013,Brorsen2017jpcl,Hait2018jctc,hait2018jcp} 
{Due to the} SIE and the {inaccurate} adiabatic local density approximation (ALDA) in TDDFT, 
poor performance is expected for a severe charge transfer {T$_1$ state} using TT-$\omega$PBEh.
To visualize our argument, we illustrate HONTO and LUNTO for the T$_1$ states of two TADF emitters, 4CzTPN and PhCz, in the right panel of Fig. \ref{fig:somo_ts}.
4CzTPN exhibits a strong charge transfer character and its {TDDFT-evaluated }AE {using} TT-$\omega$PBEh {(0.608 eV)} is three times as much as that {using} OT-$\omega$PBEh {(0.205 eV)}, while for {a minimally charge-separated} PhCz 
the performance of TT-$\omega$PBEh is significantly better (0.038 eV versus 0.364 eV).
To {treat} molecules like 4CzTPN, one should consider a methodological modification like to incorporate the frequency dependence in $E_{\rm XC}$\cite{neepa2004jcp,maitra2005jcp,ullrich2006jcp} or to implement the multi-configurational {therapy}.\cite{Manni2014,Chen2017jpcl,Bao2018}

\subsection{Optical Band Gap}
\label{sec:es}

In spite of not being the direct tuning object, $E_{\rm S}$ is a more interesting observable than $E_{\rm T}$ {and an independent benchmark of TT-$\omega$PBEh} as it can be directly measured via absorption or fluorescence spectroscopy with abundant data available in the literature. 
For a normal closed-shell molecule, $E_{\rm S}$ is the energy difference between S$_0$ and {S$_1$} 
(which possesses an identical orbital configuration to T$_1$). Therefore {TT-$\omega$PBEh is} expected to provide an accurate prediction for $E_{\rm S}$. 

\begin{table}[h]
	\centering
	\caption{MAEs (eV) of $E_{\rm S}$'s are compared across various functionals for all test sets.$^a$}
	\label{tab:ss_summary}
	\begin{tabular}{c|cccc|cccc}
		\hline\hline
		energy & \multicolumn{4}{c|}{TDDFT} & \multicolumn{4}{c}{\it worst}\\
		XC functional & PAH & OPV & TADF & BIO & PAH & OPV & TADF & BIO \\
		\hline        
		TT-$\omega$PBEh & 0.381 & {\bf 0.316} & {\bf 0.266} & \underline{0.370} & 0.581 & 0.661 & {\bf 0.357} & \underline{0.571}  \\
		OT-$\omega$PBEh & 0.343 & 0.510 & \underline{0.338} & 0.506 & 0.563 & 0.724 & 0.465 & 0.720 \\
		mix-$\omega$PBEh & 0.331 & 0.414 & 0.352 & 0.584 & 0.518 & 0.673 & 0.504 & 0.889  \\
		HF & {\it 0.774} & {\it 0.793} & {\it 1.792} & {\it 1.292} & 0.774 & {\it 0.793} & {\it 1.792} & {\it 1.358}  \\
		B3LYP & {\bf 0.276} & \underline{0.396} & 0.390 & {\bf 0.309} & 0.525 & 0.590 & \underline{0.416} & {\bf 0.432}  \\
		PBE & 0.416 & 0.507 & 1.017 & 0.456 & 0.741 & 0.792 & 1.023 & 0.614 \\
		LRC-$\omega$PBE & 0.494 & 0.525 & 0.768 & 0.545 & {\it 0.854} & 0.716 & 0.907 & 0.773  \\
		LRC-$\omega$PBEh & 0.348 & 0.430 & 0.627 & 0.499 & {\bf 0.463} & {\bf 0.540} & 0.651 & 0.635  \\
		M06-2X & \underline{0.312} & 0.437 & 0.477 & 0.466 & \underline{0.484}	& \underline{0.569} & 0.499 & 0.605  \\
		\hline\hline
		\multicolumn{9}{l}{$^a${\bf Bold}, \underline{underlined}, and {\it italic} numbers represent the smallest, second smallest,}\\
		\multicolumn{9}{l}{and largest MAEs within each column.}
	\end{tabular}
\end{table}

\begin{figure}[h]
	\centering
	\includegraphics[width=16cm]{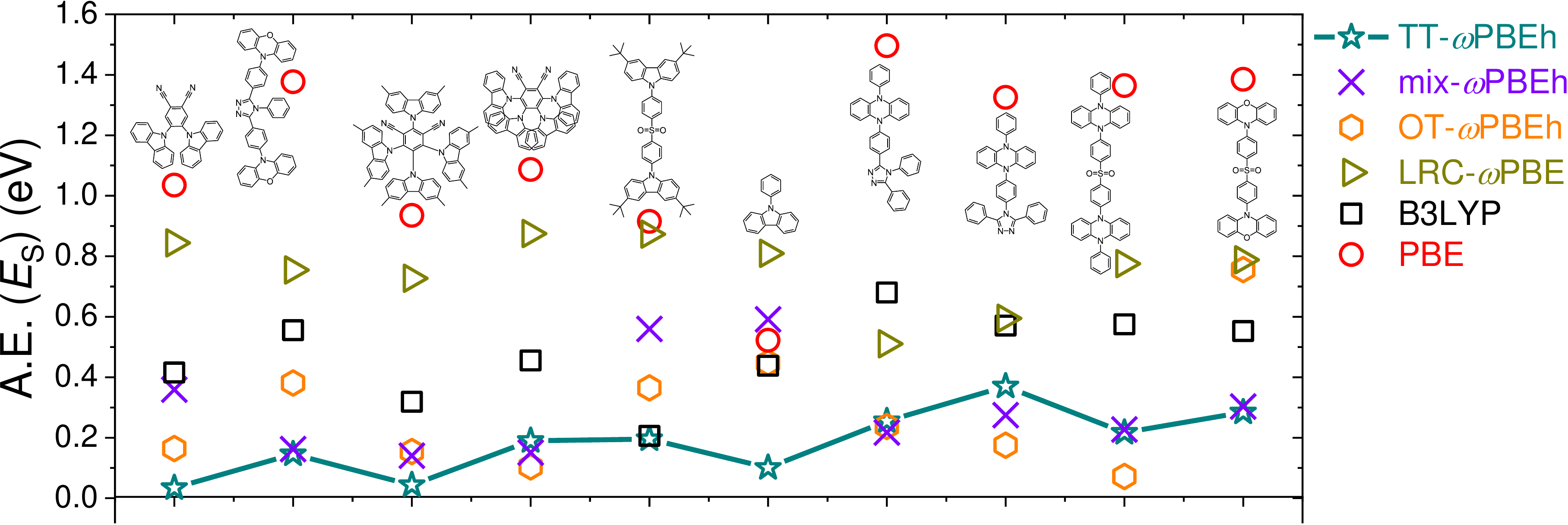}
	\caption{AE{s} of $E_{\rm S}${'s} (eV) {are} illustrated for selected {TADF emitters}. The results are compared among TT-$\omega$PBEh (dark cyan star), mix-$\omega$PBEh (violet cross), OT-$\omega$PBEh (orange hexagon), LRC-$\omega$PBE (dark yellow right triangle), B3LYP (black square), and PBE (red up triangle).}
	\label{fig:bio_ss}
\end{figure}

We summarize all MAEs for $E_{\rm S}${'s} in Tables S2, S6, S10, and S12, including the {\it best} and {\it worst} MAEs {as} defined in Sec. \ref{sec:st}, {except that ROKS is in place of $\Delta$SCF}. In Table \ref{tab:ss_summary} we list {the TDDFT-evaluated and} {\it worst} MAEs. 
In {Figs. \ref{fig:bio_ss} and S12}, we visualize the comparison of {\it best} AEs for selected {molecules with large $\Delta$AE$_{\rm B3LYP}$'s ($> 0.20$ eV except for DTC-DPS)} and $\Delta$AE$_{\rm PBE}$'s ($> 0.40$ eV). 

\subsubsection{Transferability from Triplet to Singlet}

As we can expect {in advance}, {TT-$\omega$PBEh always provides a larger MAE} {for} $E_{\rm S}$ {than} $E_{\rm T}$. {Due to the orbital similarity between S$_1$ and T$_1$},  TT-$\omega$PBEh exhibits the strongest predictive power of $E_{\rm S}$ for the {OPV and TADF} sets, and the second strongest for BIO molecules (outperformed by B3LYP only). 
{Especially, for TADF emitters TT-$\omega$PBEh displays its superiority across all DFT variants (including {\it best} and {\it worst} MAEs).}
This discovery is {\it shocking} but not completely unexpected -- it can be explained by the small $\Delta E_{\rm ST}$ {between} charge transfer S$_1$ {and T$_1$} state{s} with negligible spatial overlap{s} between HONTO and LUNTO.
These results confirm the {suggested 
	transferability} of TT-$\omega$PBEh from a difficult observable, $E_{\rm T}$, to an easy one, $E_{\rm S}$, especially for a compound with a charge transfer S$_1$ state. 
{Such a transferability sets} the stage for the usage of TT-$\omega$PBEh in large-scale screening and design of {spectroscopically and }photochemically active materials.

Similar to the assessment in Sec. \ref{sec:st}, an explicit comparison was also made {for $E_{\rm S}$} between TT-$\omega$PBEh {and popular functionals}. {Ten out of the twelve TADF emitters show positive $\Delta$AE$_{\rm B3LYP}$'s and $\Delta$AE$_{\rm PBE}$'s and are illustrated in Fig. \ref{fig:bio_ss}}. {The poor performance of B3LYP and PBE on these charge transfer molecules can also be interpreted by their significant SIEs.} 

\subsubsection{Singlet--Triplet Instability Problems}
TT-$\omega$PBEh consistently overestimates $E_{\rm S}$ for PAH molecules {using TDDFT}. Its MAE is larger than a handful of functionals including B3LYP, PBE, and LRC-$\omega$PBEh (Table \ref{tab:ss_summary}).
The singlet--triplet instability problem {is associated with the linear-response formulation} that significantly {underestimates $E_{\rm T}$ and }overestimates {$E_{\rm S}$} {and} is the probably the most substantial contribution to this error,\cite{peach2011jctc} 
as parameters {optimized} by matching {{such} underestimated} $E_{\rm T}^{\rm TDDFT}$ with $E_{\rm T}^{\rm \Delta SCF}$ can {definitely} lead to an overestimated $E_{\rm S}$. The inclusion of TDA\cite{HIRATA1999291} in TDDFT can {numerically eliminate} the instability, but it {even increases} the MAE {if TDA is not implemented in the triplet tuning procedure}. This issue indicates {a modification strategy} to replace $E_{{\rm T}_1}^{\rm TDDFT}$ in Eq. (\ref{eq:tddft}) with $E_{{\rm T}_1}^{\rm TDDFT/TDA}$ {for molecules affected by the instability problem }so that the {tuning} will arrive at a better set of parameters. 

\subsection{Singlet--Triplet Gap in a Charge Transfer System}

In a TADF system, $\Delta E_{\rm ST}$ 
serves as a direct predictor for the (reversed) intersystem crossing rate ($k_{\rm (R)ISC}$) and the quantum yield ($\phi$) of the emitter.\cite{liu2018all} 
The accuracy of $\Delta E_{\rm ST}$ proved very sensitive to the level of theory\cite{shang1992jcp,mo2006jpcb,messina2013real,huang2013jctc,rondi2015,sun2017jpcl} and is {then} an additional {benchmark} of TT-$\omega$PBEh.
In the present subsection, we will show that TT-$\omega$PBEh maintains an exceptional prediction for $\Delta E_{\rm ST}$.

\begin{table}[h]
	\centering
	\caption{MAEs (eV) of $\Delta E_{\rm ST}$'s are compared across various functionals for the TADF test set.$^{a,b}$}
	\label{tab:est_summary}
	\begin{tabular}{c|ccccc}
		\hline\hline
		XC functional & $\Delta$SCF/ROKS & TDDFT & TDDFT/TDA & {\it best} & {\it worst}  \\
		\hline        
		TT-$\omega$PBEh {(gas)}	& \underline{0.298} & 0.298 & 0.297 & 0.227 & \underline{0.362}  \\
		TT-$\omega$PBEh (PCM) & 0.309 & {\bf 0.268} & 0.278 & {\bf 0.118} & 0.447  \\     
		OT-$\omega$PBEh {(gas)} & \underline{0.298} & 0.310 & {\bf 0.208} & \underline{0.156} & 0.381  \\
		OT-$\omega$PBEh (PCM) & {\it 0.456} & 0.314 & 0.292 & 0.220 & 0.551  \\
		mix-$\omega$PBEh {(gas)} & 0.355 & 0.463 & 0.345 & 0.233 & 0.526  \\
		mix-$\omega$PBEh (PCM) & 0.392 & 0.361 & \underline{0.237} & \underline{0.156} & 0.525  \\
		HF & - & {\it 1.487} & {\it 1.487} & {\it 1.487} & {\it 1.487}  \\
		B3LYP & {\bf 0.226} & \underline{0.279} & 0.305 & 0.186 & {\bf 0.358}  \\
		LRC-$\omega$PBE & 0.360 & 0.672 & 0.349 & 0.242 & 0.727  \\
		LRC-$\omega$PBEh & - & 0.567 & 0.285 & 0.271 & 0.581  \\
		TPSS & - & 0.375 & 0.381 & 0.372 & 0.384  \\
		M06-L & - & 0.365 & 0.376 & 0.360 & 0.382  \\
		\hline\hline
		\multicolumn{6}{l}{$^a${\bf Bold}, \underline{underlined}, and {\it italic} numbers represent the smallest, second smallest,}\\
		\multicolumn{6}{l}{and largest MAEs within each column.}  \\     
		\multicolumn{6}{l}{$^b$``Gas'' and ``PCM'' represent single-point calculations performed in the gas}  \\     
		\multicolumn{6}{l}{phase and using the PCM solvent model.}  \\               
	\end{tabular}
\end{table}

In Table \ref{tab:est_summary}, we illustrate MAEs of $\Delta E_{\rm ST}$'s for TADF emitters evaluated using different DFT variants as well as {\it best} and {\it worst}.
As {was} discussed in Sec. \ref{sec:ct}, although TT-$\omega$PBEh 
does not {implement} Koopmans' theorem,\cite{KOOPMANS1934104} its long-range HF exchange allows its accuracy to approach that of OT-$\omega$PBEh.\cite{tu2017acsomega} Analogously for $\Delta E_{\rm ST}$, the {values of $|\Delta{\rm AE}_{\rm OT-\omega PBEh}| < $ 0.09 eV}
across all single-point DFT variants and TT-$\omega$PBEh exhibits the second lowest {\it worst} MAE. On the other hand, the non-LRC B3LYP functional also shows outstanding performance due to the error cancellation {between} $E_{\rm S}$ and $E_{\rm T}$, which is however {viewed} as an unreliable coincidence. Taking these {results} into account, TT-$\omega$PBEh reaches the strongest and most stable predictive power for $\Delta E_{\rm ST}$. 

\subsubsection{Solvation Effect}

Due to large {permanent} dipole moments, 
S$_1$ and T$_1$ {of a TADF emitter} can both be stabilized in a polarized solvent, leading to an unpredictable {shift of} $\Delta E_{\rm ST}$.\cite{shang1992jcp,mo2006jpcb,messina2013real,rondi2015}
Also, the experimental measurements cited in the present study were carried out in the {condensed phase (solutions or aggregates)}.\cite{zhang2014efficient,C3TC30699B,C3TC31936A}
In recent studies,\cite{sun2017jpcl,huang2013jctc} the solvation effect was modeled using the {implicit} polarizable continuum model (PCM), in which the dielectric constant {of the environment}, $\varepsilon$, rescales all electric effects. 
However, {the involvement} of even the simplest PCM significantly increases the computational cost, and therefore {it was not used in }our triplet tuning process. 

This issue motivated us to find an inexpensive way of including the solvation effect. 
Based on {gas-phase-}optimized parameters, we re-evaluated $\Delta E_{\rm ST}$ in toluene ($\varepsilon = 2.38$) using PCM (labeled accordingly in Table \ref{tab:est_summary}). {This treatment}  
improves, or at least {\it does not compromise}, the accuracy of TT-$\omega$PBEh and mix-$\omega$PBEh 
as the polarized environment increases the {charge transfer }extent that was underestimated by the gas-phase triplet tuning process. 
On the contrary, OT-$\omega$PBEh already overestimated the charge transfer extent in the gas phase, so the implementation of the solvation effect plays an adverse role. 
Our findings suggest a future direction that incorporates the solvation effect in the construction of TT-$\omega$PBEh {such as} replacing $r_{12}$ with $\varepsilon r_{12}$ in the Hamiltonian.\cite{ct5000617,ct4009975,ct400956h,ct500259m}

\subsection{Spin Contamination}
\label{sec:spin}

Although we have shown the excellent performance of TT-$\omega$PBEh in $E_{\rm T}$, $E_{\rm S}$, and $\Delta E_{\rm ST}$, the spin contamination problem of UDFT still {implies} an additional source of error.
UDFT constrains $M_S=\pm 1$ for a triplet state but {loses control of} $\langle S^2\rangle$. In the present study, we defined a UDFT-evaluated T$_1$ to be spin pure when $1.95 < \langle S^2\rangle < 2.05$. This criterion {is satisfied in our test sets, }except for a few TADF emitters {that} are {mixed with} higher spin configurations. {Such spin contamination} usually occurs at a large HF fraction (large $C_{\rm HF}$ and large $\omega$) or in a strongly-correlated or multi-radical system.\cite{wang1995jcp,fuchs2005jcp,ess2010singlet}
For instance, T$_1$ of 4CzTPN {displays} a {vast} spatial separation between HONTO and LUNTO {under TT-$\omega$PBEh ($\omega = 0.147a_0^{-1}$ and $C_{\rm HF} = 0.85$)}, and thus can be treated as a {\it diradical} structure (Fig. \ref{fig:somo_ts}). UDFT {produces} $\langle S^2\rangle = 2.3358$, equivalent to a superposition of 92\% triplet and 8\% quintet {($\langle S^2\rangle = 6$)}. Its {TDDFT-evaluated} AEs are 0.618 eV, 1.633 eV, and 1.025 eV for $E_{\rm T}$, $E_{\rm S}$, and $\Delta E_{\rm ST}$, respectively, all being huge.

In an attempt to resolve this {problem}, we re-evaluated $E_{\rm T_1}$ for {spin-contaminated molecules} using restricted open-shell DFT (RODFT) but still UDFT-tuned parameters. However, this short-term fix {raised the computational cost} by one order of magnitude {and increased the AEs by} more than 1 eV as it introduces extra difficulty to the variational {calculation} while imposing the {additional} spin symmetry. The long-term {therapy} in the frame of single-reference DFT is to replace $E_{{\rm T}_1}^{\rm UDFT}$ in Eq. (\ref{eq:dscf}) with $E_{{\rm T}_1}^{\rm RODFT}$ to {ensure} the spin purity of T$_1$, but the above-mentioned convergence difficulty makes {it} beyond the scope of the present study due to technical limitation.

\subsection{One-Electron Property}

\begin{table}[h]
	\centering
	\caption{MAEs (eV) of $-\varepsilon^{(N)}_{\rm HOMO}$'s are compared across various functionals for the test sets of PAH, OPV, and BIO.$^a$} 
	\label{tab:ip_summary}
	\begin{tabular}{c|ccc}
		\hline\hline
		XC functional & PAH & OPV & BIO \\
		\hline       
		TT-$\omega$PBEh & 1.182 & 1.737 & 2.037  \\
		OT-$\omega$PBEh & \underline{0.174} & \underline{0.213} & 0.352  \\
		mix-$\omega$PBEh & 0.191 & {\bf 0.173} & \underline{0.327}  \\
		HF & 0.394 & 0.298 & 0.429  \\
		B3LYP & 1.901 & 2.055 & 2.472  \\
		PBE & 2.446 & {\it 2.868} & {\it 3.355}  \\
		LRC-$\omega$PBE & {\bf 0.170} & 0.231 & {\bf 0.177}  \\
		TPSS & {\it 2.500} & 2.857 & 3.323  \\
		M06-L & 2.354 & 2.661 & 2.980  \\
		\hline\hline  
		\multicolumn{4}{l}{$^a${\bf Bold}, \underline{underlined}, and {\it italic} numbers}\\
		\multicolumn{4}{l}{represent the smallest, second smallest,}\\
		\multicolumn{4}{l}{and largest MAEs within each column.}\\
	\end{tabular}
\end{table}

\begin{figure}[h]
	\centering
	\includegraphics[width=16cm]{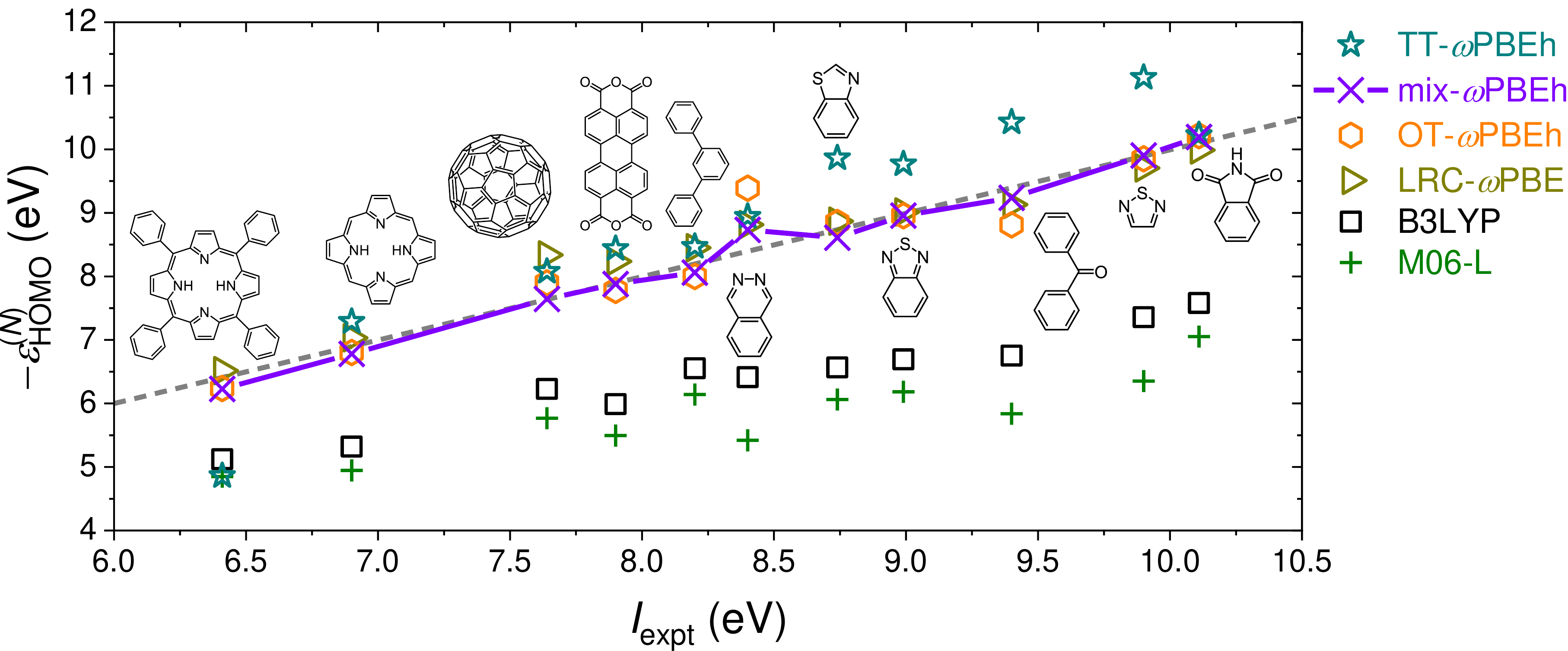}
	\caption{$-\varepsilon^{(N)}_{\rm HOMO}${'s} (eV) {are} compared against $I_{\rm expt}${'s} (eV) for selected OPV materials. The results are presented for TT-$\omega$PBEh (dark cyan star), mix-$\omega$PBEh (violet cross), OT-$\omega$PBEh (orange hexagon), LRC-$\omega$PBE (dark yellow right triangle), B3LYP (black square), and M06-L (green plus). The Koopmans' theorem \cite{KOOPMANS1934104} is illustrated as the diagonal grey dashed line.} 
	\label{fig:opv_ip}
\end{figure}

The observable one-particle property, $I_{\perp}$, {affords} an additional {benchmark} for TT-$\omega$PBEh.
In the present subsection, we will compare $-\varepsilon_{\rm HOMO}^{(N)}$ to experimental $I_{\perp}$'s ($I_{\rm expt}$) {for various functionals} and will show the superiority of TT-$\omega$PBEh over non-LRC functionals like B3LYP {and M06-L}.
Herein the difference {between} $-\varepsilon_{\rm HOMO}^{(N)}$ and $I_{\rm expt}$ is defined as the AE. All MAEs are listed in Tables S3, S7, and S13, {and those from $-\varepsilon_{\rm HOMO}^{(N)}$'s are reported}
for the PAH, OPV, and BIO sets in Table \ref{tab:ip_summary}. {The }$-\varepsilon_{\rm HOMO}^{(N)}$ versus $I_{\rm expt}$ relation for selected OPV materials {is displayed} in Fig. \ref{fig:opv_ip}. 

As we {foresaw} earlier, because of the incorrect long-range nature, $-\varepsilon^{(N)}_{\rm HOMO}$ is consistently underestimated by {all} non-LRC functionals ({B3LYP, PBE, TPSS, and M06-L}){, which also loses the predictive power for $I_{\perp}$}. 
Like all {existing} optimally tuned functionals,\cite{ct300318g,1.3656734,ct100607w,ja8087482,ct100529s,jp1057572}  OT-$\omega$PBEh was constructed based on Koopmans' theorem\cite{KOOPMANS1934104} and exhibits {expected} excellent performance.
{Meanwhile, although the accuracy of TT-$\omega$PBEh is surpassed by HF and other LRC functionals (OT-$\omega$PBEh, mix-$\omega$PBEh, and LRC-$\omega$PBE), it} is significantly better-behaving than all non-LRC functionals due to {its} correct asymptotic behavior, and it presents an overestimated $-\varepsilon^{(N)}_{\rm HOMO}$ rather {than} an underestimation{ when the overall {HF} fraction is higher than usual}. 

More interestingly, mix-$\omega$PBEh has an {\it unexpectedly} comparable performance to OT-$\omega$PBEh 
and its MAEs are even smaller than OT-$\omega$PBEh for OPV and BIO sets, {indicating that using mix-$\omega$PBEh we can obtain equally good $I_\perp$ as those using OT-$\omega$PBEh without sacrificing the accuracy for $E_{\rm T}$ and $E_{\rm S}$}. 
This result suggests the combination of the ideas of {Koopmans}-based optimal tuning and our proposed triplet tuning leads to a correct direction to {the mutual treatment} of electron-hole interactions and one-particle properties -- within a two-dimensional parameter space, these two targets are {no longer incompatible with each other.} 

The discussion we have conducted so far in Sec. \ref{sec:result} allows us to safely assert the {value} of the fully and partially triplet tuned XC functionals like TT-$\omega$PBEh and mix-$\omega$PBEh in the accurate prediction of spectroscopically and photochemically important excited states.

\section{Conclusion and Future Directions}
\label{sec:conc}

In the present study, we proposed triplet tuning, a novel prescription that allows us to {construct} an approximate XC functional based on the energy of T$_1$ ($E_{\rm T}$), in a {\it non-empirical} manner. 
The first triplet-tuned functional, TT-$\omega$PBEh, was constructed by an internal matching of $E_{\rm T}$ from two {DFT variants, $\Delta$SCF and TDDFT}.
To evaluate the behavior of TT-$\omega$PBEh, we compared its errors for {$E_{\rm T}$, $E_{\rm S}$, $\Delta E_{\rm ST}$, and $-\varepsilon_{\rm HOMO}^{(N)}$} against existing functionals, including {the most popular B3LYP, PBE, PBE0, and }one using the conventional optimal tuning scheme, OT-$\omega$PBEh. 
Without fitting to {\it any} experimental data, TT-$\omega$PBEh provides an accurate prediction of electron-hole interactions in {a $\pi$-conjugated organic molecule} and is in general significantly more {powerful than other functionals} in {photochemically interesting energetics like }$E_{\rm T}$, $E_{\rm S}$, and $\Delta E_{\rm ST}$ . 
On the other hand, {TT-$\omega$PBEh} also improves the {accuracy of }one-electron propert{ies} like $-\varepsilon^{(N)}_{\rm HOMO}$ relative to {any} (semi-)local, non-LRC functional. 
In the end, given the difficulty of balancing local electron-hole interactions and non-local one-electron properties, the mix-$\omega$PBEh functional that grants both aspects has achieved {\it unexpected} {accuracy} for $-\varepsilon^{(N)}_{\rm HOMO}$.

Similar to conventional optimally tuned functionals,\cite{C4CP05470A,inorganics5020023,C6CP02648F,sun2016jpcc,zheng2016jpcl,zhuravlev2016antioxidant,bokareva2017jctc,garza2015jpcb,acs.jctc.5b00066,jz3011749}
we expected a broad application of our {triplet tuning} approach for $\pi$-conjugated organic molecules that were considered {theoretically} challenging before. Beyond the accurate predictions of spectroscopy and dynamics, the method can set the stage for the {computationally aided} design {and screening} of photoactive materials. 
In parallel projects being conducted in our group, the present version of TT-$\omega$PBEh has been applied to molecules and clusters that are involved in singlet fission and photoluminescence. 

As discussed in Sec. \ref{sec:result}, there is room for {non-trivial development} of the present triplet tuning {scheme} so as to strengthen its predictive power for difficult excited states. 
For example, the present LRC hybrid {formula} of HF and PBE\cite{PhysRevLett.77.3865} turns out inaccurate for some PAH molecules and TADF emitters, and $\omega$ and $C_{\rm HF}$  
do not necessarily span a large enough two-dimensional space {that contains} the exact exchange functional. To resolve {these issues} without touching the current framework of single-reference DFT, we can reformulate the present triplet-tuned functionals by using alternative formulas or adjustable parameters. For example, we can include the environmental factors such as {$\varepsilon$} of the solvent and the intermolecular interactions,\cite{sun2017jpcl,PhysRevB.93.115206} or we can develop functionals that depend on the frequency \cite{neepa2004jcp,maitra2005jcp} and the local electronic density.\cite{henderson2008}
Finally, to accelerate the tuning process we can implement alternative minimization algorithms and apply the data-driven idea like machine learning.\cite{kohn2018modeling,geva2018simulating}
\section{Associated Content}
In the Supporting Information, we provide the structures for all organic molecules that are included in the four test sets {under investigation}, the discussion about parameter convergence, the tables and figures that summarize AEs and MAEs, and the statistical analysis for all functionals. 
This information is available free of charge via the Internet at http://pubs.acs.org.

\section{Acknowledgment}
Z. L. and T. V. thank the National Science Foundation for the support of the present study (Grant CHE-1464804). We also thank Prof. Leeor Kronik, Prof. Roi Baer, Dr. Tianyu Zhu, {Dr. Xing Zhang, }and Mr. Diptarka Hait for inspirational and insightful discussions.





\bibliography{cite}

\providecommand{\latin}[1]{#1}
\makeatletter
\providecommand{\doi}
  {\begingroup\let\do\@makeother\dospecials
  \catcode`\{=1 \catcode`\}=2 \doi@aux}
\providecommand{\doi@aux}[1]{\endgroup\texttt{#1}}
\makeatother
\providecommand*\mcitethebibliography{\thebibliography}
\csname @ifundefined\endcsname{endmcitethebibliography}
  {\let\endmcitethebibliography\endthebibliography}{}
\begin{mcitethebibliography}{261}
\providecommand*\natexlab[1]{#1}
\providecommand*\mciteSetBstSublistMode[1]{}
\providecommand*\mciteSetBstMaxWidthForm[2]{}
\providecommand*\mciteBstWouldAddEndPuncttrue
  {\def\EndOfBibitem{\unskip.}}
\providecommand*\mciteBstWouldAddEndPunctfalse
  {\let\EndOfBibitem\relax}
\providecommand*\mciteSetBstMidEndSepPunct[3]{}
\providecommand*\mciteSetBstSublistLabelBeginEnd[3]{}
\providecommand*\EndOfBibitem{}
\mciteSetBstSublistMode{f}
\mciteSetBstMaxWidthForm{subitem}{(\alph{mcitesubitemcount})}
\mciteSetBstSublistLabelBeginEnd
  {\mcitemaxwidthsubitemform\space}
  {\relax}
  {\relax}

\bibitem[Hohenberg and Kohn(1964)Hohenberg, and Kohn]{PhysRev.136.B864}
Hohenberg,~P.; Kohn,~W. Inhomogeneous Electron Gas. \emph{Phys. Rev.}
  \textbf{1964}, \emph{136}, B864--B871\relax
\mciteBstWouldAddEndPuncttrue
\mciteSetBstMidEndSepPunct{\mcitedefaultmidpunct}
{\mcitedefaultendpunct}{\mcitedefaultseppunct}\relax
\EndOfBibitem
\bibitem[Kohn and Sham(1965)Kohn, and Sham]{Kohn1965}
Kohn,~W.; Sham,~L.~J. Self-Consistent Equations Including Exchange and
  Correlation Effects. \emph{Phys. Rev.} \textbf{1965}, \emph{140},
  A1133--A1138\relax
\mciteBstWouldAddEndPuncttrue
\mciteSetBstMidEndSepPunct{\mcitedefaultmidpunct}
{\mcitedefaultendpunct}{\mcitedefaultseppunct}\relax
\EndOfBibitem
\bibitem[Parr and Yang(1994)Parr, and Yang]{parr1994density}
Parr,~R.~G.; Yang,~W. \emph{Density-functional theory of atoms and molecules};
  International Series of Monographs on Chemistry; Oxford university press,
  1994; Vol.~16\relax
\mciteBstWouldAddEndPuncttrue
\mciteSetBstMidEndSepPunct{\mcitedefaultmidpunct}
{\mcitedefaultendpunct}{\mcitedefaultseppunct}\relax
\EndOfBibitem
\bibitem[Perdew and Schmidt(2001)Perdew, and Schmidt]{1.1390175}
Perdew,~J.~P.; Schmidt,~K. Jacob's ladder of density functional approximations
  for the exchange-correlation energy. \emph{AIP Conference Proceedings}
  \textbf{2001}, \emph{577}, 1--20\relax
\mciteBstWouldAddEndPuncttrue
\mciteSetBstMidEndSepPunct{\mcitedefaultmidpunct}
{\mcitedefaultendpunct}{\mcitedefaultseppunct}\relax
\EndOfBibitem
\bibitem[Runge and Gross(1984)Runge, and Gross]{PhysRevLett.52.997}
Runge,~E.; Gross,~E. K.~U. Density-Functional Theory for Time-Dependent
  Systems. \emph{Phys. Rev. Lett.} \textbf{1984}, \emph{52}, 997--1000\relax
\mciteBstWouldAddEndPuncttrue
\mciteSetBstMidEndSepPunct{\mcitedefaultmidpunct}
{\mcitedefaultendpunct}{\mcitedefaultseppunct}\relax
\EndOfBibitem
\bibitem[Shao \latin{et~al.}(2003)Shao, Head-Gordon, and Krylov]{SPDFT}
Shao,~Y.; Head-Gordon,~M.; Krylov,~A.~I. The spin-flip approach within
  time-dependent density functional theory: Theory and applications to
  diradicals. \emph{J. Chem. Phys.} \textbf{2003}, \emph{118}, 4807--4818\relax
\mciteBstWouldAddEndPuncttrue
\mciteSetBstMidEndSepPunct{\mcitedefaultmidpunct}
{\mcitedefaultendpunct}{\mcitedefaultseppunct}\relax
\EndOfBibitem
\bibitem[Gavnholt \latin{et~al.}(2008)Gavnholt, Olsen, Engelund, and
  Schi\o{}tz]{PhysRevB.78.075441}
Gavnholt,~J.; Olsen,~T.; Engelund,~M.; Schi\o{}tz,~J. $\Delta$ self-consistent
  field method to obtain potential energy surfaces of excited molecules on
  surfaces. \emph{Phys. Rev. B} \textbf{2008}, \emph{78}, 075441\relax
\mciteBstWouldAddEndPuncttrue
\mciteSetBstMidEndSepPunct{\mcitedefaultmidpunct}
{\mcitedefaultendpunct}{\mcitedefaultseppunct}\relax
\EndOfBibitem
\bibitem[Kowalczyk \latin{et~al.}(2013)Kowalczyk, Tsuchimochi, Chen, Top, and
  {Van Voorhis}]{kowalczyk2013jcp}
Kowalczyk,~T.; Tsuchimochi,~T.; Chen,~P.-T.; Top,~L.; {Van Voorhis},~T.
  Excitation energies and Stokes shifts from a restricted open-shell Kohn--Sham
  approach. \emph{J. Chem. Phys.} \textbf{2013}, \emph{138}, 164101\relax
\mciteBstWouldAddEndPuncttrue
\mciteSetBstMidEndSepPunct{\mcitedefaultmidpunct}
{\mcitedefaultendpunct}{\mcitedefaultseppunct}\relax
\EndOfBibitem
\bibitem[Perdew and Zunger(1981)Perdew, and Zunger]{PhysRevB.23.5048}
Perdew,~J.~P.; Zunger,~A. Self-interaction correction to density-functional
  approximations for many-electron systems. \emph{Phys. Rev. B} \textbf{1981},
  \emph{23}, 5048--5079\relax
\mciteBstWouldAddEndPuncttrue
\mciteSetBstMidEndSepPunct{\mcitedefaultmidpunct}
{\mcitedefaultendpunct}{\mcitedefaultseppunct}\relax
\EndOfBibitem
\bibitem[Dreuw and Head-Gordon(2004)Dreuw, and Head-Gordon]{dreuw2004jacs}
Dreuw,~A.; Head-Gordon,~M. Failure of Time-Dependent Density Functional Theory
  for Long-Range Charge-Transfer Excited States: The
  Zincbacteriochlorin--Bacteriochlorin and Bacteriochlorophyll-Spheroidene
  Complexes. \emph{J. Am. Chem. Soc.} \textbf{2004}, \emph{126},
  4007--4016\relax
\mciteBstWouldAddEndPuncttrue
\mciteSetBstMidEndSepPunct{\mcitedefaultmidpunct}
{\mcitedefaultendpunct}{\mcitedefaultseppunct}\relax
\EndOfBibitem
\bibitem[Dreuw and Head-Gordon(2005)Dreuw, and Head-Gordon]{dreuw2005cr}
Dreuw,~A.; Head-Gordon,~M. Single-Reference {\it ab initio} Methods for the
  Calculation of Excited States of Large Molecules. \emph{Chem. Rev.}
  \textbf{2005}, \emph{105}, 4009--4037\relax
\mciteBstWouldAddEndPuncttrue
\mciteSetBstMidEndSepPunct{\mcitedefaultmidpunct}
{\mcitedefaultendpunct}{\mcitedefaultseppunct}\relax
\EndOfBibitem
\bibitem[Vydrov and Scuseria(2005)Vydrov, and Scuseria]{vydrov2005}
Vydrov,~O.~A.; Scuseria,~G.~E. Ionization potentials and electron affinities in
  the Perdew--Zunger self-interaction corrected density-functional theory.
  \emph{J. Chem. Phys.} \textbf{2005}, \emph{122}, 184107\relax
\mciteBstWouldAddEndPuncttrue
\mciteSetBstMidEndSepPunct{\mcitedefaultmidpunct}
{\mcitedefaultendpunct}{\mcitedefaultseppunct}\relax
\EndOfBibitem
\bibitem[Medvedev \latin{et~al.}(2017)Medvedev, Bushmarinov, Sun, Perdew, and
  Lyssenko]{Medvedev49}
Medvedev,~M.~G.; Bushmarinov,~I.~S.; Sun,~J.; Perdew,~J.~P.; Lyssenko,~K.~A.
  Density functional theory is straying from the path toward the exact
  functional. \emph{Science} \textbf{2017}, \emph{355}, 49--52\relax
\mciteBstWouldAddEndPuncttrue
\mciteSetBstMidEndSepPunct{\mcitedefaultmidpunct}
{\mcitedefaultendpunct}{\mcitedefaultseppunct}\relax
\EndOfBibitem
\bibitem[Brorsen \latin{et~al.}(2017)Brorsen, Yang, Pak, and
  Hammes-Schiffer]{Brorsen2017jpcl}
Brorsen,~K.~R.; Yang,~Y.; Pak,~M.~V.; Hammes-Schiffer,~S. Is the Accuracy of
  Density Functional Theory for Atomization Energies and Densities in Bonding
  Regions Correlated? \emph{J. Phys. Chem. Lett.} \textbf{2017}, \emph{8},
  2076--2081\relax
\mciteBstWouldAddEndPuncttrue
\mciteSetBstMidEndSepPunct{\mcitedefaultmidpunct}
{\mcitedefaultendpunct}{\mcitedefaultseppunct}\relax
\EndOfBibitem
\bibitem[Hait and Head-Gordon(2018)Hait, and Head-Gordon]{Hait2018jctc}
Hait,~D.; Head-Gordon,~M. How Accurate Is Density Functional Theory at
  Predicting Dipole Moments? An Assessment Using a New Database of 200
  Benchmark Values. \emph{J. Chem. Theory Comput.} \textbf{2018}, \emph{14},
  1969--1981\relax
\mciteBstWouldAddEndPuncttrue
\mciteSetBstMidEndSepPunct{\mcitedefaultmidpunct}
{\mcitedefaultendpunct}{\mcitedefaultseppunct}\relax
\EndOfBibitem
\bibitem[Hait and Head-Gordon(2018)Hait, and Head-Gordon]{hait2018jcp}
Hait,~D.; Head-Gordon,~M. Communication: xDH double hybrid functionals can be
  qualitatively incorrect for non-equilibrium geometries: Dipole moment
  inversion and barriers to radical-radical association using XYG3 and XYGJ-OS.
  \emph{J. Chem. Phys.} \textbf{2018}, \emph{148}, 171102\relax
\mciteBstWouldAddEndPuncttrue
\mciteSetBstMidEndSepPunct{\mcitedefaultmidpunct}
{\mcitedefaultendpunct}{\mcitedefaultseppunct}\relax
\EndOfBibitem
\bibitem[Bao \latin{et~al.}(2018)Bao, Gagliardi, and Truhlar]{Bao2018}
Bao,~J.~L.; Gagliardi,~L.; Truhlar,~D.~G. Self-Interaction Error in Density
  Functional Theory: An Appraisal. \emph{J. Phys. Chem. Lett.} \textbf{2018},
  \emph{9}, 2353--2358\relax
\mciteBstWouldAddEndPuncttrue
\mciteSetBstMidEndSepPunct{\mcitedefaultmidpunct}
{\mcitedefaultendpunct}{\mcitedefaultseppunct}\relax
\EndOfBibitem
\bibitem[Polo \latin{et~al.}(2002)Polo, Kraka, and Cremer]{Polo2002}
Polo,~V.; Kraka,~E.; Cremer,~D. Electron correlation and the self-interaction
  error of density functional theory. \emph{Mol. Phys.} \textbf{2002},
  \emph{100}, 1771--1790\relax
\mciteBstWouldAddEndPuncttrue
\mciteSetBstMidEndSepPunct{\mcitedefaultmidpunct}
{\mcitedefaultendpunct}{\mcitedefaultseppunct}\relax
\EndOfBibitem
\bibitem[Ciofini \latin{et~al.}(2005)Ciofini, Adamo, and
  Chermette]{CIOFINI200567}
Ciofini,~I.; Adamo,~C.; Chermette,~H. Self-interaction error in density
  functional theory: a mean-field correction for molecules and large systems.
  \emph{Chem. Phys.} \textbf{2005}, \emph{309}, 67--76\relax
\mciteBstWouldAddEndPuncttrue
\mciteSetBstMidEndSepPunct{\mcitedefaultmidpunct}
{\mcitedefaultendpunct}{\mcitedefaultseppunct}\relax
\EndOfBibitem
\bibitem[Wu and {Van Voorhis}(2006)Wu, and {Van Voorhis}]{wu2006jpca}
Wu,~Q.; {Van Voorhis},~T. Direct Calculation of Electron Transfer Parameters
  through Constrained Density Functional Theory. \emph{J. Phys. Chem. A}
  \textbf{2006}, \emph{110}, 9212--9218\relax
\mciteBstWouldAddEndPuncttrue
\mciteSetBstMidEndSepPunct{\mcitedefaultmidpunct}
{\mcitedefaultendpunct}{\mcitedefaultseppunct}\relax
\EndOfBibitem
\bibitem[Park \latin{et~al.}(2016)Park, Senn, Krykunov, and Ziegler]{Park2016}
Park,~Y.~C.; Senn,~F.; Krykunov,~M.; Ziegler,~T. Self-Consistent Constricted
  Variational Theory {RSCF-CV($\infty$)-DFT} and Its Restrictions To Obtain a
  Numerically Stable $\Delta$SCF-DFT-like Method: Theory and Calculations for
  Triplet States. \emph{J. Chem. Theory Comput.} \textbf{2016}, \emph{12},
  5438--5452\relax
\mciteBstWouldAddEndPuncttrue
\mciteSetBstMidEndSepPunct{\mcitedefaultmidpunct}
{\mcitedefaultendpunct}{\mcitedefaultseppunct}\relax
\EndOfBibitem
\bibitem[Tao \latin{et~al.}(2003)Tao, Perdew, Staroverov, and
  Scuseria]{PhysRevLett.91.146401}
Tao,~J.; Perdew,~J.~P.; Staroverov,~V.~N.; Scuseria,~G.~E. Climbing the Density
  Functional Ladder: Nonempirical Meta--Generalized Gradient Approximation
  Designed for Molecules and Solids. \emph{Phys. Rev. Lett.} \textbf{2003},
  \emph{91}, 146401\relax
\mciteBstWouldAddEndPuncttrue
\mciteSetBstMidEndSepPunct{\mcitedefaultmidpunct}
{\mcitedefaultendpunct}{\mcitedefaultseppunct}\relax
\EndOfBibitem
\bibitem[Sun \latin{et~al.}(2015)Sun, Ruzsinszky, and
  Perdew]{PhysRevLett.115.036402}
Sun,~J.; Ruzsinszky,~A.; Perdew,~J.~P. Strongly Constrained and Appropriately
  Normed Semilocal Density Functional. \emph{Phys. Rev. Lett.} \textbf{2015},
  \emph{115}, 036402\relax
\mciteBstWouldAddEndPuncttrue
\mciteSetBstMidEndSepPunct{\mcitedefaultmidpunct}
{\mcitedefaultendpunct}{\mcitedefaultseppunct}\relax
\EndOfBibitem
\bibitem[Zhao and Truhlar(2006)Zhao, and Truhlar]{zhao2006jcp}
Zhao,~Y.; Truhlar,~D.~G. A new local density functional for main-group
  thermochemistry, transition metal bonding, thermochemical kinetics, and
  noncovalent interactions. \emph{J. Chem. Phys.} \textbf{2006}, \emph{125},
  194101\relax
\mciteBstWouldAddEndPuncttrue
\mciteSetBstMidEndSepPunct{\mcitedefaultmidpunct}
{\mcitedefaultendpunct}{\mcitedefaultseppunct}\relax
\EndOfBibitem
\bibitem[Zhao and Truhlar(2008)Zhao, and Truhlar]{Zhao2008}
Zhao,~Y.; Truhlar,~D.~G. The M06 suite of density functionals for main group
  thermochemistry, thermochemical kinetics, noncovalent interactions, excited
  states, and transition elements: two new functionals and systematic testing
  of four M06-class functionals and 12 other functionals. \emph{Theor. Chem.
  Acc.} \textbf{2008}, \emph{120}, 215--241\relax
\mciteBstWouldAddEndPuncttrue
\mciteSetBstMidEndSepPunct{\mcitedefaultmidpunct}
{\mcitedefaultendpunct}{\mcitedefaultseppunct}\relax
\EndOfBibitem
\bibitem[Yanai \latin{et~al.}(2004)Yanai, Tew, and Handy]{YANAI200451}
Yanai,~T.; Tew,~D.~P.; Handy,~N.~C. A new hybrid exchange-correlation
  functional using the {Coulomb}-attenuating method ({CAM-B3LYP}). \emph{Chem.
  Phys. Lett.} \textbf{2004}, \emph{393}, 51--57\relax
\mciteBstWouldAddEndPuncttrue
\mciteSetBstMidEndSepPunct{\mcitedefaultmidpunct}
{\mcitedefaultendpunct}{\mcitedefaultseppunct}\relax
\EndOfBibitem
\bibitem[Jin and Bartlett(2016)Jin, and Bartlett]{jin2016jcp}
Jin,~Y.; Bartlett,~R.~J. The {QTP} family of consistent functionals and
  potentials in Kohn--Sham density functional theory. \emph{J. Chem. Phys.}
  \textbf{2016}, \emph{145}, 034107\relax
\mciteBstWouldAddEndPuncttrue
\mciteSetBstMidEndSepPunct{\mcitedefaultmidpunct}
{\mcitedefaultendpunct}{\mcitedefaultseppunct}\relax
\EndOfBibitem
\bibitem[Mardirossian and Head-Gordon(2014)Mardirossian, and
  Head-Gordon]{C3CP54374A}
Mardirossian,~N.; Head-Gordon,~M. $\omega$B97X-V: A 10-parameter{,}
  range-separated hybrid{,} generalized gradient approximation density
  functional with nonlocal correlation{,} designed by a survival-of-the-fittest
  strategy. \emph{Phys. Chem. Chem. Phys.} \textbf{2014}, \emph{16},
  9904--9924\relax
\mciteBstWouldAddEndPuncttrue
\mciteSetBstMidEndSepPunct{\mcitedefaultmidpunct}
{\mcitedefaultendpunct}{\mcitedefaultseppunct}\relax
\EndOfBibitem
\bibitem[Mardirossian and Head-Gordon(2016)Mardirossian, and
  Head-Gordon]{Mardirossian2016jcp}
Mardirossian,~N.; Head-Gordon,~M. "$\omega$B97M-V: A combinatorially optimized,
  range-separated hybrid, meta-GGA density functional with VV10 nonlocal
  correlation. \emph{J. Chem. Phys.} \textbf{2016}, \emph{144}, 214110\relax
\mciteBstWouldAddEndPuncttrue
\mciteSetBstMidEndSepPunct{\mcitedefaultmidpunct}
{\mcitedefaultendpunct}{\mcitedefaultseppunct}\relax
\EndOfBibitem
\bibitem[Vydrov and Scuseria(2006)Vydrov, and Scuseria]{vydrov2006jcp}
Vydrov,~O.~A.; Scuseria,~G.~E. Assessment of a long-range corrected hybrid
  functional. \emph{J. Chem. Phys.} \textbf{2006}, \emph{125}, 234109\relax
\mciteBstWouldAddEndPuncttrue
\mciteSetBstMidEndSepPunct{\mcitedefaultmidpunct}
{\mcitedefaultendpunct}{\mcitedefaultseppunct}\relax
\EndOfBibitem
\bibitem[Rohrdanz and Herbert(2008)Rohrdanz, and Herbert]{Rohrdanz2008jcp}
Rohrdanz,~M.~A.; Herbert,~J.~M. Simultaneous benchmarking of ground- and
  excited-state properties with long-range-corrected density functional theory.
  \emph{J. Chem. Phys.} \textbf{2008}, \emph{129}, 034107\relax
\mciteBstWouldAddEndPuncttrue
\mciteSetBstMidEndSepPunct{\mcitedefaultmidpunct}
{\mcitedefaultendpunct}{\mcitedefaultseppunct}\relax
\EndOfBibitem
\bibitem[Vydrov \latin{et~al.}(2006)Vydrov, Heyd, Krukau, and
  Scuseria]{vydrov2006jcp2}
Vydrov,~O.~A.; Heyd,~J.; Krukau,~A.~V.; Scuseria,~G.~E. Importance of
  short-range versus long-range Hartree--Fock exchange for the performance of
  hybrid density functionals. \emph{J. Chem. Phys.} \textbf{2006}, \emph{125},
  074106\relax
\mciteBstWouldAddEndPuncttrue
\mciteSetBstMidEndSepPunct{\mcitedefaultmidpunct}
{\mcitedefaultendpunct}{\mcitedefaultseppunct}\relax
\EndOfBibitem
\bibitem[Rohrdanz \latin{et~al.}(2009)Rohrdanz, Martins, and
  Herbert]{Rohrdanz2009}
Rohrdanz,~M.~A.; Martins,~K.~M.; Herbert,~J.~M. A long-range-corrected density
  functional that performs well for both ground-state properties and
  time-dependent density functional theory excitation energies, including
  charge-transfer excited states. \emph{J. Chem. Phys.} \textbf{2009},
  \emph{130}, 054112\relax
\mciteBstWouldAddEndPuncttrue
\mciteSetBstMidEndSepPunct{\mcitedefaultmidpunct}
{\mcitedefaultendpunct}{\mcitedefaultseppunct}\relax
\EndOfBibitem
\bibitem[Livshits and Baer(2007)Livshits, and Baer]{B617919C}
Livshits,~E.; Baer,~R. A well-tempered density functional theory of electrons
  in molecules. \emph{Phys. Chem. Chem. Phys.} \textbf{2007}, \emph{9},
  2932--2941\relax
\mciteBstWouldAddEndPuncttrue
\mciteSetBstMidEndSepPunct{\mcitedefaultmidpunct}
{\mcitedefaultendpunct}{\mcitedefaultseppunct}\relax
\EndOfBibitem
\bibitem[Salzner and Aydin(2011)Salzner, and Aydin]{ct2003447}
Salzner,~U.; Aydin,~A. Improved Prediction of Properties of $\pi$-Conjugated
  Oligomers with Range-Separated Hybrid Density Functionals. \emph{J. Chem.
  Theory Comput.} \textbf{2011}, \emph{7}, 2568--2583\relax
\mciteBstWouldAddEndPuncttrue
\mciteSetBstMidEndSepPunct{\mcitedefaultmidpunct}
{\mcitedefaultendpunct}{\mcitedefaultseppunct}\relax
\EndOfBibitem
\bibitem[K\"orzd\"orfer \latin{et~al.}(2012)K\"orzd\"orfer, Parrish, Marom,
  Sears, Sherrill, and Br\'edas]{PhysRevB.86.205110}
K\"orzd\"orfer,~T.; Parrish,~R.~M.; Marom,~N.; Sears,~J.~S.; Sherrill,~C.~D.;
  Br\'edas,~J.-L. Assessment of the performance of tuned range-separated hybrid
  density functionals in predicting accurate quasiparticle spectra. \emph{Phys.
  Rev. B} \textbf{2012}, \emph{86}, 205110\relax
\mciteBstWouldAddEndPuncttrue
\mciteSetBstMidEndSepPunct{\mcitedefaultmidpunct}
{\mcitedefaultendpunct}{\mcitedefaultseppunct}\relax
\EndOfBibitem
\bibitem[Stein \latin{et~al.}(2009)Stein, Kronik, and Baer]{ja8087482}
Stein,~T.; Kronik,~L.; Baer,~R. Reliable Prediction of Charge Transfer
  Excitations in Molecular Complexes Using Time-Dependent Density Functional
  Theory. \emph{J. Am. Chem. Soc.} \textbf{2009}, \emph{131}, 2818--2820\relax
\mciteBstWouldAddEndPuncttrue
\mciteSetBstMidEndSepPunct{\mcitedefaultmidpunct}
{\mcitedefaultendpunct}{\mcitedefaultseppunct}\relax
\EndOfBibitem
\bibitem[Baer \latin{et~al.}(2010)Baer, Livshits, and Salzner]{baer2010tuned}
Baer,~R.; Livshits,~E.; Salzner,~U. Tuned range-separated hybrids in density
  functional theory. \emph{Annu. Rev. Phys. Chem.} \textbf{2010}, \emph{61},
  85--109\relax
\mciteBstWouldAddEndPuncttrue
\mciteSetBstMidEndSepPunct{\mcitedefaultmidpunct}
{\mcitedefaultendpunct}{\mcitedefaultseppunct}\relax
\EndOfBibitem
\bibitem[Iikura \latin{et~al.}(2001)Iikura, Tsuneda, Yanai, and
  Hirao]{1.1383587}
Iikura,~H.; Tsuneda,~T.; Yanai,~T.; Hirao,~K. A long-range correction scheme
  for generalized-gradient-approximation exchange functionals. \emph{J. Chem.
  Phys.} \textbf{2001}, \emph{115}, 3540--3544\relax
\mciteBstWouldAddEndPuncttrue
\mciteSetBstMidEndSepPunct{\mcitedefaultmidpunct}
{\mcitedefaultendpunct}{\mcitedefaultseppunct}\relax
\EndOfBibitem
\bibitem[Koopmans(1934)]{KOOPMANS1934104}
Koopmans,~T. \'{U}ber die Zuordnung von Wellenfunktionen und Eigenwerten zu den
  Einzelnen Elektronen Eines Atoms. \emph{Physica} \textbf{1934}, \emph{1},
  104--113\relax
\mciteBstWouldAddEndPuncttrue
\mciteSetBstMidEndSepPunct{\mcitedefaultmidpunct}
{\mcitedefaultendpunct}{\mcitedefaultseppunct}\relax
\EndOfBibitem
\bibitem[Livshits \latin{et~al.}(2011)Livshits, Granot, and Baer]{jp1057572}
Livshits,~E.; Granot,~R.~S.; Baer,~R. A Density Functional Theory for Studying
  Ionization Processes in Water Clusters. \emph{J. Phys. Chem. A}
  \textbf{2011}, \emph{115}, 5735--5744\relax
\mciteBstWouldAddEndPuncttrue
\mciteSetBstMidEndSepPunct{\mcitedefaultmidpunct}
{\mcitedefaultendpunct}{\mcitedefaultseppunct}\relax
\EndOfBibitem
\bibitem[Sears \latin{et~al.}(2011)Sears, K\"{o}rzd\"{o}rfer, Zhang, and
  Br\'{e}das]{1.3656734}
Sears,~J.~S.; K\"{o}rzd\"{o}rfer,~T.; Zhang,~C.-R.; Br\'{e}das,~J.-L.
  Communication: Orbital instabilities and triplet states from time-dependent
  density functional theory and long-range corrected functionals. \emph{J.
  Chem. Phys.} \textbf{2011}, \emph{135}, 151103\relax
\mciteBstWouldAddEndPuncttrue
\mciteSetBstMidEndSepPunct{\mcitedefaultmidpunct}
{\mcitedefaultendpunct}{\mcitedefaultseppunct}\relax
\EndOfBibitem
\bibitem[Refaely-Abramson \latin{et~al.}(2011)Refaely-Abramson, Baer, and
  Kronik]{PhysRevB.84.075144}
Refaely-Abramson,~S.; Baer,~R.; Kronik,~L. Fundamental and excitation gaps in
  molecules of relevance for organic photovoltaics from an optimally tuned
  range-separated hybrid functional. \emph{Phys. Rev. B} \textbf{2011},
  \emph{84}, 075144\relax
\mciteBstWouldAddEndPuncttrue
\mciteSetBstMidEndSepPunct{\mcitedefaultmidpunct}
{\mcitedefaultendpunct}{\mcitedefaultseppunct}\relax
\EndOfBibitem
\bibitem[Kuritz \latin{et~al.}(2011)Kuritz, Stein, Baer, and Kronik]{ct2002804}
Kuritz,~N.; Stein,~T.; Baer,~R.; Kronik,~L. Charge-Transfer-Like
  $\pi\rightarrow\pi^\ast$ Excitations in Time-Dependent Density Functional
  Theory: A Conundrum and Its Solution. \emph{J. Chem. Theory Comput.}
  \textbf{2011}, \emph{7}, 2408--2415\relax
\mciteBstWouldAddEndPuncttrue
\mciteSetBstMidEndSepPunct{\mcitedefaultmidpunct}
{\mcitedefaultendpunct}{\mcitedefaultseppunct}\relax
\EndOfBibitem
\bibitem[Kronik \latin{et~al.}(2012)Kronik, Stein, Refaely-Abramson, and
  Baer]{ct2009363}
Kronik,~L.; Stein,~T.; Refaely-Abramson,~S.; Baer,~R. Excitation Gaps of
  Finite-Sized Systems from Optimally Tuned Range-Separated Hybrid Functionals.
  \emph{J. Chem. Theory Comput.} \textbf{2012}, \emph{8}, 1515--1531\relax
\mciteBstWouldAddEndPuncttrue
\mciteSetBstMidEndSepPunct{\mcitedefaultmidpunct}
{\mcitedefaultendpunct}{\mcitedefaultseppunct}\relax
\EndOfBibitem
\bibitem[K\"{o}rzd\"{o}rfer and Br\'{e}das(2014)K\"{o}rzd\"{o}rfer, and
  Br\'{e}das]{ar500021t}
K\"{o}rzd\"{o}rfer,~T.; Br\'{e}das,~J.-L. Organic Electronic Materials: Recent
  Advances in the DFT Description of the Ground and Excited States Using Tuned
  Range-Separated Hybrid Functionals. \emph{Acc. Chem. Res.} \textbf{2014},
  \emph{47}, 3284--3291\relax
\mciteBstWouldAddEndPuncttrue
\mciteSetBstMidEndSepPunct{\mcitedefaultmidpunct}
{\mcitedefaultendpunct}{\mcitedefaultseppunct}\relax
\EndOfBibitem
\bibitem[Jacquemin \latin{et~al.}(2014)Jacquemin, Moore, Planchat, Adamo, and
  Autschbach]{ct5000617}
Jacquemin,~D.; Moore,~B.; Planchat,~A.; Adamo,~C.; Autschbach,~J. Performance
  of an Optimally Tuned Range-Separated Hybrid Functional for 0--0 Electronic
  Excitation Energies. \emph{J. Chem. Theory Comput.} \textbf{2014}, \emph{10},
  1677--1685\relax
\mciteBstWouldAddEndPuncttrue
\mciteSetBstMidEndSepPunct{\mcitedefaultmidpunct}
{\mcitedefaultendpunct}{\mcitedefaultseppunct}\relax
\EndOfBibitem
\bibitem[Manna \latin{et~al.}(2018)Manna, Refaely-Abramson, Reilly, Tkatchenko,
  Neaton, and Kronik]{acs.jctc.7b01058}
Manna,~A.~K.; Refaely-Abramson,~S.; Reilly,~A.~M.; Tkatchenko,~A.;
  Neaton,~J.~B.; Kronik,~L. Quantitative Prediction of Optical Absorption in
  Molecular Solids from an Optimally Tuned Screened Range-Separated Hybrid
  Functional. \emph{J. Chem. Theory Comput.} \textbf{2018}, \emph{14},
  2919--2929\relax
\mciteBstWouldAddEndPuncttrue
\mciteSetBstMidEndSepPunct{\mcitedefaultmidpunct}
{\mcitedefaultendpunct}{\mcitedefaultseppunct}\relax
\EndOfBibitem
\bibitem[G\"orling(1993)]{PhysRevA.47.2783}
G\"orling,~A. Symmetry in density-functional theory. \emph{Phys. Rev. A}
  \textbf{1993}, \emph{47}, 2783--2799\relax
\mciteBstWouldAddEndPuncttrue
\mciteSetBstMidEndSepPunct{\mcitedefaultmidpunct}
{\mcitedefaultendpunct}{\mcitedefaultseppunct}\relax
\EndOfBibitem
\bibitem[Pople and Nesbet(1954)Pople, and Nesbet]{pople1954jcp}
Pople,~J.~A.; Nesbet,~R.~K. Self-Consistent Orbitals for Radicals. \emph{J.
  Chem. Phys.} \textbf{1954}, \emph{22}, 571--572\relax
\mciteBstWouldAddEndPuncttrue
\mciteSetBstMidEndSepPunct{\mcitedefaultmidpunct}
{\mcitedefaultendpunct}{\mcitedefaultseppunct}\relax
\EndOfBibitem
\bibitem[McWeeny(1992)]{mcweeny1992methods}
McWeeny,~R. \emph{Methods of molecular quantum mechanics}; Academic press,
  1992\relax
\mciteBstWouldAddEndPuncttrue
\mciteSetBstMidEndSepPunct{\mcitedefaultmidpunct}
{\mcitedefaultendpunct}{\mcitedefaultseppunct}\relax
\EndOfBibitem
\bibitem[Marques and Gross(2004)Marques, and Gross]{Marques2004}
Marques,~M.; Gross,~E. Time-Dependent Density Functional Theory. \emph{Annu.
  Rev. Phys. Chem.} \textbf{2004}, \emph{55}, 427--455\relax
\mciteBstWouldAddEndPuncttrue
\mciteSetBstMidEndSepPunct{\mcitedefaultmidpunct}
{\mcitedefaultendpunct}{\mcitedefaultseppunct}\relax
\EndOfBibitem
\bibitem[Jensen(2017)]{jensen2017introduction}
Jensen,~F. \emph{Introduction to computational chemistry}; John Wiley \& Sons,
  2017\relax
\mciteBstWouldAddEndPuncttrue
\mciteSetBstMidEndSepPunct{\mcitedefaultmidpunct}
{\mcitedefaultendpunct}{\mcitedefaultseppunct}\relax
\EndOfBibitem
\bibitem[Baer and Neuhauser(2005)Baer, and Neuhauser]{PhysRevLett.94.043002}
Baer,~R.; Neuhauser,~D. Density Functional Theory with Correct Long-Range
  Asymptotic Behavior. \emph{Phys. Rev. Lett.} \textbf{2005}, \emph{94},
  043002\relax
\mciteBstWouldAddEndPuncttrue
\mciteSetBstMidEndSepPunct{\mcitedefaultmidpunct}
{\mcitedefaultendpunct}{\mcitedefaultseppunct}\relax
\EndOfBibitem
\bibitem[Toulouse \latin{et~al.}(2004)Toulouse, Colonna, and
  Savin]{PhysRevA.70.062505}
Toulouse,~J.; Colonna,~F.; Savin,~A. Long-range--short-range separation of the
  electron-electron interaction in density-functional theory. \emph{Phys. Rev.
  A} \textbf{2004}, \emph{70}, 062505\relax
\mciteBstWouldAddEndPuncttrue
\mciteSetBstMidEndSepPunct{\mcitedefaultmidpunct}
{\mcitedefaultendpunct}{\mcitedefaultseppunct}\relax
\EndOfBibitem
\bibitem[Chai and Head-Gordon(2008)Chai, and Head-Gordon]{chai2008jcp}
Chai,~J.-D.; Head-Gordon,~M. Systematic optimization of long-range corrected
  hybrid density functionals. \emph{J. Chem. Phys.} \textbf{2008}, \emph{128},
  084106\relax
\mciteBstWouldAddEndPuncttrue
\mciteSetBstMidEndSepPunct{\mcitedefaultmidpunct}
{\mcitedefaultendpunct}{\mcitedefaultseppunct}\relax
\EndOfBibitem
\bibitem[Szabo and Ostlund(2012)Szabo, and Ostlund]{szabo2012modern}
Szabo,~A.; Ostlund,~N.~S. \emph{Modern quantum chemistry: introduction to
  advanced electronic structure theory}; Courier Corporation, 2012\relax
\mciteBstWouldAddEndPuncttrue
\mciteSetBstMidEndSepPunct{\mcitedefaultmidpunct}
{\mcitedefaultendpunct}{\mcitedefaultseppunct}\relax
\EndOfBibitem
\bibitem[Abramowitz and Stegun(1964)Abramowitz, and
  Stegun]{abramowitz1964handbook}
Abramowitz,~M.; Stegun,~I.~A. \emph{Handbook of mathematical functions: with
  formulas, graphs, and mathematical tables}; Courier Corporation, 1964;
  Vol.~55\relax
\mciteBstWouldAddEndPuncttrue
\mciteSetBstMidEndSepPunct{\mcitedefaultmidpunct}
{\mcitedefaultendpunct}{\mcitedefaultseppunct}\relax
\EndOfBibitem
\bibitem[Mori-S\'anchez \latin{et~al.}(2008)Mori-S\'anchez, Cohen, and
  Yang]{PhysRevLett.100.146401}
Mori-S\'anchez,~P.; Cohen,~A.~J.; Yang,~W. Localization and Delocalization
  Errors in Density Functional Theory and Implications for Band-Gap Prediction.
  \emph{Phys. Rev. Lett.} \textbf{2008}, \emph{100}, 146401\relax
\mciteBstWouldAddEndPuncttrue
\mciteSetBstMidEndSepPunct{\mcitedefaultmidpunct}
{\mcitedefaultendpunct}{\mcitedefaultseppunct}\relax
\EndOfBibitem
\bibitem[Becke(1988)]{PhysRevA.38.3098}
Becke,~A.~D. Density-functional exchange-energy approximation with correct
  asymptotic behavior. \emph{Phys. Rev. A} \textbf{1988}, \emph{38},
  3098--3100\relax
\mciteBstWouldAddEndPuncttrue
\mciteSetBstMidEndSepPunct{\mcitedefaultmidpunct}
{\mcitedefaultendpunct}{\mcitedefaultseppunct}\relax
\EndOfBibitem
\bibitem[Langreth and Mehl(1983)Langreth, and Mehl]{PhysRevB.28.1809}
Langreth,~D.~C.; Mehl,~M.~J. Beyond the local-density approximation in
  calculations of ground-state electronic properties. \emph{Phys. Rev. B}
  \textbf{1983}, \emph{28}, 1809--1834\relax
\mciteBstWouldAddEndPuncttrue
\mciteSetBstMidEndSepPunct{\mcitedefaultmidpunct}
{\mcitedefaultendpunct}{\mcitedefaultseppunct}\relax
\EndOfBibitem
\bibitem[Perdew \latin{et~al.}(1992)Perdew, Chevary, Vosko, Jackson, Pederson,
  Singh, and Fiolhais]{PhysRevB.46.6671}
Perdew,~J.~P.; Chevary,~J.~A.; Vosko,~S.~H.; Jackson,~K.~A.; Pederson,~M.~R.;
  Singh,~D.~J.; Fiolhais,~C. Atoms, molecules, solids, and surfaces:
  Applications of the generalized gradient approximation for exchange and
  correlation. \emph{Phys. Rev. B} \textbf{1992}, \emph{46}, 6671--6687\relax
\mciteBstWouldAddEndPuncttrue
\mciteSetBstMidEndSepPunct{\mcitedefaultmidpunct}
{\mcitedefaultendpunct}{\mcitedefaultseppunct}\relax
\EndOfBibitem
\bibitem[Perdew \latin{et~al.}(1996)Perdew, Burke, and
  Ernzerhof]{PhysRevLett.77.3865}
Perdew,~J.~P.; Burke,~K.; Ernzerhof,~M. Generalized Gradient Approximation Made
  Simple. \emph{Phys. Rev. Lett.} \textbf{1996}, \emph{77}, 3865--3868\relax
\mciteBstWouldAddEndPuncttrue
\mciteSetBstMidEndSepPunct{\mcitedefaultmidpunct}
{\mcitedefaultendpunct}{\mcitedefaultseppunct}\relax
\EndOfBibitem
\bibitem[Murov \latin{et~al.}(1993)Murov, Carmichael, and
  Hug]{murov1993handbook}
Murov,~S.~L.; Carmichael,~I.; Hug,~G.~L. \emph{Handbook of photochemistry}; CRC
  Press, 1993\relax
\mciteBstWouldAddEndPuncttrue
\mciteSetBstMidEndSepPunct{\mcitedefaultmidpunct}
{\mcitedefaultendpunct}{\mcitedefaultseppunct}\relax
\EndOfBibitem
\bibitem[Doering(1969)]{doering1969jcp}
Doering,~J.~P. Low-Energy Electron -- Impact Study of the First, Second, and
  Third Triplet States of Benzene. \emph{J. Chem. Phys.} \textbf{1969},
  \emph{51}, 2866--2870\relax
\mciteBstWouldAddEndPuncttrue
\mciteSetBstMidEndSepPunct{\mcitedefaultmidpunct}
{\mcitedefaultendpunct}{\mcitedefaultseppunct}\relax
\EndOfBibitem
\bibitem[Bolovinos \latin{et~al.}(1984)Bolovinos, Tsekeris, Philis, Pantos, and
  Andritsopoulos]{BOLOVINOS1984240}
Bolovinos,~A.; Tsekeris,~P.; Philis,~J.; Pantos,~E.; Andritsopoulos,~G.
  Absolute vacuum ultraviolet absorption spectra of some gaseous azabenzenes.
  \emph{J. Mol. Spect.} \textbf{1984}, \emph{103}, 240--256\relax
\mciteBstWouldAddEndPuncttrue
\mciteSetBstMidEndSepPunct{\mcitedefaultmidpunct}
{\mcitedefaultendpunct}{\mcitedefaultseppunct}\relax
\EndOfBibitem
\bibitem[Hellner and Vermeil(1976)Hellner, and Vermeil]{HELLNER197671}
Hellner,~L.; Vermeil,~C. VUV excitation of benzene. \emph{J. Mol. Spect.}
  \textbf{1976}, \emph{60}, 71--92\relax
\mciteBstWouldAddEndPuncttrue
\mciteSetBstMidEndSepPunct{\mcitedefaultmidpunct}
{\mcitedefaultendpunct}{\mcitedefaultseppunct}\relax
\EndOfBibitem
\bibitem[Clar(1950)]{CLAR1950116}
Clar,~E. Absorption spectra of aromatic hydrocarbons at low temperatures.
  LV-Aromatic hydrocarbons. \emph{Spectrochim. Acta} \textbf{1950}, \emph{4},
  116--121\relax
\mciteBstWouldAddEndPuncttrue
\mciteSetBstMidEndSepPunct{\mcitedefaultmidpunct}
{\mcitedefaultendpunct}{\mcitedefaultseppunct}\relax
\EndOfBibitem
\bibitem[Allan(1989)]{ALLAN1989219}
Allan,~M. Study of triplet states and short-lived negative ions by means of
  electron impact spectroscopy. \emph{J. Electron Spectrosc. Relat. Phenom.}
  \textbf{1989}, \emph{48}, 219--351\relax
\mciteBstWouldAddEndPuncttrue
\mciteSetBstMidEndSepPunct{\mcitedefaultmidpunct}
{\mcitedefaultendpunct}{\mcitedefaultseppunct}\relax
\EndOfBibitem
\bibitem[McClure(1951)]{mcclure1951jcp}
McClure,~D.~S. Excited Triplet States of Some Polyatomic Molecules. I. \emph{J.
  Chem. Phys.} \textbf{1951}, \emph{19}, 670--675\relax
\mciteBstWouldAddEndPuncttrue
\mciteSetBstMidEndSepPunct{\mcitedefaultmidpunct}
{\mcitedefaultendpunct}{\mcitedefaultseppunct}\relax
\EndOfBibitem
\bibitem[McClure(1949)]{mcclure1949jcp}
McClure,~D.~S. Triplet--Singlet Transitions in Organic Molecules. Lifetime
  Measurements of the Triplet State. \emph{J. Chem. Phys.} \textbf{1949},
  \emph{17}, 905--913\relax
\mciteBstWouldAddEndPuncttrue
\mciteSetBstMidEndSepPunct{\mcitedefaultmidpunct}
{\mcitedefaultendpunct}{\mcitedefaultseppunct}\relax
\EndOfBibitem
\bibitem[Morgan \latin{et~al.}(1977)Morgan, Warshawsky, and
  Atkinson]{PHP:PHP31}
Morgan,~D.~D.; Warshawsky,~D.; Atkinson,~T. The Relationship between
  Carcinogenic Activities of Polycyclic Aromatic Hydrocarbons and Their
  Singlet, Triplet and Singlet--Triplet Splitting Energies and Phosphorescence
  Lifetimes. \emph{Photochem. Photobiol.} \textbf{1977}, \emph{25},
  31--38\relax
\mciteBstWouldAddEndPuncttrue
\mciteSetBstMidEndSepPunct{\mcitedefaultmidpunct}
{\mcitedefaultendpunct}{\mcitedefaultseppunct}\relax
\EndOfBibitem
\bibitem[Moodie and Reid(1954)Moodie, and Reid]{moodie1954jcp}
Moodie,~M.~M.; Reid,~C. Inter- and Intramolecular Energy Transfer Processes. 3.
  Phosphorescence Bands of Some Carcinogenic Aromatic Hydrocarbons. \emph{J.
  Chem. Phys.} \textbf{1954}, \emph{22}, 252--254\relax
\mciteBstWouldAddEndPuncttrue
\mciteSetBstMidEndSepPunct{\mcitedefaultmidpunct}
{\mcitedefaultendpunct}{\mcitedefaultseppunct}\relax
\EndOfBibitem
\bibitem[McGlynn \latin{et~al.}(1964)McGlynn, Azumi, and Kasha]{mcglynn1964jcp}
McGlynn,~S.~P.; Azumi,~T.; Kasha,~M. External Heavy-Atom Spin--Orbital Coupling
  Effect. V. Absorption Studies of Triplet States. \emph{J. Chem. Phys.}
  \textbf{1964}, \emph{40}, 507--515\relax
\mciteBstWouldAddEndPuncttrue
\mciteSetBstMidEndSepPunct{\mcitedefaultmidpunct}
{\mcitedefaultendpunct}{\mcitedefaultseppunct}\relax
\EndOfBibitem
\bibitem[Perkampus(1992)]{perkampus1992uv}
Perkampus,~H.-H. \emph{UV-VIS atlas of organic compounds}; VCH, 1992\relax
\mciteBstWouldAddEndPuncttrue
\mciteSetBstMidEndSepPunct{\mcitedefaultmidpunct}
{\mcitedefaultendpunct}{\mcitedefaultseppunct}\relax
\EndOfBibitem
\bibitem[Birks(1970)]{birks1970photophysics}
Birks,~J.~B. \emph{Photophysics of aromatic molecules}; Wiley monographs in
  chemical physics; John Wiley \& Sons Ltd, 1970\relax
\mciteBstWouldAddEndPuncttrue
\mciteSetBstMidEndSepPunct{\mcitedefaultmidpunct}
{\mcitedefaultendpunct}{\mcitedefaultseppunct}\relax
\EndOfBibitem
\bibitem[Angliker \latin{et~al.}(1982)Angliker, Rommel, and
  Wirz]{ANGLIKER1982208}
Angliker,~H.; Rommel,~E.; Wirz,~J. Electronic spectra of hexacene in solution
  (ground state, triplet state, dication and dianion). \emph{Chem. Phys. Lett.}
  \textbf{1982}, \emph{87}, 208--212\relax
\mciteBstWouldAddEndPuncttrue
\mciteSetBstMidEndSepPunct{\mcitedefaultmidpunct}
{\mcitedefaultendpunct}{\mcitedefaultseppunct}\relax
\EndOfBibitem
\bibitem[Clarke and Hochstrasser(1969)Clarke, and Hochstrasser]{CLARKE1969309}
Clarke,~R.~H.; Hochstrasser,~R.~M. Location and assignment of the lowest
  triplet state of perylene. \emph{J. Mol. Spect.} \textbf{1969}, \emph{32},
  309--319\relax
\mciteBstWouldAddEndPuncttrue
\mciteSetBstMidEndSepPunct{\mcitedefaultmidpunct}
{\mcitedefaultendpunct}{\mcitedefaultseppunct}\relax
\EndOfBibitem
\bibitem[Paris \latin{et~al.}(1961)Paris, Hirt, and Schmitt]{paris1961jcp}
Paris,~J.~P.; Hirt,~R.~C.; Schmitt,~R.~G. Observed Phosphorescence and
  Singlet--Triplet Absorption in $s$-Triazine and Trimethyl-$s$-Triazine.
  \emph{J. Chem. Phys.} \textbf{1961}, \emph{34}, 1851--1852\relax
\mciteBstWouldAddEndPuncttrue
\mciteSetBstMidEndSepPunct{\mcitedefaultmidpunct}
{\mcitedefaultendpunct}{\mcitedefaultseppunct}\relax
\EndOfBibitem
\bibitem[Duncan \latin{et~al.}(1981)Duncan, Dietz, and Smalley]{duncan1981jcp}
Duncan,~M.~A.; Dietz,~T.~G.; Smalley,~R.~E. Two-color photoionization of
  naphthalene and benzene at threshold. \emph{J. Chem. Phys.} \textbf{1981},
  \emph{75}, 2118--2125\relax
\mciteBstWouldAddEndPuncttrue
\mciteSetBstMidEndSepPunct{\mcitedefaultmidpunct}
{\mcitedefaultendpunct}{\mcitedefaultseppunct}\relax
\EndOfBibitem
\bibitem[Jalbout \latin{et~al.}(2007)Jalbout, Trzaskowski, Chen, Chen, and
  Adamowicz]{QUA:QUA21237}
Jalbout,~A.~F.; Trzaskowski,~B.; Chen,~E. C.~M.; Chen,~E.~S.; Adamowicz,~L.
  Electron affinities, gas phase acidities, and potential energy curves:
  Benzene. \emph{Int. J. Quant. Chem.} \textbf{2007}, \emph{107},
  1115--1125\relax
\mciteBstWouldAddEndPuncttrue
\mciteSetBstMidEndSepPunct{\mcitedefaultmidpunct}
{\mcitedefaultendpunct}{\mcitedefaultseppunct}\relax
\EndOfBibitem
\bibitem[Lyapustina \latin{et~al.}(2000)Lyapustina, Xu, Nilles, and {Bowen
  Jr.}]{Lyapustina2000jcp}
Lyapustina,~S.~A.; Xu,~S.; Nilles,~J.~M.; {Bowen Jr.},~K.~H. Solvent-induced
  stabilization of the naphthalene anion by water molecules: A negative cluster
  ion photoelectron spectroscopic study. \emph{J. Chem. Phys.} \textbf{2000},
  \emph{112}, 6643--6648\relax
\mciteBstWouldAddEndPuncttrue
\mciteSetBstMidEndSepPunct{\mcitedefaultmidpunct}
{\mcitedefaultendpunct}{\mcitedefaultseppunct}\relax
\EndOfBibitem
\bibitem[Sch\"{a}fer \latin{et~al.}(1975)Sch\"{a}fer, Schweig, Vermeer,
  Bickelhaupt, and Graaf]{SCHAFER197591}
Sch\"{a}fer,~W.; Schweig,~A.; Vermeer,~H.; Bickelhaupt,~F.; Graaf,~H.~D. On the
  nature of the ``free electron pair'' on phosphorus in aromatic phosphorus
  compounds: The photoelectron spectrum of 2-phosphanaphthalene. \emph{J.
  Electron Spectrosc. Relat. Phenom.} \textbf{1975}, \emph{6}, 91--98\relax
\mciteBstWouldAddEndPuncttrue
\mciteSetBstMidEndSepPunct{\mcitedefaultmidpunct}
{\mcitedefaultendpunct}{\mcitedefaultseppunct}\relax
\EndOfBibitem
\bibitem[Ando \latin{et~al.}(2007)Ando, Mitsui, and Nakajima]{ando2007jcp}
Ando,~N.; Mitsui,~M.; Nakajima,~A. Comprehensive photoelectron spectroscopic
  study of anionic clusters of anthracene and its alkyl derivatives: Electronic
  structures bridging molecules to bulk. \emph{J. Chem. Phys.} \textbf{2007},
  \emph{127}, 234305\relax
\mciteBstWouldAddEndPuncttrue
\mciteSetBstMidEndSepPunct{\mcitedefaultmidpunct}
{\mcitedefaultendpunct}{\mcitedefaultseppunct}\relax
\EndOfBibitem
\bibitem[Eland(1972)]{ELAND1972214}
Eland,~J. Photoelectron spectra and ionization potentials of aromatic
  hydrocarbons. \emph{Int. J. Mass Spectrom. Ion Phys.} \textbf{1972},
  \emph{9}, 214--219\relax
\mciteBstWouldAddEndPuncttrue
\mciteSetBstMidEndSepPunct{\mcitedefaultmidpunct}
{\mcitedefaultendpunct}{\mcitedefaultseppunct}\relax
\EndOfBibitem
\bibitem[Becker and Chen(1966)Becker, and Chen]{becker1966jcp}
Becker,~R.~S.; Chen,~E. Extension of Electron Affinities and Ionization
  Potentials of Aromatic Hydrocarbons. \emph{J. Chem. Phys.} \textbf{1966},
  \emph{45}, 2403--2410\relax
\mciteBstWouldAddEndPuncttrue
\mciteSetBstMidEndSepPunct{\mcitedefaultmidpunct}
{\mcitedefaultendpunct}{\mcitedefaultseppunct}\relax
\EndOfBibitem
\bibitem[Schmidt(1977)]{schmidt1977jcp}
Schmidt,~W. Photoelectron spectra of polynuclear aromatics. V. Correlations
  with ultraviolet absorption spectra in the catacondensed series. \emph{J.
  Chem. Phys.} \textbf{1977}, \emph{66}, 828--845\relax
\mciteBstWouldAddEndPuncttrue
\mciteSetBstMidEndSepPunct{\mcitedefaultmidpunct}
{\mcitedefaultendpunct}{\mcitedefaultseppunct}\relax
\EndOfBibitem
\bibitem[Hager and Wallace(1988)Hager, and Wallace]{hagen1988ac}
Hager,~J.~W.; Wallace,~S.~C. Two-laser photoionization supersonic jet mass
  spectrometry of aromatic molecules. \emph{Anal. Chem.} \textbf{1988},
  \emph{60}, 5--10\relax
\mciteBstWouldAddEndPuncttrue
\mciteSetBstMidEndSepPunct{\mcitedefaultmidpunct}
{\mcitedefaultendpunct}{\mcitedefaultseppunct}\relax
\EndOfBibitem
\bibitem[Mitsui \latin{et~al.}(2007)Mitsui, Ando, and
  Nakajima]{masaaki2007jpca}
Mitsui,~M.; Ando,~N.; Nakajima,~A. Mass Spectrometry and Photoelectron
  Spectroscopy of Tetracene Cluster Anions, (Tetracene)$_n^-$ ($n = 1-100$):
  Evidence for the Highly Localized Nature of Polarization in a Cluster
  Analogue of Oligoacene Crystals. \emph{J. Phys. Chem. A} \textbf{2007},
  \emph{111}, 9644--9648\relax
\mciteBstWouldAddEndPuncttrue
\mciteSetBstMidEndSepPunct{\mcitedefaultmidpunct}
{\mcitedefaultendpunct}{\mcitedefaultseppunct}\relax
\EndOfBibitem
\bibitem[Stahl and Maquin(1984)Stahl, and Maquin]{STAHL1984613}
Stahl,~D.; Maquin,~F. Charge-stripping mass spectrometry of molecular ions from
  polyacenes and molecular orbital theory. \emph{Chem. Phys. Lett.}
  \textbf{1984}, \emph{108}, 613--617\relax
\mciteBstWouldAddEndPuncttrue
\mciteSetBstMidEndSepPunct{\mcitedefaultmidpunct}
{\mcitedefaultendpunct}{\mcitedefaultseppunct}\relax
\EndOfBibitem
\bibitem[Crocker \latin{et~al.}(1993)Crocker, Wang, and
  Kebarle]{crocker1993jacs}
Crocker,~L.; Wang,~T.; Kebarle,~P. Electron affinities of some polycyclic
  aromatic hydrocarbons, obtained from electron-transfer equilibria. \emph{J.
  Am. Chem. Soc.} \textbf{1993}, \emph{115}, 7818--7822\relax
\mciteBstWouldAddEndPuncttrue
\mciteSetBstMidEndSepPunct{\mcitedefaultmidpunct}
{\mcitedefaultendpunct}{\mcitedefaultseppunct}\relax
\EndOfBibitem
\bibitem[Schiedt and Weinkauf(1997)Schiedt, and Weinkauf]{SCHIEDT199718}
Schiedt,~J.; Weinkauf,~R. Photodetachment photoelectron spectroscopy of
  perylene and $\ce{CS2}$: two extreme cases. \emph{J. Chem. Phys.}
  \textbf{1997}, \emph{274}, 18--22\relax
\mciteBstWouldAddEndPuncttrue
\mciteSetBstMidEndSepPunct{\mcitedefaultmidpunct}
{\mcitedefaultendpunct}{\mcitedefaultseppunct}\relax
\EndOfBibitem
\bibitem[Shchuka \latin{et~al.}(1989)Shchuka, Motyka, and Topp]{SHCHUKA198987}
Shchuka,~M.~I.; Motyka,~A.~L.; Topp,~M.~R. Two-photon threshold ionization
  spectroscopy of perylene and van der {Waals} complexes. \emph{J. Chem. Phys.}
  \textbf{1989}, \emph{164}, 87--95\relax
\mciteBstWouldAddEndPuncttrue
\mciteSetBstMidEndSepPunct{\mcitedefaultmidpunct}
{\mcitedefaultendpunct}{\mcitedefaultseppunct}\relax
\EndOfBibitem
\bibitem[Boschi \latin{et~al.}(1974)Boschi, Clar, and Schmidt]{boschi1974jcp}
Boschi,~R.; Clar,~E.; Schmidt,~W. Photoelectron spectra of polynuclear
  aromatics. III. The effect of nonplanarity in sterically overcrowded aromatic
  hydrocarbons. \emph{J. Chem. Phys.} \textbf{1974}, \emph{60},
  4406--4418\relax
\mciteBstWouldAddEndPuncttrue
\mciteSetBstMidEndSepPunct{\mcitedefaultmidpunct}
{\mcitedefaultendpunct}{\mcitedefaultseppunct}\relax
\EndOfBibitem
\bibitem[Clar \latin{et~al.}(1981)Clar, Robertson, Schloegl, and
  Schmidt]{clar1981jacs}
Clar,~E.; Robertson,~J.~M.; Schloegl,~R.; Schmidt,~W. Photoelectron spectra of
  polynuclear aromatics. 6. Applications to structural elucidation:
  ``circumanthracene''. \emph{J. Am. Chem. Soc.} \textbf{1981}, \emph{103},
  1320--1328\relax
\mciteBstWouldAddEndPuncttrue
\mciteSetBstMidEndSepPunct{\mcitedefaultmidpunct}
{\mcitedefaultendpunct}{\mcitedefaultseppunct}\relax
\EndOfBibitem
\bibitem[Hajgat\'{o} \latin{et~al.}(2008)Hajgat\'{o}, Deleuze, Tozer, and
  Proft]{hajgato2008jcp}
Hajgat\'{o},~B.; Deleuze,~M.~S.; Tozer,~D.~J.; Proft,~F.~D. A benchmark
  theoretical study of the electron affinities of benzene and linear acenes.
  \emph{J. Chem. Phys.} \textbf{2008}, \emph{129}, 084308\relax
\mciteBstWouldAddEndPuncttrue
\mciteSetBstMidEndSepPunct{\mcitedefaultmidpunct}
{\mcitedefaultendpunct}{\mcitedefaultseppunct}\relax
\EndOfBibitem
\bibitem[Obenland and Schmidt(1975)Obenland, and Schmidt]{obenland1975jacs}
Obenland,~S.; Schmidt,~W. Photoelectron spectra of polynuclear aromatics. IV.
  Helicenes. \emph{J. Am. Chem. Soc.} \textbf{1975}, \emph{97},
  6633--6638\relax
\mciteBstWouldAddEndPuncttrue
\mciteSetBstMidEndSepPunct{\mcitedefaultmidpunct}
{\mcitedefaultendpunct}{\mcitedefaultseppunct}\relax
\EndOfBibitem
\bibitem[Hung and Grabowski(1991)Hung, and Grabowski]{hung1991jpc}
Hung,~R.~R.; Grabowski,~J.~J. A precise determination of the triplet energy of
  carbon ($\ce{C60}$) by photoacoustic calorimetry. \emph{J. Phys. Chem.}
  \textbf{1991}, \emph{95}, 6073--6075\relax
\mciteBstWouldAddEndPuncttrue
\mciteSetBstMidEndSepPunct{\mcitedefaultmidpunct}
{\mcitedefaultendpunct}{\mcitedefaultseppunct}\relax
\EndOfBibitem
\bibitem[Arbogast \latin{et~al.}(1991)Arbogast, Darmanyan, Foote, Diederich,
  Whetten, Rubin, Alvarez, and Anz]{arbogast1991jpc}
Arbogast,~J.~W.; Darmanyan,~A.~P.; Foote,~C.~S.; Diederich,~F.~N.;
  Whetten,~R.~L.; Rubin,~Y.; Alvarez,~M.~M.; Anz,~S.~J. Photophysical
  properties of sixty atom carbon molecule ($\ce{C60}$). \emph{J. Phys. Chem.}
  \textbf{1991}, \emph{95}, 11--12\relax
\mciteBstWouldAddEndPuncttrue
\mciteSetBstMidEndSepPunct{\mcitedefaultmidpunct}
{\mcitedefaultendpunct}{\mcitedefaultseppunct}\relax
\EndOfBibitem
\bibitem[McVie \latin{et~al.}(1978)McVie, Sinclair, and Truscott]{F29787401870}
McVie,~J.; Sinclair,~R.~S.; Truscott,~T.~G. Triplet states of copper and
  metal-free phthalocyanines. \emph{J. Chem. Soc.{,} Faraday Trans. 2}
  \textbf{1978}, \emph{74}, 1870--1879\relax
\mciteBstWouldAddEndPuncttrue
\mciteSetBstMidEndSepPunct{\mcitedefaultmidpunct}
{\mcitedefaultendpunct}{\mcitedefaultseppunct}\relax
\EndOfBibitem
\bibitem[Gouterman and Khalil(1974)Gouterman, and Khalil]{GOUTERMAN197488}
Gouterman,~M.; Khalil,~G.-E. Porphyrin free base phosphorescence. \emph{J. Mol.
  Spect.} \textbf{1974}, \emph{53}, 88--100\relax
\mciteBstWouldAddEndPuncttrue
\mciteSetBstMidEndSepPunct{\mcitedefaultmidpunct}
{\mcitedefaultendpunct}{\mcitedefaultseppunct}\relax
\EndOfBibitem
\bibitem[Vincett \latin{et~al.}(1971)Vincett, Voigt, and
  Rieckhoff]{vincett1971jcp}
Vincett,~P.~S.; Voigt,~E.~M.; Rieckhoff,~K.~E. Phosphorescence and Fluorescence
  of Phthalocyanines. \emph{J. Chem. Phys.} \textbf{1971}, \emph{55},
  4131--4140\relax
\mciteBstWouldAddEndPuncttrue
\mciteSetBstMidEndSepPunct{\mcitedefaultmidpunct}
{\mcitedefaultendpunct}{\mcitedefaultseppunct}\relax
\EndOfBibitem
\bibitem[Petruska(1961)]{petruska1961jcp}
Petruska,~J. Changes in the Electronic Transitions of Aromatic Hydrocarbons on
  Chemical Substitution. {II}. Application of Perturbation Theory to
  Substituted-Benzene Spectra. \emph{J. Chem. Phys.} \textbf{1961}, \emph{34},
  1120--1136\relax
\mciteBstWouldAddEndPuncttrue
\mciteSetBstMidEndSepPunct{\mcitedefaultmidpunct}
{\mcitedefaultendpunct}{\mcitedefaultseppunct}\relax
\EndOfBibitem
\bibitem[Palummo \latin{et~al.}(2009)Palummo, Hogan, Sottile, Bagal\'{a}, and
  Rubio]{palummo2009jcp}
Palummo,~M.; Hogan,~C.; Sottile,~F.; Bagal\'{a},~P.; Rubio,~A. {\it Ab initio}
  electronic and optical spectra of free-base porphyrins: The role of
  electronic correlation. \emph{J. Chem. Phys.} \textbf{2009}, \emph{131},
  084102\relax
\mciteBstWouldAddEndPuncttrue
\mciteSetBstMidEndSepPunct{\mcitedefaultmidpunct}
{\mcitedefaultendpunct}{\mcitedefaultseppunct}\relax
\EndOfBibitem
\bibitem[Dvorak \latin{et~al.}(2012)Dvorak, M\"{u}ller, Knoblauch,
  B\"{u}nermann, Rydlo, Minniberger, Harbich, and Stienkemeier]{dvorak2012jcp}
Dvorak,~M.; M\"{u}ller,~M.; Knoblauch,~T.; B\"{u}nermann,~O.; Rydlo,~A.;
  Minniberger,~S.; Harbich,~W.; Stienkemeier,~F. Spectroscopy of 3, 4, 9,
  10-perylenetetracarboxylic dianhydride ({PTCDA}) attached to rare gas
  samples: Clusters vs. bulk matrices. I. Absorption spectroscopy. \emph{J.
  Chem. Phys.} \textbf{2012}, \emph{137}, 164301\relax
\mciteBstWouldAddEndPuncttrue
\mciteSetBstMidEndSepPunct{\mcitedefaultmidpunct}
{\mcitedefaultendpunct}{\mcitedefaultseppunct}\relax
\EndOfBibitem
\bibitem[Becker \latin{et~al.}(1996)Becker, Seixas~de Melo, Ma\c{c}anita, and
  Elisei]{becker1996jpc}
Becker,~R.~S.; Seixas~de Melo,~J.; Ma\c{c}anita,~A.~L.; Elisei,~F.
  Comprehensive Evaluation of the Absorption, Photophysical, Energy Transfer,
  Structural, and Theoretical Properties of $\alpha$-Oligothiophenes with One
  to Seven Rings. \emph{J. Phys. Chem.} \textbf{1996}, \emph{100},
  18683--18695\relax
\mciteBstWouldAddEndPuncttrue
\mciteSetBstMidEndSepPunct{\mcitedefaultmidpunct}
{\mcitedefaultendpunct}{\mcitedefaultseppunct}\relax
\EndOfBibitem
\bibitem[{Seixas de Melo} \latin{et~al.}(1999){Seixas de Melo}, Silva, Arnaut,
  and Becker]{seixas1999jcp}
{Seixas de Melo},~J.; Silva,~L.~M.; Arnaut,~L.~G.; Becker,~R.~S. Singlet and
  triplet energies of $\alpha$-oligothiophenes: A spectroscopic, theoretical,
  and photoacoustic study: Extrapolation to polythiophene. \emph{J. Chem.
  Phys.} \textbf{1999}, \emph{111}, 5427--5433\relax
\mciteBstWouldAddEndPuncttrue
\mciteSetBstMidEndSepPunct{\mcitedefaultmidpunct}
{\mcitedefaultendpunct}{\mcitedefaultseppunct}\relax
\EndOfBibitem
\bibitem[Shizuka \latin{et~al.}(1982)Shizuka, Ueki, Iizuka, and
  Kanamaru]{shizuka1982jpc}
Shizuka,~H.; Ueki,~Y.; Iizuka,~T.; Kanamaru,~N. Radiative and radiationless
  transitions in the excited state of methyl- and methylene-substituted
  benzenes in condensed media. \emph{J. Phys. Chem.} \textbf{1982}, \emph{86},
  3327--3333\relax
\mciteBstWouldAddEndPuncttrue
\mciteSetBstMidEndSepPunct{\mcitedefaultmidpunct}
{\mcitedefaultendpunct}{\mcitedefaultseppunct}\relax
\EndOfBibitem
\bibitem[Marchetti and Kearns(1967)Marchetti, and Kearns]{marchetti1967jacs}
Marchetti,~A.~P.; Kearns,~D.~R. Investigation of Singlet--Triplet Transitions
  by the Phosphorescence Excitation Method. IV. The Singlet--Triplet Absorption
  Spectra of Aromatic Hydrocarbons. \emph{J. Am. Chem. Soc.} \textbf{1967},
  \emph{89}, 768--777\relax
\mciteBstWouldAddEndPuncttrue
\mciteSetBstMidEndSepPunct{\mcitedefaultmidpunct}
{\mcitedefaultendpunct}{\mcitedefaultseppunct}\relax
\EndOfBibitem
\bibitem[Kearns and Case(1966)Kearns, and Case]{kearns1966jacs}
Kearns,~D.~R.; Case,~W.~A. Investigation of Singlet $\rightarrow$ Triplet
  Transitions by the Phosphorescence Excitation Method. III. Aromatic Ketones
  and Aldehydes. \emph{J. Am. Chem. Soc.} \textbf{1966}, \emph{88},
  5087--5097\relax
\mciteBstWouldAddEndPuncttrue
\mciteSetBstMidEndSepPunct{\mcitedefaultmidpunct}
{\mcitedefaultendpunct}{\mcitedefaultseppunct}\relax
\EndOfBibitem
\bibitem[Ghoshal \latin{et~al.}(1981)Ghoshal, Sarkar, and
  Kastha]{ghoshal1981effects}
Ghoshal,~S.~K.; Sarkar,~S.~K.; Kastha,~G.~S. Effects of Intermolecular
  Hydrogen-bonding on the Luminescence Properties of Acetophenone,
  Characterization of Emission States. \emph{Bull. Chem. Soc. Jpn.}
  \textbf{1981}, \emph{54}, 3556--3561\relax
\mciteBstWouldAddEndPuncttrue
\mciteSetBstMidEndSepPunct{\mcitedefaultmidpunct}
{\mcitedefaultendpunct}{\mcitedefaultseppunct}\relax
\EndOfBibitem
\bibitem[Kuboyama and Yabe(1967)Kuboyama, and
  Yabe]{kuboyama1967phosphorescence}
Kuboyama,~A.; Yabe,~S. Phosphorescence bands of quinones and
  $\alpha$-diketones. \emph{Bull. Chem. Soc. Jpn.} \textbf{1967}, \emph{40},
  2475--2479\relax
\mciteBstWouldAddEndPuncttrue
\mciteSetBstMidEndSepPunct{\mcitedefaultmidpunct}
{\mcitedefaultendpunct}{\mcitedefaultseppunct}\relax
\EndOfBibitem
\bibitem[Borkman and Kearns(1967)Borkman, and Kearns]{borkman1967jcp}
Borkman,~R.~F.; Kearns,~D.~R. Heavy-Atom and Substituent Effects on $S-T$
  Transitions of Halogenated Carbonyl Compounds. \emph{J. Chem. Phys.}
  \textbf{1967}, \emph{46}, 2333--2341\relax
\mciteBstWouldAddEndPuncttrue
\mciteSetBstMidEndSepPunct{\mcitedefaultmidpunct}
{\mcitedefaultendpunct}{\mcitedefaultseppunct}\relax
\EndOfBibitem
\bibitem[Kanda \latin{et~al.}(1961)Kanda, Shimada, and Sakai]{KANDA19611}
Kanda,~Y.; Shimada,~R.; Sakai,~Y. The phosphorescence spectrum of biphenyl at
  90 {\degree}K. \emph{Spectrochim. Acta} \textbf{1961}, \emph{17}, 1--6\relax
\mciteBstWouldAddEndPuncttrue
\mciteSetBstMidEndSepPunct{\mcitedefaultmidpunct}
{\mcitedefaultendpunct}{\mcitedefaultseppunct}\relax
\EndOfBibitem
\bibitem[Berlman(1971)]{berlman1971handbook}
Berlman,~I.~B. \emph{Handbook of Fluorescence Spectra of Aromatic Molecules},
  2nd ed.; Academic Press, 1971\relax
\mciteBstWouldAddEndPuncttrue
\mciteSetBstMidEndSepPunct{\mcitedefaultmidpunct}
{\mcitedefaultendpunct}{\mcitedefaultseppunct}\relax
\EndOfBibitem
\bibitem[Lim and Li(1970)Lim, and Li]{lim1970jcp}
Lim,~E.~C.; Li,~Y.~H. Luminescence of Biphenyl and Geometry of the Molecule in
  Excited Electronic States. \emph{J. Chem. Phys.} \textbf{1970}, \emph{52},
  6416--6422\relax
\mciteBstWouldAddEndPuncttrue
\mciteSetBstMidEndSepPunct{\mcitedefaultmidpunct}
{\mcitedefaultendpunct}{\mcitedefaultseppunct}\relax
\EndOfBibitem
\bibitem[Bulliard \latin{et~al.}(1999)Bulliard, Allan, Wirtz, Haselbach,
  Zachariasse, Detzer, and Grimme]{bulliard1999jpca}
Bulliard,~C.; Allan,~M.; Wirtz,~G.; Haselbach,~E.; Zachariasse,~K.~A.;
  Detzer,~N.; Grimme,~S. Electron Energy Loss and {DFT/SCI} Study of the
  Singlet and Triplet Excited States of Aminobenzonitriles and
  Benzoquinuclidines: Role of the Amino Group Twist Angle. \emph{J. Phys. Chem.
  A} \textbf{1999}, \emph{103}, 7766--7772\relax
\mciteBstWouldAddEndPuncttrue
\mciteSetBstMidEndSepPunct{\mcitedefaultmidpunct}
{\mcitedefaultendpunct}{\mcitedefaultseppunct}\relax
\EndOfBibitem
\bibitem[Adams \latin{et~al.}(1973)Adams, Mantulin, and Huber]{adams1973jacs}
Adams,~J.~E.; Mantulin,~W.~W.; Huber,~J.~R. Effect of molecular geometry on
  spin--orbit coupling of aromatic amines in solution. Diphenylamine,
  iminobibenzyl, acridan, and carbazole. \emph{J. Am. Chem. Soc.}
  \textbf{1973}, \emph{95}, 5477--5481\relax
\mciteBstWouldAddEndPuncttrue
\mciteSetBstMidEndSepPunct{\mcitedefaultmidpunct}
{\mcitedefaultendpunct}{\mcitedefaultseppunct}\relax
\EndOfBibitem
\bibitem[Grimme and Waletzke(1999)Grimme, and Waletzke]{grimme1999jcp}
Grimme,~S.; Waletzke,~M. A combination of Kohn--Sham density functional theory
  and multi-reference configuration interaction methods. \emph{J. Chem. Phys.}
  \textbf{1999}, \emph{111}, 5645--5655\relax
\mciteBstWouldAddEndPuncttrue
\mciteSetBstMidEndSepPunct{\mcitedefaultmidpunct}
{\mcitedefaultendpunct}{\mcitedefaultseppunct}\relax
\EndOfBibitem
\bibitem[Carsey \latin{et~al.}(1979)Carsey, Findley, and
  McGlynn]{carsey1979jacs}
Carsey,~T.~P.; Findley,~G.~L.; McGlynn,~S.~P. Systematics in the electronic
  spectra of polar molecules. 1. $Para$-disubstituted benzenes. \emph{J. Am.
  Chem. Soc.} \textbf{1979}, \emph{101}, 4502--4510\relax
\mciteBstWouldAddEndPuncttrue
\mciteSetBstMidEndSepPunct{\mcitedefaultmidpunct}
{\mcitedefaultendpunct}{\mcitedefaultseppunct}\relax
\EndOfBibitem
\bibitem[Fujitsuka \latin{et~al.}(1998)Fujitsuka, Sato, Sezaki, Tanaka,
  Watanabe, and Ito]{A806072J}
Fujitsuka,~M.; Sato,~T.; Sezaki,~F.; Tanaka,~K.; Watanabe,~A.; Ito,~O. Laser
  flash photolysis study on the photoinduced reactions of
  3{,}3$^\prime$-bridged bithiophenes. \emph{J. Chem. Soc.{,} Faraday Trans.}
  \textbf{1998}, \emph{94}, 3331--3337\relax
\mciteBstWouldAddEndPuncttrue
\mciteSetBstMidEndSepPunct{\mcitedefaultmidpunct}
{\mcitedefaultendpunct}{\mcitedefaultseppunct}\relax
\EndOfBibitem
\bibitem[Wasserberg \latin{et~al.}(2005)Wasserberg, Marsal, Meskers, Janssen,
  and Beljonne]{wasserberg2005jpcb}
Wasserberg,~D.; Marsal,~P.; Meskers,~S. C.~J.; Janssen,~R. A.~J.; Beljonne,~D.
  Phosphorescence and Triplet State Energies of Oligothiophenes. \emph{J. Phys.
  Chem. B} \textbf{2005}, \emph{109}, 4410--4415\relax
\mciteBstWouldAddEndPuncttrue
\mciteSetBstMidEndSepPunct{\mcitedefaultmidpunct}
{\mcitedefaultendpunct}{\mcitedefaultseppunct}\relax
\EndOfBibitem
\bibitem[Dyck and McClure(1962)Dyck, and McClure]{dyck1962jcp}
Dyck,~R.~H.; McClure,~D.~S. Ultraviolet Spectra of Stilbene, $p$-Monohalogen
  Stilbenes, and Azobenzene and the $trans$ to $cis$ Photoisomerization
  Process. \emph{J. Chem. Phys.} \textbf{1962}, \emph{36}, 2326--2345\relax
\mciteBstWouldAddEndPuncttrue
\mciteSetBstMidEndSepPunct{\mcitedefaultmidpunct}
{\mcitedefaultendpunct}{\mcitedefaultseppunct}\relax
\EndOfBibitem
\bibitem[Saltiel \latin{et~al.}(1980)Saltiel, Khalil, and
  Schanze]{SALTIEL1980233}
Saltiel,~J.; Khalil,~G.-E.; Schanze,~K. $Trans$-stilbene phosphorescence.
  \emph{Chem. Phys. Lett.} \textbf{1980}, \emph{70}, 233--235\relax
\mciteBstWouldAddEndPuncttrue
\mciteSetBstMidEndSepPunct{\mcitedefaultmidpunct}
{\mcitedefaultendpunct}{\mcitedefaultseppunct}\relax
\EndOfBibitem
\bibitem[Beljonne \latin{et~al.}(1996)Beljonne, Cornil, Friend, Janssen, and
  Br\'{e}das]{beljonne1996jacs}
Beljonne,~D.; Cornil,~J.; Friend,~R.~H.; Janssen,~R. A.~J.; Br\'{e}das,~J.~L.
  Influence of Chain Length and Derivatization on the Lowest Singlet and
  Triplet States and Intersystem Crossing in Oligothiophenes. \emph{J. Am.
  Chem. Soc.} \textbf{1996}, \emph{118}, 6453--6461\relax
\mciteBstWouldAddEndPuncttrue
\mciteSetBstMidEndSepPunct{\mcitedefaultmidpunct}
{\mcitedefaultendpunct}{\mcitedefaultseppunct}\relax
\EndOfBibitem
\bibitem[Scaiano \latin{et~al.}(1990)Scaiano, Redmond, Mehta, and
  Arnason]{PHP:PHP655}
Scaiano,~J.~C.; Redmond,~R.~W.; Mehta,~B.; Arnason,~J.~T. Efficiency of the
  Photoprocesses Leading to Singlet Oxygen ($^1\Delta_g$) Generation by
  $\alpha$-Terthinenyl: Optical Absorption, Optoacoustic Calorimetry and
  Infrared Luminescence Studies. \emph{Photochem. Photobiol.} \textbf{1990},
  \emph{52}, 655--659\relax
\mciteBstWouldAddEndPuncttrue
\mciteSetBstMidEndSepPunct{\mcitedefaultmidpunct}
{\mcitedefaultendpunct}{\mcitedefaultseppunct}\relax
\EndOfBibitem
\bibitem[Burke \latin{et~al.}(1973)Burke, Small, Braun, and Lin]{BURKE1973574}
Burke,~F.~P.; Small,~G.~J.; Braun,~J.~R.; Lin,~T.-S. The polarized absorption,
  fluorescence and phosphorescence spectra of 1,3-diazaazulene. \emph{Chem.
  Phys. Lett.} \textbf{1973}, \emph{19}, 574--579\relax
\mciteBstWouldAddEndPuncttrue
\mciteSetBstMidEndSepPunct{\mcitedefaultmidpunct}
{\mcitedefaultendpunct}{\mcitedefaultseppunct}\relax
\EndOfBibitem
\bibitem[Herkstroeter(1975)]{Herkstroeter1975jacs}
Herkstroeter,~W.~G. Triplet energies of azulene, $\beta$-carotene, and
  ferrocene. \emph{J. Am. Chem. Soc.} \textbf{1975}, \emph{97},
  4161--4167\relax
\mciteBstWouldAddEndPuncttrue
\mciteSetBstMidEndSepPunct{\mcitedefaultmidpunct}
{\mcitedefaultendpunct}{\mcitedefaultseppunct}\relax
\EndOfBibitem
\bibitem[Tway and Love(1982)Tway, and Love]{tway1982jpc}
Tway,~P.~C.; Love,~L. J.~C. Photophysical properties of benzimidazole and
  thiabendazole and their homologs. Effect of substituents and solvent on the
  nature of the transition. \emph{J. Phys. Chem.} \textbf{1982}, \emph{86},
  5223--5226\relax
\mciteBstWouldAddEndPuncttrue
\mciteSetBstMidEndSepPunct{\mcitedefaultmidpunct}
{\mcitedefaultendpunct}{\mcitedefaultseppunct}\relax
\EndOfBibitem
\bibitem[Zander(1985)]{zander1985z}
Zander,~M. Zur Photolumineszenz von Benzologen des Thiophens. \emph{Z.
  Naturforsch.} \textbf{1985}, \emph{40A}, 497--502\relax
\mciteBstWouldAddEndPuncttrue
\mciteSetBstMidEndSepPunct{\mcitedefaultmidpunct}
{\mcitedefaultendpunct}{\mcitedefaultseppunct}\relax
\EndOfBibitem
\bibitem[Goodman and Harrell(1959)Goodman, and Harrell]{goodman1959jcp}
Goodman,~L.; Harrell,~R.~W. Calculation of $n\rightarrow\pi^\star$ Transition
  Energies in N-Heterocyclic Molecules by a One-Electron Approximation.
  \emph{J. Chem. Phys.} \textbf{1959}, \emph{30}, 1131--1138\relax
\mciteBstWouldAddEndPuncttrue
\mciteSetBstMidEndSepPunct{\mcitedefaultmidpunct}
{\mcitedefaultendpunct}{\mcitedefaultseppunct}\relax
\EndOfBibitem
\bibitem[Wintgens \latin{et~al.}(1994)Wintgens, Valat, Kossanyi, Biczok,
  Demeter, and Berces]{FT9949000411}
Wintgens,~V.; Valat,~P.; Kossanyi,~J.; Biczok,~L.; Demeter,~A.; Berces,~T.
  Spectroscopic properties of aromatic dicarboximides. Part 1.-N-H and
  $N$-methyl-substituted naphthalimides. \emph{J. Chem. Soc.{,} Faraday Trans.}
  \textbf{1994}, \emph{90}, 411--421\relax
\mciteBstWouldAddEndPuncttrue
\mciteSetBstMidEndSepPunct{\mcitedefaultmidpunct}
{\mcitedefaultendpunct}{\mcitedefaultseppunct}\relax
\EndOfBibitem
\bibitem[Evans(1959)]{JR9590002753}
Evans,~D.~F. Magnetic perturbation of singlet--triplet transitions. Part III.
  Benzene derivatives and heterocyclic compounds. \emph{J. Chem. Soc.}
  \textbf{1959}, 2753--2757\relax
\mciteBstWouldAddEndPuncttrue
\mciteSetBstMidEndSepPunct{\mcitedefaultmidpunct}
{\mcitedefaultendpunct}{\mcitedefaultseppunct}\relax
\EndOfBibitem
\bibitem[Najbar \latin{et~al.}(1980)Najbar, Trzcinska, Urbanek, and
  Proniewicz]{najbar1980japp}
Najbar,~J.; Trzcinska,~B.~M.; Urbanek,~Z.~H.; Proniewicz,~L.~M. \emph{Acta
  Phys. Pol., A} \textbf{1980}, \emph{A58}, 331--344\relax
\mciteBstWouldAddEndPuncttrue
\mciteSetBstMidEndSepPunct{\mcitedefaultmidpunct}
{\mcitedefaultendpunct}{\mcitedefaultseppunct}\relax
\EndOfBibitem
\bibitem[Suga and Kinoshita(1982)Suga, and Kinoshita]{suga1982static}
Suga,~K.; Kinoshita,~M. Static and Dynamic Properties of Quinoxalines in the
  Phosphorescent Triplet State from Optically Detected Magnetic Resonance.
  \emph{Bull. Chem. Soc. Jpn.} \textbf{1982}, \emph{55}, 1695--1704\relax
\mciteBstWouldAddEndPuncttrue
\mciteSetBstMidEndSepPunct{\mcitedefaultmidpunct}
{\mcitedefaultendpunct}{\mcitedefaultseppunct}\relax
\EndOfBibitem
\bibitem[Anderson \latin{et~al.}(1994)Anderson, An, Rubin, and
  Foote]{anderson1994jacs}
Anderson,~J.~L.; An,~Y.-Z.; Rubin,~Y.; Foote,~C.~S. Photophysical
  Characterization and Singlet Oxygen Yield of a Dihydrofullerene. \emph{J. Am.
  Chem. Soc.} \textbf{1994}, \emph{116}, 9763--9764\relax
\mciteBstWouldAddEndPuncttrue
\mciteSetBstMidEndSepPunct{\mcitedefaultmidpunct}
{\mcitedefaultendpunct}{\mcitedefaultseppunct}\relax
\EndOfBibitem
\bibitem[Johnstone and Mellon(1973)Johnstone, and Mellon]{F29736901155}
Johnstone,~R. A.~W.; Mellon,~F.~A. Photoelectron spectroscopy of
  sulphur-containing heteroaromatics and molecular orbital calculations.
  \emph{J. Chem. Soc.{,} Faraday Trans. 2} \textbf{1973}, \emph{69},
  1155--1163\relax
\mciteBstWouldAddEndPuncttrue
\mciteSetBstMidEndSepPunct{\mcitedefaultmidpunct}
{\mcitedefaultendpunct}{\mcitedefaultseppunct}\relax
\EndOfBibitem
\bibitem[Eland(1969)]{ELAND1969471}
Eland,~J. Photoelectron spectra of conjugated hydrocarbons and heteromolecules.
  \emph{Int. J. Mass Spectrom. Ion Phys.} \textbf{1969}, \emph{2},
  471--484\relax
\mciteBstWouldAddEndPuncttrue
\mciteSetBstMidEndSepPunct{\mcitedefaultmidpunct}
{\mcitedefaultendpunct}{\mcitedefaultseppunct}\relax
\EndOfBibitem
\bibitem[Muigg \latin{et~al.}(1996)Muigg, Scheier, Becker, and
  M\"{a}rk]{muigg1996jpb}
Muigg,~D.; Scheier,~P.; Becker,~K.; M\"{a}rk,~T.~D. Measured appearance
  energies of $\ce{C_{\it n}^+}$ ($3 \le n \le 10$) fragment ions produced by
  electron impact on $\ce{C60}$. \emph{J. Phys. B} \textbf{1996}, \emph{29},
  5193--5198\relax
\mciteBstWouldAddEndPuncttrue
\mciteSetBstMidEndSepPunct{\mcitedefaultmidpunct}
{\mcitedefaultendpunct}{\mcitedefaultseppunct}\relax
\EndOfBibitem
\bibitem[Huang \latin{et~al.}(2014)Huang, Dau, Liu, and Wang]{huang2014jcp}
Huang,~D.-L.; Dau,~P.~D.; Liu,~H.-T.; Wang,~L.-S. High-resolution photoelectron
  imaging of cold $\ce{C60}$ anions and accurate determination of the electron
  affinity of $\ce{C60}$. \emph{J. Chem. Phys.} \textbf{2014}, \emph{140},
  224315\relax
\mciteBstWouldAddEndPuncttrue
\mciteSetBstMidEndSepPunct{\mcitedefaultmidpunct}
{\mcitedefaultendpunct}{\mcitedefaultseppunct}\relax
\EndOfBibitem
\bibitem[Dewar \latin{et~al.}(1970)Dewar, Haselbach, and Worley]{Dewar431}
Dewar,~M. J.~S.; Haselbach,~E.; Worley,~S.~D. Calculated and Observed
  Ionization Potentials of Unsaturated Polycyclic Hydrocarbons; Calculated
  Heats of Formation by Several Semiempirical S.C.F. M.O. Methods.
  \emph{Proceedings of the Royal Society of London A: Mathematical, Physical
  and Engineering Sciences} \textbf{1970}, \emph{315}, 431--442\relax
\mciteBstWouldAddEndPuncttrue
\mciteSetBstMidEndSepPunct{\mcitedefaultmidpunct}
{\mcitedefaultendpunct}{\mcitedefaultseppunct}\relax
\EndOfBibitem
\bibitem[Wojn\'{a}rovits and F\"{o}ldi\'{a}k(1981)Wojn\'{a}rovits, and
  F\"{o}ldi\'{a}k]{WOJNAROVITS1981511}
Wojn\'{a}rovits,~L.; F\"{o}ldi\'{a}k,~G. Electron-capture detection of aromatic
  hydrocarbons. \emph{J. Chromatogr. A} \textbf{1981}, \emph{206},
  511--519\relax
\mciteBstWouldAddEndPuncttrue
\mciteSetBstMidEndSepPunct{\mcitedefaultmidpunct}
{\mcitedefaultendpunct}{\mcitedefaultseppunct}\relax
\EndOfBibitem
\bibitem[Eley \latin{et~al.}(1973)Eley, Hazeldine, and Palmer]{F29736901808}
Eley,~D.~D.; Hazeldine,~D.~J.; Palmer,~T.~F. Mass spectra, ionisation
  potentials and related properties of metal-free and transition metal
  phthalocyanines. \emph{J. Chem. Soc.{,} Faraday Trans. 2} \textbf{1973},
  \emph{69}, 1808--1814\relax
\mciteBstWouldAddEndPuncttrue
\mciteSetBstMidEndSepPunct{\mcitedefaultmidpunct}
{\mcitedefaultendpunct}{\mcitedefaultseppunct}\relax
\EndOfBibitem
\bibitem[Khandelwal and Roebber(1975)Khandelwal, and
  Roebber]{KHANDELWAL1975355}
Khandelwal,~S.~C.; Roebber,~J.~L. The photoelectron spectra of
  tetraphenylporphine and some metallotetraphenylporphyrins. \emph{Chem. Phys.
  Lett.} \textbf{1975}, \emph{34}, 355--359\relax
\mciteBstWouldAddEndPuncttrue
\mciteSetBstMidEndSepPunct{\mcitedefaultmidpunct}
{\mcitedefaultendpunct}{\mcitedefaultseppunct}\relax
\EndOfBibitem
\bibitem[Butler and Baer(1980)Butler, and Baer]{butler1980jacs}
Butler,~J.~J.; Baer,~T. Thermochemistry and dissociation dynamics of state
  selected $\ce{C4H4X}$ ions. 1. Thiophene. \emph{J. Am. Chem. Soc.}
  \textbf{1980}, \emph{102}, 6764--6769\relax
\mciteBstWouldAddEndPuncttrue
\mciteSetBstMidEndSepPunct{\mcitedefaultmidpunct}
{\mcitedefaultendpunct}{\mcitedefaultseppunct}\relax
\EndOfBibitem
\bibitem[Hunter and Lias(1998)Hunter, and Lias]{edward1998jpcrd}
Hunter,~E. P.~L.; Lias,~S.~G. Evaluated Gas Phase Basicities and Proton
  Affinities of Molecules: An Update. \emph{J. Phys. Chem. Ref. Data}
  \textbf{1998}, \emph{27}, 413--656\relax
\mciteBstWouldAddEndPuncttrue
\mciteSetBstMidEndSepPunct{\mcitedefaultmidpunct}
{\mcitedefaultendpunct}{\mcitedefaultseppunct}\relax
\EndOfBibitem
\bibitem[Sato \latin{et~al.}(1981)Sato, Seki, and Inokuchi]{F29817701621}
Sato,~N.; Seki,~K.; Inokuchi,~H. Polarization energies of organic solids
  determined by ultraviolet photoelectron spectroscopy. \emph{J. Chem. Soc.{,}
  Faraday Trans. 2} \textbf{1981}, \emph{77}, 1621--1633\relax
\mciteBstWouldAddEndPuncttrue
\mciteSetBstMidEndSepPunct{\mcitedefaultmidpunct}
{\mcitedefaultendpunct}{\mcitedefaultseppunct}\relax
\EndOfBibitem
\bibitem[Chen \latin{et~al.}(1991)Chen, Pan, Groh, Hagan, and
  Ridge]{chen1991jacs}
Chen,~H.~L.; Pan,~Y.~H.; Groh,~S.; Hagan,~T.~E.; Ridge,~D.~P. Gas-phase
  charge-transfer reactions and electron affinities of macrocyclic, anionic
  nickel complexes: Ni(SALEN), Ni(tetraphenylporphyrin), and derivatives.
  \emph{J. Am. Chem. Soc.} \textbf{1991}, \emph{113}, 2766--2767\relax
\mciteBstWouldAddEndPuncttrue
\mciteSetBstMidEndSepPunct{\mcitedefaultmidpunct}
{\mcitedefaultendpunct}{\mcitedefaultseppunct}\relax
\EndOfBibitem
\bibitem[Pasinszki \latin{et~al.}(2010)Pasinszki, Krebsz, and
  Vass]{PASINSZKI201085}
Pasinszki,~T.; Krebsz,~M.; Vass,~G. Ground and ionic states of
  1,2,5-thiadiazoles: An UV-photoelectron spectroscopic and theoretical study.
  \emph{J. Mol. Spect.} \textbf{2010}, \emph{966}, 85--91\relax
\mciteBstWouldAddEndPuncttrue
\mciteSetBstMidEndSepPunct{\mcitedefaultmidpunct}
{\mcitedefaultendpunct}{\mcitedefaultseppunct}\relax
\EndOfBibitem
\bibitem[Berkowitz(1979)]{berkowitz1979jcp}
Berkowitz,~J. Photoelectron spectroscopy of phthalocyanine vapors. \emph{J.
  Chem. Phys.} \textbf{1979}, \emph{70}, 2819--2828\relax
\mciteBstWouldAddEndPuncttrue
\mciteSetBstMidEndSepPunct{\mcitedefaultmidpunct}
{\mcitedefaultendpunct}{\mcitedefaultseppunct}\relax
\EndOfBibitem
\bibitem[Stenuit \latin{et~al.}(2010)Stenuit, Castellarin-Cudia, Plekan, Feyer,
  Prince, Goldoni, and Umari]{C004332J}
Stenuit,~G.; Castellarin-Cudia,~C.; Plekan,~O.; Feyer,~V.; Prince,~K.~C.;
  Goldoni,~A.; Umari,~P. Valence electronic properties of porphyrin
  derivatives. \emph{Phys. Chem. Chem. Phys.} \textbf{2010}, \emph{12},
  10812--10817\relax
\mciteBstWouldAddEndPuncttrue
\mciteSetBstMidEndSepPunct{\mcitedefaultmidpunct}
{\mcitedefaultendpunct}{\mcitedefaultseppunct}\relax
\EndOfBibitem
\bibitem[Dori \latin{et~al.}(2006)Dori, Menon, Kilian, Sokolowski, Kronik, and
  Umbach]{PhysRevB.73.195208}
Dori,~N.; Menon,~M.; Kilian,~L.; Sokolowski,~M.; Kronik,~L.; Umbach,~E. Valence
  electronic structure of gas-phase 3,4,9,10-perylene tetracarboxylic acid
  dianhydride: Experiment and theory. \emph{Phys. Rev. B} \textbf{2006},
  \emph{73}, 195208\relax
\mciteBstWouldAddEndPuncttrue
\mciteSetBstMidEndSepPunct{\mcitedefaultmidpunct}
{\mcitedefaultendpunct}{\mcitedefaultseppunct}\relax
\EndOfBibitem
\bibitem[Dallinga \latin{et~al.}(1981)Dallinga, Nibbering, and
  Louter]{OMS:OMS1210160409}
Dallinga,~J.~W.; Nibbering,~N. M.~M.; Louter,~G.~J. Formation and structure of
  $\ce{[C8H8O]+\cdot}$ ions, generated from gas phase ions of
  phenyl-cyclopropylcarbinol and 1-phenyl-1-(hydroxymethyl)cyclopropane.
  \emph{Org. Mass Spectrom.} \textbf{1981}, \emph{16}, 183--187\relax
\mciteBstWouldAddEndPuncttrue
\mciteSetBstMidEndSepPunct{\mcitedefaultmidpunct}
{\mcitedefaultendpunct}{\mcitedefaultseppunct}\relax
\EndOfBibitem
\bibitem[Wentworth \latin{et~al.}(1975)Wentworth, Kao, and
  Becker]{wentworth1975jpc}
Wentworth,~W.~E.; Kao,~L.~W.; Becker,~R.~S. Electron affinities of substituted
  aromatic compounds. \emph{J. Phys. Chem.} \textbf{1975}, \emph{79},
  1161--1169\relax
\mciteBstWouldAddEndPuncttrue
\mciteSetBstMidEndSepPunct{\mcitedefaultmidpunct}
{\mcitedefaultendpunct}{\mcitedefaultseppunct}\relax
\EndOfBibitem
\bibitem[Paul and Kebarle(1989)Paul, and Kebarle]{paul1989jacs}
Paul,~G.; Kebarle,~P. Electron affinities of cyclic unsaturated dicarbonyls:
  maleic anhydrides, maleimides, and cyclopentenedione. \emph{J. Am. Chem.
  Soc.} \textbf{1989}, \emph{111}, 464--470\relax
\mciteBstWouldAddEndPuncttrue
\mciteSetBstMidEndSepPunct{\mcitedefaultmidpunct}
{\mcitedefaultendpunct}{\mcitedefaultseppunct}\relax
\EndOfBibitem
\bibitem[Polevoi \latin{et~al.}(1987)Polevoi, Matyuk, Grigor'eva, and
  Potapov]{polevoi1987formation}
Polevoi,~A.~V.; Matyuk,~V.~M.; Grigor'eva,~G.~A.; Potapov,~V.~K. Formation of
  intermediate products during the resonance stepwise polarization of dibenzyl
  ketone and benzil molecules. \emph{High Energy Chem.(Engl. Transl.);(United
  States)} \textbf{1987}, \emph{21}, 17--21\relax
\mciteBstWouldAddEndPuncttrue
\mciteSetBstMidEndSepPunct{\mcitedefaultmidpunct}
{\mcitedefaultendpunct}{\mcitedefaultseppunct}\relax
\EndOfBibitem
\bibitem[Gr\"{u}tzmacher and Schubert(1979)Gr\"{u}tzmacher, and
  Schubert]{OMS:OMS1210141011}
Gr\"{u}tzmacher,~H.-F.; Schubert,~R. Substituent effects in the mass spectra of
  benzoyl hetarenes. \emph{Org. Mass Spectrom.} \textbf{1979}, \emph{14},
  567--570\relax
\mciteBstWouldAddEndPuncttrue
\mciteSetBstMidEndSepPunct{\mcitedefaultmidpunct}
{\mcitedefaultendpunct}{\mcitedefaultseppunct}\relax
\EndOfBibitem
\bibitem[Maeyama \latin{et~al.}(2008)Maeyama, Yagi, Fujii, and
  Mikami]{MAEYAMA200818}
Maeyama,~T.; Yagi,~I.; Fujii,~A.; Mikami,~N. Photoelectron spectroscopy of
  microsolvated benzophenone radical anions to reveal the origin of
  solvatochromic shifts in alcoholic media. \emph{Chem. Phys. Lett.}
  \textbf{2008}, \emph{457}, 18--22\relax
\mciteBstWouldAddEndPuncttrue
\mciteSetBstMidEndSepPunct{\mcitedefaultmidpunct}
{\mcitedefaultendpunct}{\mcitedefaultseppunct}\relax
\EndOfBibitem
\bibitem[Loudon and Mazengo(1974)Loudon, and Mazengo]{OMS:OMS1210080121}
Loudon,~A.~G.; Mazengo,~R.~Z. Steric strain and electron-impact. The behaviour
  of some $n$,$n^\prime$-dimethyl-1,1-binaphthyls, some
  $n$,$n^\prime$-dimethylbiphenyls and model compounds. \emph{Org. Mass
  Spectrom.} \textbf{1974}, \emph{8}, 179--187\relax
\mciteBstWouldAddEndPuncttrue
\mciteSetBstMidEndSepPunct{\mcitedefaultmidpunct}
{\mcitedefaultendpunct}{\mcitedefaultseppunct}\relax
\EndOfBibitem
\bibitem[Kobayashi(1983)]{kobayashi1983conformational}
Kobayashi,~T. Conformational analysis of terphenyls by photoelectron
  spectroscopy. \emph{Bull. Chem. Soc. Jpn.} \textbf{1983}, \emph{56},
  3224--3229\relax
\mciteBstWouldAddEndPuncttrue
\mciteSetBstMidEndSepPunct{\mcitedefaultmidpunct}
{\mcitedefaultendpunct}{\mcitedefaultseppunct}\relax
\EndOfBibitem
\bibitem[Haink \latin{et~al.}(1974)Haink, Adams, and
  Huber]{BBPC:BBPC19740780503}
Haink,~H.~J.; Adams,~J.~E.; Huber,~J.~R. The Electronic Structure of Aromatic
  Amines: Photoelectron Spectroscopy of Diphenylamine, Iminobibenzyl, Acridan
  and Carbazole. \emph{Ber. Bunsen-Ges. Phys. Chem.} \textbf{1974}, \emph{78},
  436--440\relax
\mciteBstWouldAddEndPuncttrue
\mciteSetBstMidEndSepPunct{\mcitedefaultmidpunct}
{\mcitedefaultendpunct}{\mcitedefaultseppunct}\relax
\EndOfBibitem
\bibitem[Debies and Rabalais(1974)Debies, and Rabalais]{debies1974ic}
Debies,~T.~P.; Rabalais,~J.~W. Photoelectron spectra of substituted benzenes.
  III. Bonding with Group V substituents. \emph{Inorg. Chem.} \textbf{1974},
  \emph{13}, 308--312\relax
\mciteBstWouldAddEndPuncttrue
\mciteSetBstMidEndSepPunct{\mcitedefaultmidpunct}
{\mcitedefaultendpunct}{\mcitedefaultseppunct}\relax
\EndOfBibitem
\bibitem[Potapov and Sorokin(1971)Potapov, and
  Sorokin]{potapov1971photoionization}
Potapov,~V.~K.; Sorokin,~V.~V. Photoionization and ion-molecule reactions in
  quinones and alcohols. \emph{High Energy Chem.(Engl. Transl.);(United
  States)} \textbf{1971}, \emph{5}, 435--487\relax
\mciteBstWouldAddEndPuncttrue
\mciteSetBstMidEndSepPunct{\mcitedefaultmidpunct}
{\mcitedefaultendpunct}{\mcitedefaultseppunct}\relax
\EndOfBibitem
\bibitem[Lipert and Colson(1990)Lipert, and Colson]{lipert1990jcp}
Lipert,~R.~J.; Colson,~S.~D. Accurate ionization potentials of phenol and
  phenol-($\ce{H2O}$) from the electric field dependence of the pump--probe
  photoionization threshold. \emph{J. Chem. Phys.} \textbf{1990}, \emph{92},
  3240--3241\relax
\mciteBstWouldAddEndPuncttrue
\mciteSetBstMidEndSepPunct{\mcitedefaultmidpunct}
{\mcitedefaultendpunct}{\mcitedefaultseppunct}\relax
\EndOfBibitem
\bibitem[Hudson \latin{et~al.}(1976)Hudson, Ridyard, and
  Diamond]{hudson1976jacs}
Hudson,~B.~S.; Ridyard,~J. N.~A.; Diamond,~J. Polyene spectroscopy.
  Photoelectron spectra of the diphenylpolyenes. \emph{J. Am. Chem. Soc.}
  \textbf{1976}, \emph{98}, 1126--1129\relax
\mciteBstWouldAddEndPuncttrue
\mciteSetBstMidEndSepPunct{\mcitedefaultmidpunct}
{\mcitedefaultendpunct}{\mcitedefaultseppunct}\relax
\EndOfBibitem
\bibitem[Siegert \latin{et~al.}(2011)Siegert, Vogeler, Marian, and
  Weinkauf]{C0CP02712J}
Siegert,~S.; Vogeler,~F.; Marian,~C.~M.; Weinkauf,~R. Throwing light on dark
  states of $\alpha$-oligothiophenes of chain lengths 2 to 6: radical anion
  photoelectron spectroscopy and excited-state theory. \emph{Phys. Chem. Chem.
  Phys.} \textbf{2011}, \emph{13}, 10350--10363\relax
\mciteBstWouldAddEndPuncttrue
\mciteSetBstMidEndSepPunct{\mcitedefaultmidpunct}
{\mcitedefaultendpunct}{\mcitedefaultseppunct}\relax
\EndOfBibitem
\bibitem[Lu \latin{et~al.}(1992)Lu, Eiden, and Weisshaar]{lu1992jpc}
Lu,~K.~T.; Eiden,~G.~C.; Weisshaar,~J.~C. Toluene cation: nearly free rotation
  of the methyl group. \emph{J. Phys. Chem.} \textbf{1992}, \emph{96},
  9742--9748\relax
\mciteBstWouldAddEndPuncttrue
\mciteSetBstMidEndSepPunct{\mcitedefaultmidpunct}
{\mcitedefaultendpunct}{\mcitedefaultseppunct}\relax
\EndOfBibitem
\bibitem[Schiedt \latin{et~al.}(2000)Schiedt, Knott, {Le Barbu}, Schlag, and
  Weinkauf]{shiedt2000jcp}
Schiedt,~J.; Knott,~W.~J.; {Le Barbu},~K.; Schlag,~E.~W.; Weinkauf,~R.
  Microsolvation of similar-sized aromatic molecules: Photoelectron
  spectroscopy of bithiophene--, azulene--, and naphthalene--water anion
  clusters. \emph{J. Chem. Phys.} \textbf{2000}, \emph{113}, 9470--9478\relax
\mciteBstWouldAddEndPuncttrue
\mciteSetBstMidEndSepPunct{\mcitedefaultmidpunct}
{\mcitedefaultendpunct}{\mcitedefaultseppunct}\relax
\EndOfBibitem
\bibitem[Jochims \latin{et~al.}(1992)Jochims, Rasekh, R\"{u}hl, Baumg\"{a}rtel,
  and Leach]{JOCHIMS1992159}
Jochims,~H.~W.; Rasekh,~H.; R\"{u}hl,~E.; Baumg\"{a}rtel,~H.; Leach,~S. The
  photofragmentation of naphthalene and azulene monocations in the energy range
  7--22 eV. \emph{Chem. Phys.} \textbf{1992}, \emph{168}, 159--184\relax
\mciteBstWouldAddEndPuncttrue
\mciteSetBstMidEndSepPunct{\mcitedefaultmidpunct}
{\mcitedefaultendpunct}{\mcitedefaultseppunct}\relax
\EndOfBibitem
\bibitem[Klasinc \latin{et~al.}(1980)Klasinc, Trinajsti\'{c}a, and
  Knop]{QUA:QUA560180739}
Klasinc,~L.; Trinajsti\'{c}a,~N.; Knop,~J.~V. Application of photoelectron
  spectroscopy to biologically active molecules and their constituent parts.
  VIII. Thalidomide. \emph{Int. J. Quant. Chem.} \textbf{1980}, \emph{18},
  403--409\relax
\mciteBstWouldAddEndPuncttrue
\mciteSetBstMidEndSepPunct{\mcitedefaultmidpunct}
{\mcitedefaultendpunct}{\mcitedefaultseppunct}\relax
\EndOfBibitem
\bibitem[Ham and Meer(1972)Ham, and Meer]{VANDENHAM1972447}
Ham,~D. V.~D.; Meer,~D. V.~D. The photoelectron spectra of the
  diazanaphthalenes. \emph{Chem. Phys. Lett.} \textbf{1972}, \emph{12},
  447--453\relax
\mciteBstWouldAddEndPuncttrue
\mciteSetBstMidEndSepPunct{\mcitedefaultmidpunct}
{\mcitedefaultendpunct}{\mcitedefaultseppunct}\relax
\EndOfBibitem
\bibitem[Dillow and Kebarle(1989)Dillow, and Kebarle]{dillow1989cjc}
Dillow,~G.~W.; Kebarle,~P. Electron affinities of aza-substituted polycyclic
  aromatic hydrocarbons. \emph{Can. J. Chem.} \textbf{1989}, \emph{67},
  1628--1631\relax
\mciteBstWouldAddEndPuncttrue
\mciteSetBstMidEndSepPunct{\mcitedefaultmidpunct}
{\mcitedefaultendpunct}{\mcitedefaultseppunct}\relax
\EndOfBibitem
\bibitem[Brogli \latin{et~al.}(1972)Brogli, Heilbronner, and
  Kobayashi]{HLCA:HLCA19720550131}
Brogli,~F.; Heilbronner,~E.; Kobayashi,~T. Photoelectron Spektra of Azabenzenes
  and Azanaphthalenes: II. A Reinvestigation of Azanaphthalenes by
  High-Resolution Photoelectron Spectroscopy. \emph{Helv. Chim. Acta}
  \textbf{1972}, \emph{55}, 274--288\relax
\mciteBstWouldAddEndPuncttrue
\mciteSetBstMidEndSepPunct{\mcitedefaultmidpunct}
{\mcitedefaultendpunct}{\mcitedefaultseppunct}\relax
\EndOfBibitem
\bibitem[Huang \latin{et~al.}(2013)Huang, Zhang, Shiota, Nakagawa, Kuwabara,
  Yoshizawa, and Adachi]{huang2013jctc}
Huang,~S.; Zhang,~Q.; Shiota,~Y.; Nakagawa,~T.; Kuwabara,~K.; Yoshizawa,~K.;
  Adachi,~C. Computational Prediction for Singlet- and Triplet-Transition
  Energies of Charge-Transfer Compounds. \emph{J. Chem. Theory Comput.}
  \textbf{2013}, \emph{9}, 3872--3877\relax
\mciteBstWouldAddEndPuncttrue
\mciteSetBstMidEndSepPunct{\mcitedefaultmidpunct}
{\mcitedefaultendpunct}{\mcitedefaultseppunct}\relax
\EndOfBibitem
\bibitem[Lee \latin{et~al.}(2013)Lee, Shizu, Tanaka, Nomura, Yasuda, and
  Adachi]{C3TC30699B}
Lee,~J.; Shizu,~K.; Tanaka,~H.; Nomura,~H.; Yasuda,~T.; Adachi,~C. Oxadiazole-
  and triazole-based highly-efficient thermally activated delayed fluorescence
  emitters for organic light-emitting diodes. \emph{J. Mater. Chem. C}
  \textbf{2013}, \emph{1}, 4599--4604\relax
\mciteBstWouldAddEndPuncttrue
\mciteSetBstMidEndSepPunct{\mcitedefaultmidpunct}
{\mcitedefaultendpunct}{\mcitedefaultseppunct}\relax
\EndOfBibitem
\bibitem[Wu \latin{et~al.}(2014)Wu, Aonuma, Zhang, Huang, Nakagawa, Kuwabara,
  and Adachi]{C3TC31936A}
Wu,~S.; Aonuma,~M.; Zhang,~Q.; Huang,~S.; Nakagawa,~T.; Kuwabara,~K.;
  Adachi,~C. High-efficiency deep-blue organic light-emitting diodes based on a
  thermally activated delayed fluorescence emitter. \emph{J. Mater. Chem. C}
  \textbf{2014}, \emph{2}, 421--424\relax
\mciteBstWouldAddEndPuncttrue
\mciteSetBstMidEndSepPunct{\mcitedefaultmidpunct}
{\mcitedefaultendpunct}{\mcitedefaultseppunct}\relax
\EndOfBibitem
\bibitem[Zhang \latin{et~al.}(2014)Zhang, Li, Huang, Nomura, Tanaka, and
  Adachi]{zhang2014efficient}
Zhang,~Q.; Li,~B.; Huang,~S.; Nomura,~H.; Tanaka,~H.; Adachi,~C. Efficient blue
  organic light-emitting diodes employing thermally activated delayed
  fluorescence. \emph{Nat. Photon.} \textbf{2014}, \emph{8}, 326--332\relax
\mciteBstWouldAddEndPuncttrue
\mciteSetBstMidEndSepPunct{\mcitedefaultmidpunct}
{\mcitedefaultendpunct}{\mcitedefaultseppunct}\relax
\EndOfBibitem
\bibitem[Hait \latin{et~al.}(2016)Hait, Zhu, McMahon, and {Van
  Voorhis}]{hait2016}
Hait,~D.; Zhu,~T.; McMahon,~D.~P.; {Van Voorhis},~T. Prediction of
  Excited-State Energies and Singlet--Triplet Gaps of Charge-Transfer States
  Using a Restricted Open-Shell Kohn--Sham Approach. \emph{J. Chem. Theory
  Comput.} \textbf{2016}, \emph{12}, 3353--3359\relax
\mciteBstWouldAddEndPuncttrue
\mciteSetBstMidEndSepPunct{\mcitedefaultmidpunct}
{\mcitedefaultendpunct}{\mcitedefaultseppunct}\relax
\EndOfBibitem
\bibitem[Longworth \latin{et~al.}(1966)Longworth, Rahn, and
  Shulman]{longworth1966jcp}
Longworth,~J.~W.; Rahn,~R.~O.; Shulman,~R.~G. Luminescence of Pyrimidines,
  Purines, Nucleosides, and Nucleotides at 77 {\degree}K. The Effect of
  Ionization and Tautomerization. \emph{J. Chem. Phys.} \textbf{1966},
  \emph{45}, 2930--2939\relax
\mciteBstWouldAddEndPuncttrue
\mciteSetBstMidEndSepPunct{\mcitedefaultmidpunct}
{\mcitedefaultendpunct}{\mcitedefaultseppunct}\relax
\EndOfBibitem
\bibitem[Daniels and Hauswirth(1971)Daniels, and Hauswirth]{Daniels675}
Daniels,~M.; Hauswirth,~W. Fluorescence of the Purine and Pyrimidine Bases of
  the Nucleic Acids in Neutral Aqueous Solution at 300 {\degree}K.
  \emph{Science} \textbf{1971}, \emph{171}, 675--677\relax
\mciteBstWouldAddEndPuncttrue
\mciteSetBstMidEndSepPunct{\mcitedefaultmidpunct}
{\mcitedefaultendpunct}{\mcitedefaultseppunct}\relax
\EndOfBibitem
\bibitem[Lin \latin{et~al.}(1980)Lin, Yu, Peng, Akiyama, Li, Lee, and
  LeBreton]{lin1980jacs}
Lin,~J.; Yu,~C.; Peng,~S.; Akiyama,~I.; Li,~K.; Lee,~L.~K.; LeBreton,~P.~R.
  Ultraviolet photoelectron studies of the ground-state electronic structure
  and gas-phase tautomerism of purine and adenine. \emph{J. Am. Chem. Soc.}
  \textbf{1980}, \emph{102}, 4627--4631\relax
\mciteBstWouldAddEndPuncttrue
\mciteSetBstMidEndSepPunct{\mcitedefaultmidpunct}
{\mcitedefaultendpunct}{\mcitedefaultseppunct}\relax
\EndOfBibitem
\bibitem[Aflatooni \latin{et~al.}(1998)Aflatooni, Gallup, and
  Burrow]{aflatooni1998jpca}
Aflatooni,~K.; Gallup,~G.~A.; Burrow,~P.~D. Electron Attachment Energies of the
  DNA Bases. \emph{J. Phys. Chem. A} \textbf{1998}, \emph{102},
  6205--6207\relax
\mciteBstWouldAddEndPuncttrue
\mciteSetBstMidEndSepPunct{\mcitedefaultmidpunct}
{\mcitedefaultendpunct}{\mcitedefaultseppunct}\relax
\EndOfBibitem
\bibitem[Gu\'{e}ron \latin{et~al.}(1967)Gu\'{e}ron, Eisinger, and
  Shulman]{gueron1967jcp}
Gu\'{e}ron,~M.; Eisinger,~J.; Shulman,~R.~G. Excited States of Nucleotides and
  Singlet Energy Transfer in Polynucleotides. \emph{J. Chem. Phys.}
  \textbf{1967}, \emph{47}, 4077--4091\relax
\mciteBstWouldAddEndPuncttrue
\mciteSetBstMidEndSepPunct{\mcitedefaultmidpunct}
{\mcitedefaultendpunct}{\mcitedefaultseppunct}\relax
\EndOfBibitem
\bibitem[Nguyen \latin{et~al.}(2004)Nguyen, Zhang, Nam, and
  Ceulemans]{nguyen2004jpca}
Nguyen,~M.~T.; Zhang,~R.; Nam,~P.-C.; Ceulemans,~A. Singlet--Triplet Energy
  Gaps of Gas-Phase RNA and DNA Bases. A Quantum Chemical Study. \emph{J. Phys.
  Chem. A} \textbf{2004}, \emph{108}, 6554--6561\relax
\mciteBstWouldAddEndPuncttrue
\mciteSetBstMidEndSepPunct{\mcitedefaultmidpunct}
{\mcitedefaultendpunct}{\mcitedefaultseppunct}\relax
\EndOfBibitem
\bibitem[Li \latin{et~al.}(2007)Li, Bowen, Haranczyk, Bachorz, Mazurkiewicz,
  Rak, and Gutowski]{li2006jcp}
Li,~X.; Bowen,~K.~H.; Haranczyk,~M.; Bachorz,~R.~A.; Mazurkiewicz,~K.; Rak,~J.;
  Gutowski,~M. Photoelectron spectroscopy of adiabatically bound valence anions
  of rare tautomers of the nucleic acid bases. \emph{J. Chem. Phys.}
  \textbf{2007}, \emph{127}, 174309\relax
\mciteBstWouldAddEndPuncttrue
\mciteSetBstMidEndSepPunct{\mcitedefaultmidpunct}
{\mcitedefaultendpunct}{\mcitedefaultseppunct}\relax
\EndOfBibitem
\bibitem[Schiedt \latin{et~al.}(1998)Schiedt, Weinkauf, Neumark, and
  Schlag]{SCHIEDT1998511}
Schiedt,~J.; Weinkauf,~R.; Neumark,~D.~M.; Schlag,~E.~W. Anion spectroscopy of
  uracil, thymine and the amino-oxo and amino-hydroxy tautomers of cytosine and
  their water clusters. \emph{Chem. Phys.} \textbf{1998}, \emph{239},
  511--524\relax
\mciteBstWouldAddEndPuncttrue
\mciteSetBstMidEndSepPunct{\mcitedefaultmidpunct}
{\mcitedefaultendpunct}{\mcitedefaultseppunct}\relax
\EndOfBibitem
\bibitem[Dougherty \latin{et~al.}(1978)Dougherty, Younathan, Voll, Abdulnur,
  and McGlynn]{DOUGHERTY1978379}
Dougherty,~D.; Younathan,~E.; Voll,~R.; Abdulnur,~S.; McGlynn,~S. Photoelectron
  spectroscopy of some biological molecules. \emph{J. Electron Spectrosc.
  Relat. Phenom.} \textbf{1978}, \emph{13}, 379--393\relax
\mciteBstWouldAddEndPuncttrue
\mciteSetBstMidEndSepPunct{\mcitedefaultmidpunct}
{\mcitedefaultendpunct}{\mcitedefaultseppunct}\relax
\EndOfBibitem
\bibitem[Dougherty \latin{et~al.}(1976)Dougherty, Wittel, Meeks, and
  McGlynn]{dougherty1976jacs}
Dougherty,~D.; Wittel,~K.; Meeks,~J.; McGlynn,~S.~P. Photoelectron spectroscopy
  of carbonyls. Ureas, uracils, and thymine. \emph{J. Am. Chem. Soc.}
  \textbf{1976}, \emph{98}, 3815--3820\relax
\mciteBstWouldAddEndPuncttrue
\mciteSetBstMidEndSepPunct{\mcitedefaultmidpunct}
{\mcitedefaultendpunct}{\mcitedefaultseppunct}\relax
\EndOfBibitem
\bibitem[Hendricks \latin{et~al.}(1996)Hendricks, Lyapustina, {de Clercq},
  Snodgrass, and Bowen]{hendricks1996jcp}
Hendricks,~J.~H.; Lyapustina,~S.~A.; {de Clercq},~H.~L.; Snodgrass,~J.~T.;
  Bowen,~K.~H. Dipole bound, nucleic acid base anions studied via negative ion
  photoelectron spectroscopy. \emph{J. Chem. Phys.} \textbf{1996}, \emph{104},
  7788--7791\relax
\mciteBstWouldAddEndPuncttrue
\mciteSetBstMidEndSepPunct{\mcitedefaultmidpunct}
{\mcitedefaultendpunct}{\mcitedefaultseppunct}\relax
\EndOfBibitem
\bibitem[Yu \latin{et~al.}(1981)Yu, O'Donnell, and LeBreton]{yu1981jpc}
Yu,~C.; O'Donnell,~T.~J.; LeBreton,~P.~R. Ultraviolet photoelectron studies of
  volatile nucleoside models. Vertical ionization potential measurements of
  methylated uridine, thymidine, cytidine, and adenosine. \emph{J. Phys. Chem.}
  \textbf{1981}, \emph{85}, 3851--3855\relax
\mciteBstWouldAddEndPuncttrue
\mciteSetBstMidEndSepPunct{\mcitedefaultmidpunct}
{\mcitedefaultendpunct}{\mcitedefaultseppunct}\relax
\EndOfBibitem
\bibitem[Kearns \latin{et~al.}(1971)Kearns, Marsh, and
  Schaffner]{kearns1971jacs}
Kearns,~D.~R.; Marsh,~G.; Schaffner,~K. Investigation of singlet $\rightarrow$
  triplet transitions by phosphorescence excitation spectroscopy. IX.
  Conjugated enones. \emph{J. Am. Chem. Soc.} \textbf{1971}, \emph{93},
  3129--3137\relax
\mciteBstWouldAddEndPuncttrue
\mciteSetBstMidEndSepPunct{\mcitedefaultmidpunct}
{\mcitedefaultendpunct}{\mcitedefaultseppunct}\relax
\EndOfBibitem
\bibitem[Dvornikov \latin{et~al.}(1979)Dvornikov, Knyukshto, Solovev, and
  Tsvirko]{dvornikov1979phosphorescence}
Dvornikov,~S.~S.; Knyukshto,~V.~N.; Solovev,~K.~N.; Tsvirko,~M.~P.
  Phosphorescence of chlorophyllis $a$ and $b$ and their pheophytins.
  \emph{Opt. Spect. (USSR)} \textbf{1979}, \emph{46}, 385--388\relax
\mciteBstWouldAddEndPuncttrue
\mciteSetBstMidEndSepPunct{\mcitedefaultmidpunct}
{\mcitedefaultendpunct}{\mcitedefaultseppunct}\relax
\EndOfBibitem
\bibitem[McLendon and Miller(1980)McLendon, and Miller]{C39800000533}
McLendon,~G.; Miller,~D.~S. Metalloporphyrins catalyse the photo-reduction of
  water to $\ce{H2}$. \emph{J. Chem. Soc.{,} Chem. Commun.} \textbf{1980},
  533--534\relax
\mciteBstWouldAddEndPuncttrue
\mciteSetBstMidEndSepPunct{\mcitedefaultmidpunct}
{\mcitedefaultendpunct}{\mcitedefaultseppunct}\relax
\EndOfBibitem
\bibitem[Chattopadhyay \latin{et~al.}(1984)Chattopadhyay, Kumar, and
  Das]{F19848001151}
Chattopadhyay,~S.~K.; Kumar,~C.~V.; Das,~P.~K. Triplet-state photophysics of
  retinal analogues. Interaction of polyene triplets with the di-t-butylnitroxy
  radical. \emph{J. Chem. Soc.{,} Faraday Trans. 1} \textbf{1984}, \emph{80},
  1151--1161\relax
\mciteBstWouldAddEndPuncttrue
\mciteSetBstMidEndSepPunct{\mcitedefaultmidpunct}
{\mcitedefaultendpunct}{\mcitedefaultseppunct}\relax
\EndOfBibitem
\bibitem[Becker \latin{et~al.}(1971)Becker, Inuzuka, and Balke]{becker1971jacs}
Becker,~R.~S.; Inuzuka,~K.; Balke,~D.~E. Spectroscopy and photochemistry of
  retinals. I. Theoretical and experimental considerations of absorption
  spectra. \emph{J. Am. Chem. Soc.} \textbf{1971}, \emph{93}, 38--42\relax
\mciteBstWouldAddEndPuncttrue
\mciteSetBstMidEndSepPunct{\mcitedefaultmidpunct}
{\mcitedefaultendpunct}{\mcitedefaultseppunct}\relax
\EndOfBibitem
\bibitem[Thomson(1969)]{thomson1969jcp}
Thomson,~A.~J. Fluorescence Spectra of Some Retinyl Polyenes. \emph{J. Chem.
  Phys.} \textbf{1969}, \emph{51}, 4106--4116\relax
\mciteBstWouldAddEndPuncttrue
\mciteSetBstMidEndSepPunct{\mcitedefaultmidpunct}
{\mcitedefaultendpunct}{\mcitedefaultseppunct}\relax
\EndOfBibitem
\bibitem[Haley \latin{et~al.}(1992)Haley, Fitch, Goyal, Lambert, Truscott,
  Chacon, Stirling, and Schalch]{C39920001175}
Haley,~J.~L.; Fitch,~A.~N.; Goyal,~R.; Lambert,~C.; Truscott,~T.~G.;
  Chacon,~J.~N.; Stirling,~D.; Schalch,~W. The S$_1$ and T$_1$ energy levels of
  all-$trans$-$\beta$-carotene. \emph{J. Chem. Soc.{,} Chem. Commun.}
  \textbf{1992}, 1175--1176\relax
\mciteBstWouldAddEndPuncttrue
\mciteSetBstMidEndSepPunct{\mcitedefaultmidpunct}
{\mcitedefaultendpunct}{\mcitedefaultseppunct}\relax
\EndOfBibitem
\bibitem[Chattopadhyay \latin{et~al.}(1985)Chattopadhyay, Kumar, and
  Das]{PHP:PHP17}
Chattopadhyay,~S.~K.; Kumar,~C.~V.; Das,~P.~K. Triplet Excitation Transfer
  Involving $\beta$-Ionone. A Kinetic Study by Laser Flash Photolysis.
  \emph{Photochem. Photobiol.} \textbf{1985}, \emph{42}, 17--24\relax
\mciteBstWouldAddEndPuncttrue
\mciteSetBstMidEndSepPunct{\mcitedefaultmidpunct}
{\mcitedefaultendpunct}{\mcitedefaultseppunct}\relax
\EndOfBibitem
\bibitem[Marsh \latin{et~al.}(1970)Marsh, Kearns, and Fisch]{marsh1970jacs}
Marsh,~G.; Kearns,~D.~R.; Fisch,~M. Investigation of singlet $\rightarrow$
  triplet and singlet $\rightarrow$ singlet transitions by phosphorescence
  excitation spectroscopy. VIII. Santonins. \emph{J. Am. Chem. Soc.}
  \textbf{1970}, \emph{92}, 2252--2257\relax
\mciteBstWouldAddEndPuncttrue
\mciteSetBstMidEndSepPunct{\mcitedefaultmidpunct}
{\mcitedefaultendpunct}{\mcitedefaultseppunct}\relax
\EndOfBibitem
\bibitem[Hansen and Undheim(1975)Hansen, and Undheim]{hansen1975mass}
Hansen,~P.~E.; Undheim,~K. Mass spectrometry of onium compounds. XXIX.
  Ionisation potential in structure analysis of valence isomers. \emph{Acta
  Chem. Scand., Ser. B} \textbf{1975}, 221--223\relax
\mciteBstWouldAddEndPuncttrue
\mciteSetBstMidEndSepPunct{\mcitedefaultmidpunct}
{\mcitedefaultendpunct}{\mcitedefaultseppunct}\relax
\EndOfBibitem
\bibitem[Case and Kearns(1970)Case, and Kearns]{case1970jcp}
Case,~W.~A.; Kearns,~D.~R. Investigation of S $\rightarrow$ T and S
  $\rightarrow$ S Transitions by Phosphorescence Excitation Spectroscopy VII.
  1-Indanone and Other Aromatic Ketones. \emph{J. Chem. Phys.} \textbf{1970},
  \emph{52}, 2175--2191\relax
\mciteBstWouldAddEndPuncttrue
\mciteSetBstMidEndSepPunct{\mcitedefaultmidpunct}
{\mcitedefaultendpunct}{\mcitedefaultseppunct}\relax
\EndOfBibitem
\bibitem[Matsushima and Sakai(1986)Matsushima, and Sakai]{P29860001217}
Matsushima,~R.; Sakai,~K. Specific photoreactions of flavanones typical of
  $n${,}$\pi^*$ and $\pi${,}$\pi^*$ characters in lowest triplet states.
  \emph{J. Chem. Soc.{,} Perkin Trans. 2} \textbf{1986}, 1217--1222\relax
\mciteBstWouldAddEndPuncttrue
\mciteSetBstMidEndSepPunct{\mcitedefaultmidpunct}
{\mcitedefaultendpunct}{\mcitedefaultseppunct}\relax
\EndOfBibitem
\bibitem[Bhattacharyya \latin{et~al.}(1986)Bhattacharyya, Das, Ramamurthy, and
  Rao]{F29868200135}
Bhattacharyya,~K.; Das,~P.~K.; Ramamurthy,~V.; Rao,~V.~P. Triplet-state
  photophysics and transient photochemistry of cyclic enethiones. A laser flash
  photolysis study. \emph{J. Chem. Soc.{,} Faraday Trans. 2} \textbf{1986},
  \emph{82}, 135--147\relax
\mciteBstWouldAddEndPuncttrue
\mciteSetBstMidEndSepPunct{\mcitedefaultmidpunct}
{\mcitedefaultendpunct}{\mcitedefaultseppunct}\relax
\EndOfBibitem
\bibitem[Mantulin and Song(1973)Mantulin, and Song]{mantulin1973jacs}
Mantulin,~W.~W.; Song,~P.-S. Excited states of skin-sensitizing coumarins and
  psoralens. Spectroscopic studies. \emph{J. Am. Chem. Soc.} \textbf{1973},
  \emph{95}, 5122--5129\relax
\mciteBstWouldAddEndPuncttrue
\mciteSetBstMidEndSepPunct{\mcitedefaultmidpunct}
{\mcitedefaultendpunct}{\mcitedefaultseppunct}\relax
\EndOfBibitem
\bibitem[Usacheva \latin{et~al.}(1984)Usacheva, Osipov, Drozdenko, and
  Dilung]{usacheva1984rjpc}
Usacheva,~M.~N.; Osipov,~V.~V.; Drozdenko,~I.~V.; Dilung,~I. \emph{Russ. J.
  Phys. Chem.} \textbf{1984}, \emph{58}, 1550--1553\relax
\mciteBstWouldAddEndPuncttrue
\mciteSetBstMidEndSepPunct{\mcitedefaultmidpunct}
{\mcitedefaultendpunct}{\mcitedefaultseppunct}\relax
\EndOfBibitem
\bibitem[Maier \latin{et~al.}(1975)Maier, Muller, Kubota, and
  Yamakawa]{HLCA:HLCA19750580619}
Maier,~J.~P.; Muller,~J.-F.; Kubota,~T.; Yamakawa,~M. Ionisation Energies and
  the Electronic Structures of the N-oxides of Azanaphthalenes and
  azaanthracenes. \emph{Helv. Chim. Acta} \textbf{1975}, \emph{58},
  1641--1648\relax
\mciteBstWouldAddEndPuncttrue
\mciteSetBstMidEndSepPunct{\mcitedefaultmidpunct}
{\mcitedefaultendpunct}{\mcitedefaultseppunct}\relax
\EndOfBibitem
\bibitem[Kokubo \latin{et~al.}(2004)Kokubo, Ando, Koyasu, Mitsui, and
  Nakajima]{shunsuke2004jcp}
Kokubo,~S.; Ando,~N.; Koyasu,~K.; Mitsui,~M.; Nakajima,~A. Negative ion
  photoelectron spectroscopy of acridine molecular anion and its monohydrate.
  \emph{J. Chem. Phys.} \textbf{2004}, \emph{121}, 11112--11117\relax
\mciteBstWouldAddEndPuncttrue
\mciteSetBstMidEndSepPunct{\mcitedefaultmidpunct}
{\mcitedefaultendpunct}{\mcitedefaultseppunct}\relax
\EndOfBibitem
\bibitem[Chambers and Kearns(1969)Chambers, and Kearns]{PHP:PHP215}
Chambers,~R.~W.; Kearns,~D.~R. Triplet States of Some Common Photosensitizing
  Dyes. \emph{Photochem. Photobiol.} \textbf{1969}, \emph{10}, 215--219\relax
\mciteBstWouldAddEndPuncttrue
\mciteSetBstMidEndSepPunct{\mcitedefaultmidpunct}
{\mcitedefaultendpunct}{\mcitedefaultseppunct}\relax
\EndOfBibitem
\bibitem[Korobov and Chibisov(1983)Korobov, and Chibisov]{0036-021X-52-1-R03}
Korobov,~V.~E.; Chibisov,~A.~K. Primary Photoprocesses in Colorant Molecules.
  \emph{Russ. Chem. Rev.} \textbf{1983}, \emph{52}, 27\relax
\mciteBstWouldAddEndPuncttrue
\mciteSetBstMidEndSepPunct{\mcitedefaultmidpunct}
{\mcitedefaultendpunct}{\mcitedefaultseppunct}\relax
\EndOfBibitem
\bibitem[Sikorska \latin{et~al.}(2004)Sikorska, Khmelinskii, Pruka{\l}a,
  Williams, Patel, Worrall, Bourdelande, Koput, and Sikorski]{sikorska2004jpca}
Sikorska,~E.; Khmelinskii,~I.~V.; Pruka{\l}a,~W.; Williams,~S.~L.; Patel,~M.;
  Worrall,~D.~R.; Bourdelande,~J.~L.; Koput,~J.; Sikorski,~M. Spectroscopy and
  Photophysics of Lumiflavins and Lumichromes. \emph{J. Phys. Chem. A}
  \textbf{2004}, \emph{108}, 1501--1508\relax
\mciteBstWouldAddEndPuncttrue
\mciteSetBstMidEndSepPunct{\mcitedefaultmidpunct}
{\mcitedefaultendpunct}{\mcitedefaultseppunct}\relax
\EndOfBibitem
\bibitem[Palmer \latin{et~al.}(1980)Palmer, Simpson, and
  Platenkamp]{PALMER1980243}
Palmer,~M.~H.; Simpson,~I.; Platenkamp,~R.~J. The electronic structure of
  flavin derivatives. \emph{J. Mol. Struct.} \textbf{1980}, \emph{66},
  243--263\relax
\mciteBstWouldAddEndPuncttrue
\mciteSetBstMidEndSepPunct{\mcitedefaultmidpunct}
{\mcitedefaultendpunct}{\mcitedefaultseppunct}\relax
\EndOfBibitem
\bibitem[Timoshenko \latin{et~al.}(1981)Timoshenko, Korkoshko, Kleimenov,
  Petrachenko, Chizhov, Rylkov, and Akopian]{timoshenko1981ionization}
Timoshenko,~M.~M.; Korkoshko,~I.~V.; Kleimenov,~V.~I.; Petrachenko,~N.~E.;
  Chizhov,~I.~V.; Rylkov,~V.~V.; Akopian,~M.~E. Ionization potentials of
  rhodamine dyes. \emph{Dokl. Phys. Chem. (USSR)} \textbf{1981}, \emph{260},
  138--140\relax
\mciteBstWouldAddEndPuncttrue
\mciteSetBstMidEndSepPunct{\mcitedefaultmidpunct}
{\mcitedefaultendpunct}{\mcitedefaultseppunct}\relax
\EndOfBibitem
\bibitem[Lee \latin{et~al.}(1988)Lee, Yang, and Parr]{PhysRevB.37.785}
Lee,~C.; Yang,~W.; Parr,~R.~G. Development of the {Colle--Salvetti}
  correlation-energy formula into a functional of the electron density.
  \emph{Phys. Rev. B} \textbf{1988}, \emph{37}, 785--789\relax
\mciteBstWouldAddEndPuncttrue
\mciteSetBstMidEndSepPunct{\mcitedefaultmidpunct}
{\mcitedefaultendpunct}{\mcitedefaultseppunct}\relax
\EndOfBibitem
\bibitem[Becke(1993)]{becke1993jcp}
Becke,~A.~D. A new mixing of {Hartree--Fock} and local density-functional
  theories. \emph{J. Chem. Phys.} \textbf{1993}, \emph{98}, 1372--1377\relax
\mciteBstWouldAddEndPuncttrue
\mciteSetBstMidEndSepPunct{\mcitedefaultmidpunct}
{\mcitedefaultendpunct}{\mcitedefaultseppunct}\relax
\EndOfBibitem
\bibitem[Kiefer(1953)]{kiefer1953sequential}
Kiefer,~J. Sequential minimax search for a maximum. \emph{Proc. Am. Math. Soc.}
  \textbf{1953}, \emph{4}, 502--506\relax
\mciteBstWouldAddEndPuncttrue
\mciteSetBstMidEndSepPunct{\mcitedefaultmidpunct}
{\mcitedefaultendpunct}{\mcitedefaultseppunct}\relax
\EndOfBibitem
\bibitem[Press(2007)]{press2007numerical}
Press,~W.~H. \emph{Numerical recipes 3rd edition: The art of scientific
  computing}; Cambridge university press, 2007\relax
\mciteBstWouldAddEndPuncttrue
\mciteSetBstMidEndSepPunct{\mcitedefaultmidpunct}
{\mcitedefaultendpunct}{\mcitedefaultseppunct}\relax
\EndOfBibitem
\bibitem[Hirata and Head-Gordon(1999)Hirata, and Head-Gordon]{HIRATA1999291}
Hirata,~S.; Head-Gordon,~M. Time-dependent density functional theory within the
  Tamm--Dancoff approximation. \emph{Chem. Phys. Lett.} \textbf{1999},
  \emph{314}, 291--299\relax
\mciteBstWouldAddEndPuncttrue
\mciteSetBstMidEndSepPunct{\mcitedefaultmidpunct}
{\mcitedefaultendpunct}{\mcitedefaultseppunct}\relax
\EndOfBibitem
\bibitem[Becke(1993)]{beckejcp1993-2}
Becke,~A.~D. Density-functional thermochemistry. III. The role of exact
  exchange. \emph{J. Chem. Phys.} \textbf{1993}, \emph{98}, 5648--5652\relax
\mciteBstWouldAddEndPuncttrue
\mciteSetBstMidEndSepPunct{\mcitedefaultmidpunct}
{\mcitedefaultendpunct}{\mcitedefaultseppunct}\relax
\EndOfBibitem
\bibitem[Adamo and Barone(1999)Adamo, and Barone]{adamo1999jcp}
Adamo,~C.; Barone,~V. Toward reliable density functional methods without
  adjustable parameters: The {PBE0} model. \emph{J. Chem. Phys.} \textbf{1999},
  \emph{110}, 6158--6170\relax
\mciteBstWouldAddEndPuncttrue
\mciteSetBstMidEndSepPunct{\mcitedefaultmidpunct}
{\mcitedefaultendpunct}{\mcitedefaultseppunct}\relax
\EndOfBibitem
\bibitem[{Dunning Jr.}(1989)]{dunning1989jcp}
{Dunning Jr.},~T.~H. Gaussian basis sets for use in correlated molecular
  calculations. {I}. The atoms boron through neon and hydrogen. \emph{J. Chem.
  Phys.} \textbf{1989}, \emph{90}, 1007--1023\relax
\mciteBstWouldAddEndPuncttrue
\mciteSetBstMidEndSepPunct{\mcitedefaultmidpunct}
{\mcitedefaultendpunct}{\mcitedefaultseppunct}\relax
\EndOfBibitem
\bibitem[Shao \latin{et~al.}(2015)Shao, Gan, Epifanovsky, Gilbert, Wormit,
  Kussmann, Lange, Behn, Deng, Feng, Ghosh, Goldey, Horn, Jacobson, Kaliman,
  Khaliullin, K\'us, Landau, Liu, Proynov, Rhee, Richard, Rohrdanz, Steele,
  Sundstrom, {Woodcock III}, Zimmerman, Zuev, Albrecht, Alguire, Austin, Beran,
  Bernard, Berquist, Brandhorst, Bravaya, Brown, Casanova, Chang, Chen, Chien,
  Closser, Crittenden, Diedenhofen, {DiStasio Jr.}, Dop, Dutoi, Edgar, Fatehi,
  {Fusti-Molnar}, Ghysels, {Golubeva-Zadorozhnaya}, Gomes, {Hanson-Heine},
  Harbach, Hauser, Hohenstein, Holden, Jagau, Ji, Kaduk, Khistyaev, Kim, Kim,
  King, Klunzinger, Kosenkov, Kowalczyk, Krauter, Lao, Laurent, Lawler,
  Levchenko, Lin, Liu, Livshits, Lochan, Luenser, Manohar, Manzer, Mao,
  Mardirossian, Marenich, Maurer, Mayhall, Oana, {Olivares-Amaya}, O'Neill,
  Parkhill, Perrine, Peverati, Pieniazek, Prociuk, Rehn, Rosta, Russ, Sergueev,
  Sharada, Sharmaa, Small, Sodt, Stein, St\"uck, Su, Thom, Tsuchimochi, Vogt,
  Vydrov, Wang, Watson, Wenzel, White, Williams, Vanovschi, Yeganeh, Yost, You,
  Zhang, Zhang, Zhou, Brooks, Chan, Chipman, Cramer, {Goddard III}, Gordon,
  Hehre, Klamt, {Schaefer III}, Schmidt, Sherrill, Truhlar, Warshel, Xua,
  {Aspuru-Guzik}, Baer, Bell, Besley, Chai, Dreuw, Dunietz, Furlani, Gwaltney,
  Hsu, Jung, Kong, Lambrecht, Liang, Ochsenfeld, Rassolov, Slipchenko,
  Subotnik, {{Van Voorhis}}, Herbert, Krylov, Gill, and {Head-Gordon}]{QCHEM4}
Shao,~Y.; Gan,~Z.; Epifanovsky,~E.; Gilbert,~A. T.~B.; Wormit,~M.;
  Kussmann,~J.; Lange,~A.~W.; Behn,~A.; Deng,~J.; Feng,~X.; Ghosh,~D.;
  Goldey,~M.; Horn,~P.~R.; Jacobson,~L.~D.; Kaliman,~I.; Khaliullin,~R.~Z.;
  K\'us,~T.; Landau,~A.; Liu,~J.; Proynov,~E.~I.; Rhee,~Y.~M.; Richard,~R.~M.;
  Rohrdanz,~M.~A.; Steele,~R.~P.; Sundstrom,~E.~J.; {Woodcock III},~H.~L.;
  Zimmerman,~P.~M.; Zuev,~D.; Albrecht,~B.; Alguire,~E.; Austin,~B.; Beran,~G.
  J.~O.; Bernard,~Y.~A.; Berquist,~E.; Brandhorst,~K.; Bravaya,~K.~B.;
  Brown,~S.~T.; Casanova,~D.; Chang,~C.-M.; Chen,~Y.; Chien,~S.~H.;
  Closser,~K.~D.; Crittenden,~D.~L.; Diedenhofen,~M.; {DiStasio Jr.},~R.~A.;
  Dop,~H.; Dutoi,~A.~D.; Edgar,~R.~G.; Fatehi,~S.; {Fusti-Molnar},~L.;
  Ghysels,~A.; {Golubeva-Zadorozhnaya},~A.; Gomes,~J.; {Hanson-Heine},~M.
  W.~D.; Harbach,~P. H.~P.; Hauser,~A.~W.; Hohenstein,~E.~G.; Holden,~Z.~C.;
  Jagau,~T.-C.; Ji,~H.; Kaduk,~B.; Khistyaev,~K.; Kim,~J.; Kim,~J.;
  King,~R.~A.; Klunzinger,~P.; Kosenkov,~D.; Kowalczyk,~T.; Krauter,~C.~M.;
  Lao,~K.~U.; Laurent,~A.; Lawler,~K.~V.; Levchenko,~S.~V.; Lin,~C.~Y.;
  Liu,~F.; Livshits,~E.; Lochan,~R.~C.; Luenser,~A.; Manohar,~P.;
  Manzer,~S.~F.; Mao,~S.-P.; Mardirossian,~N.; Marenich,~A.~V.; Maurer,~S.~A.;
  Mayhall,~N.~J.; Oana,~C.~M.; {Olivares-Amaya},~R.; O'Neill,~D.~P.;
  Parkhill,~J.~A.; Perrine,~T.~M.; Peverati,~R.; Pieniazek,~P.~A.; Prociuk,~A.;
  Rehn,~D.~R.; Rosta,~E.; Russ,~N.~J.; Sergueev,~N.; Sharada,~S.~M.;
  Sharmaa,~S.; Small,~D.~W.; Sodt,~A.; Stein,~T.; St\"uck,~D.; Su,~Y.-C.;
  Thom,~A. J.~W.; Tsuchimochi,~T.; Vogt,~L.; Vydrov,~O.; Wang,~T.;
  Watson,~M.~A.; Wenzel,~J.; White,~A.; Williams,~C.~F.; Vanovschi,~V.;
  Yeganeh,~S.; Yost,~S.~R.; You,~Z.-Q.; Zhang,~I.~Y.; Zhang,~X.; Zhou,~Y.;
  Brooks,~B.~R.; Chan,~G. K.~L.; Chipman,~D.~M.; Cramer,~C.~J.; {Goddard
  III},~W.~A.; Gordon,~M.~S.; Hehre,~W.~J.; Klamt,~A.; {Schaefer III},~H.~F.;
  Schmidt,~M.~W.; Sherrill,~C.~D.; Truhlar,~D.~G.; Warshel,~A.; Xua,~X.;
  {Aspuru-Guzik},~A.; Baer,~R.; Bell,~A.~T.; Besley,~N.~A.; Chai,~J.-D.;
  Dreuw,~A.; Dunietz,~B.~D.; Furlani,~T.~R.; Gwaltney,~S.~R.; Hsu,~C.-P.;
  Jung,~Y.; Kong,~J.; Lambrecht,~D.~S.; Liang,~W.; Ochsenfeld,~C.;
  Rassolov,~V.~A.; Slipchenko,~L.~V.; Subotnik,~J.~E.; {{Van Voorhis}},~T.;
  Herbert,~J.~M.; Krylov,~A.~I.; Gill,~P. M.~W.; {Head-Gordon},~M. Advances in
  molecular quantum chemistry contained in the Q-Chem 4 program package.
  \emph{Mol. Phys.} \textbf{2015}, \emph{113}, 184--215\relax
\mciteBstWouldAddEndPuncttrue
\mciteSetBstMidEndSepPunct{\mcitedefaultmidpunct}
{\mcitedefaultendpunct}{\mcitedefaultseppunct}\relax
\EndOfBibitem
\bibitem[Tu and Liang(2017)Tu, and Liang]{tu2017acsomega}
Tu,~C.; Liang,~W. NB-Type Electronic Asymmetric Compounds as Potential
  Blue-Color TADF Emitters: Steric Hindrance, Substitution Effect, and
  Electronic Characteristics. \emph{ACS Omega} \textbf{2017}, \emph{2},
  3098--3109\relax
\mciteBstWouldAddEndPuncttrue
\mciteSetBstMidEndSepPunct{\mcitedefaultmidpunct}
{\mcitedefaultendpunct}{\mcitedefaultseppunct}\relax
\EndOfBibitem
\bibitem[Liu \latin{et~al.}(2015)Liu, Adamska, Doorn, and Tretiak]{C5CP01782C}
Liu,~J.; Adamska,~L.; Doorn,~S.~K.; Tretiak,~S. Singlet and triplet excitons
  and charge polarons in cycloparaphenylenes: a density functional theory
  study. \emph{Phys. Chem. Chem. Phys.} \textbf{2015}, \emph{17},
  14613--14622\relax
\mciteBstWouldAddEndPuncttrue
\mciteSetBstMidEndSepPunct{\mcitedefaultmidpunct}
{\mcitedefaultendpunct}{\mcitedefaultseppunct}\relax
\EndOfBibitem
\bibitem[Zhang \latin{et~al.}(2016)Zhang, Steyrleuthner, and
  Br\'{e}das]{zhang2016jpcc}
Zhang,~Y.; Steyrleuthner,~R.; Br\'{e}das,~J.-L. Charge Delocalization in
  Oligomers of Poly(2,5-bis(3-alkylthiophene-2-yl)thieno[3,2-b]thiophene)
  (PBTTT). \emph{J. Phys. Chem. C} \textbf{2016}, \emph{120}, 9671--9677\relax
\mciteBstWouldAddEndPuncttrue
\mciteSetBstMidEndSepPunct{\mcitedefaultmidpunct}
{\mcitedefaultendpunct}{\mcitedefaultseppunct}\relax
\EndOfBibitem
\bibitem[Isborn \latin{et~al.}(2013)Isborn, Mar, Curchod, Tavernelli, and
  Mart\'{i}nez]{Isborn2013}
Isborn,~C.~M.; Mar,~B.~D.; Curchod,~B. F.~E.; Tavernelli,~I.;
  Mart\'{i}nez,~T.~J. The Charge Transfer Problem in Density Functional Theory
  Calculations of Aqueously Solvated Molecules. \emph{J. Phys. Chem. B}
  \textbf{2013}, \emph{117}, 12189--12201\relax
\mciteBstWouldAddEndPuncttrue
\mciteSetBstMidEndSepPunct{\mcitedefaultmidpunct}
{\mcitedefaultendpunct}{\mcitedefaultseppunct}\relax
\EndOfBibitem
\bibitem[Maitra \latin{et~al.}(2004)Maitra, Zhang, Cave, and
  Burke]{neepa2004jcp}
Maitra,~N.~T.; Zhang,~F.; Cave,~R.~J.; Burke,~K. Double excitations within
  time-dependent density functional theory linear response. \emph{J. Chem.
  Phys.} \textbf{2004}, \emph{120}, 5932--5937\relax
\mciteBstWouldAddEndPuncttrue
\mciteSetBstMidEndSepPunct{\mcitedefaultmidpunct}
{\mcitedefaultendpunct}{\mcitedefaultseppunct}\relax
\EndOfBibitem
\bibitem[Maitra(2005)]{maitra2005jcp}
Maitra,~N.~T. Undoing static correlation: Long-range charge transfer in
  time-dependent density-functional theory. \emph{J. Chem. Phys.}
  \textbf{2005}, \emph{122}, 234104\relax
\mciteBstWouldAddEndPuncttrue
\mciteSetBstMidEndSepPunct{\mcitedefaultmidpunct}
{\mcitedefaultendpunct}{\mcitedefaultseppunct}\relax
\EndOfBibitem
\bibitem[Ullrich(2006)]{ullrich2006jcp}
Ullrich,~C.~A. Time-dependent density-functional theory beyond the adiabatic
  approximation: Insights from a two-electron model system. \emph{J. Chem.
  Phys.} \textbf{2006}, \emph{125}, 234108\relax
\mciteBstWouldAddEndPuncttrue
\mciteSetBstMidEndSepPunct{\mcitedefaultmidpunct}
{\mcitedefaultendpunct}{\mcitedefaultseppunct}\relax
\EndOfBibitem
\bibitem[Li~Manni \latin{et~al.}(2014)Li~Manni, Carlson, Luo, Ma, Olsen,
  Truhlar, and Gagliardi]{Manni2014}
Li~Manni,~G.; Carlson,~R.~K.; Luo,~S.; Ma,~D.; Olsen,~J.; Truhlar,~D.~G.;
  Gagliardi,~L. Multiconfiguration Pair-Density Functional Theory.
  \emph{Journal of Chemical Theory and Computation} \textbf{2014}, \emph{10},
  3669--3680\relax
\mciteBstWouldAddEndPuncttrue
\mciteSetBstMidEndSepPunct{\mcitedefaultmidpunct}
{\mcitedefaultendpunct}{\mcitedefaultseppunct}\relax
\EndOfBibitem
\bibitem[Chen \latin{et~al.}(2017)Chen, Zhang, Jin, Yang, Su, and
  Yang]{Chen2017jpcl}
Chen,~Z.; Zhang,~D.; Jin,~Y.; Yang,~Y.; Su,~N.~Q.; Yang,~W. Multireference
  Density Functional Theory with Generalized Auxiliary Systems for Ground and
  Excited States. \emph{J. Phys. Chem. Lett.} \textbf{2017}, \emph{8},
  4479--4485\relax
\mciteBstWouldAddEndPuncttrue
\mciteSetBstMidEndSepPunct{\mcitedefaultmidpunct}
{\mcitedefaultendpunct}{\mcitedefaultseppunct}\relax
\EndOfBibitem
\bibitem[Peach \latin{et~al.}(2011)Peach, Williamson, and Tozer]{peach2011jctc}
Peach,~M. J.~G.; Williamson,~M.~J.; Tozer,~D.~J. Influence of Triplet
  Instabilities in {TDDFT}. \emph{J. Chem. Theory Comput.} \textbf{2011},
  \emph{7}, 3578--3585\relax
\mciteBstWouldAddEndPuncttrue
\mciteSetBstMidEndSepPunct{\mcitedefaultmidpunct}
{\mcitedefaultendpunct}{\mcitedefaultseppunct}\relax
\EndOfBibitem
\bibitem[Liu \latin{et~al.}(2018)Liu, Li, Ren, Yan, and Bryce]{liu2018all}
Liu,~Y.; Li,~C.; Ren,~Z.; Yan,~S.; Bryce,~M.~R. All-organic thermally activated
  delayed fluorescence materials for organic light-emitting diodes. \emph{Nat.
  Rev. Mater.} \textbf{2018}, \emph{3}, 18020\relax
\mciteBstWouldAddEndPuncttrue
\mciteSetBstMidEndSepPunct{\mcitedefaultmidpunct}
{\mcitedefaultendpunct}{\mcitedefaultseppunct}\relax
\EndOfBibitem
\bibitem[Shang and Bernstein(1992)Shang, and Bernstein]{shang1992jcp}
Shang,~Q.; Bernstein,~E.~R. Solvation effects on the electronic structure of
  4‐$N$, $N$‐dimethylaminobenzonitrile: Mixing of the local $\pi\pi\*$ and
  charge‐transfer states. \emph{J. Chem. Phys.} \textbf{1992}, \emph{97},
  60--68\relax
\mciteBstWouldAddEndPuncttrue
\mciteSetBstMidEndSepPunct{\mcitedefaultmidpunct}
{\mcitedefaultendpunct}{\mcitedefaultseppunct}\relax
\EndOfBibitem
\bibitem[Mo and Gao(2006)Mo, and Gao]{mo2006jpcb}
Mo,~Y.; Gao,~J. Polarization and Charge-Transfer Effects in Aqueous Solution
  via {\it Ab Initio} QM/MM Simulations. \emph{J. Phys. Chem. B} \textbf{2006},
  \emph{110}, 2976--2980\relax
\mciteBstWouldAddEndPuncttrue
\mciteSetBstMidEndSepPunct{\mcitedefaultmidpunct}
{\mcitedefaultendpunct}{\mcitedefaultseppunct}\relax
\EndOfBibitem
\bibitem[Messina \latin{et~al.}(2013)Messina, Br{\"a}m, Cannizzo, and
  Chergui]{messina2013real}
Messina,~F.; Br{\"a}m,~O.; Cannizzo,~A.; Chergui,~M. Real-time observation of
  the charge transfer to solvent dynamics. \emph{Nat. Comm.} \textbf{2013},
  \emph{4}, 2119\relax
\mciteBstWouldAddEndPuncttrue
\mciteSetBstMidEndSepPunct{\mcitedefaultmidpunct}
{\mcitedefaultendpunct}{\mcitedefaultseppunct}\relax
\EndOfBibitem
\bibitem[Rondi \latin{et~al.}(2015)Rondi, Rodriguez, Feurer, and
  Cannizzo]{rondi2015}
Rondi,~A.; Rodriguez,~Y.; Feurer,~T.; Cannizzo,~A. Solvation-Driven Charge
  Transfer and Localization in Metal Complexes. \emph{Acc. Chem. Res.}
  \textbf{2015}, \emph{48}, 1432--1440\relax
\mciteBstWouldAddEndPuncttrue
\mciteSetBstMidEndSepPunct{\mcitedefaultmidpunct}
{\mcitedefaultendpunct}{\mcitedefaultseppunct}\relax
\EndOfBibitem
\bibitem[Sun \latin{et~al.}(2017)Sun, Hu, Zhong, Chen, Sun, and
  Br\'{e}das]{sun2017jpcl}
Sun,~H.; Hu,~Z.; Zhong,~C.; Chen,~X.; Sun,~Z.; Br\'{e}das,~J.-L. Impact of
  Dielectric Constant on the Singlet--Triplet Gap in Thermally Activated
  Delayed Fluorescence Materials. \emph{J. Phys. Chem. Lett.} \textbf{2017},
  \emph{8}, 2393--2398\relax
\mciteBstWouldAddEndPuncttrue
\mciteSetBstMidEndSepPunct{\mcitedefaultmidpunct}
{\mcitedefaultendpunct}{\mcitedefaultseppunct}\relax
\EndOfBibitem
\bibitem[Sun and Autschbach(2014)Sun, and Autschbach]{ct4009975}
Sun,~H.; Autschbach,~J. Electronic Energy Gaps for $\pi$-Conjugated Oligomers
  and Polymers Calculated with Density Functional Theory. \emph{J. Chem. Theory
  Comput.} \textbf{2014}, \emph{10}, 1035--1047\relax
\mciteBstWouldAddEndPuncttrue
\mciteSetBstMidEndSepPunct{\mcitedefaultmidpunct}
{\mcitedefaultendpunct}{\mcitedefaultseppunct}\relax
\EndOfBibitem
\bibitem[Egger \latin{et~al.}(2014)Egger, Weissman, Refaely-Abramson,
  Sharifzadeh, Dauth, Baer, K\"{u}mmel, Neaton, Zojer, and Kronik]{ct400956h}
Egger,~D.~A.; Weissman,~S.; Refaely-Abramson,~S.; Sharifzadeh,~S.; Dauth,~M.;
  Baer,~R.; K\"{u}mmel,~S.; Neaton,~J.~B.; Zojer,~E.; Kronik,~L. Outer-Valence
  Electron Spectra of Prototypical Aromatic Heterocycles from an Optimally
  Tuned Range-Separated Hybrid Functional. \emph{J. Chem. Theory Comput.}
  \textbf{2014}, \emph{10}, 1934--1952\relax
\mciteBstWouldAddEndPuncttrue
\mciteSetBstMidEndSepPunct{\mcitedefaultmidpunct}
{\mcitedefaultendpunct}{\mcitedefaultseppunct}\relax
\EndOfBibitem
\bibitem[Zhang \latin{et~al.}(2014)Zhang, Sears, Yang, Aziz, Coropceanu, and
  Br\'{e}das]{ct500259m}
Zhang,~C.-R.; Sears,~J.~S.; Yang,~B.; Aziz,~S.~G.; Coropceanu,~V.;
  Br\'{e}das,~J.-L. Theoretical Study of the Local and Charge-Transfer
  Excitations in Model Complexes of Pentacene--C60 Using Tuned Range-Separated
  Hybrid Functionals. \emph{J. Chem. Theory Comput.} \textbf{2014}, \emph{10},
  2379--2388\relax
\mciteBstWouldAddEndPuncttrue
\mciteSetBstMidEndSepPunct{\mcitedefaultmidpunct}
{\mcitedefaultendpunct}{\mcitedefaultseppunct}\relax
\EndOfBibitem
\bibitem[Wang \latin{et~al.}(1995)Wang, Becke, and Smith]{wang1995jcp}
Wang,~J.; Becke,~A.~D.; Smith,~V.~H. Evaluation of $\langle S^2\rangle$ in
  restricted, unrestricted Hartree--Fock, and density functional based
  theories. \emph{J. Chem. Phys.} \textbf{1995}, \emph{102}, 3477--3480\relax
\mciteBstWouldAddEndPuncttrue
\mciteSetBstMidEndSepPunct{\mcitedefaultmidpunct}
{\mcitedefaultendpunct}{\mcitedefaultseppunct}\relax
\EndOfBibitem
\bibitem[Fuchs \latin{et~al.}(2005)Fuchs, Niquet, Gonze, and
  Burke]{fuchs2005jcp}
Fuchs,~M.; Niquet,~Y.-M.; Gonze,~X.; Burke,~K. Describing static correlation in
  bond dissociation by Kohn--Sham density functional theory. \emph{J. Chem.
  Phys.} \textbf{2005}, \emph{122}, 094116\relax
\mciteBstWouldAddEndPuncttrue
\mciteSetBstMidEndSepPunct{\mcitedefaultmidpunct}
{\mcitedefaultendpunct}{\mcitedefaultseppunct}\relax
\EndOfBibitem
\bibitem[Ess \latin{et~al.}(2010)Ess, Johnson, Hu, and Yang]{ess2010singlet}
Ess,~D.~H.; Johnson,~E.~R.; Hu,~X.; Yang,~W. Singlet- Triplet Energy Gaps for
  Diradicals from Fractional-Spin Density-Functional Theory. \emph{J. Phys.
  Chem. A} \textbf{2010}, \emph{115}, 76--83\relax
\mciteBstWouldAddEndPuncttrue
\mciteSetBstMidEndSepPunct{\mcitedefaultmidpunct}
{\mcitedefaultendpunct}{\mcitedefaultseppunct}\relax
\EndOfBibitem
\bibitem[Phillips \latin{et~al.}(2012)Phillips, Geva, and Dunietz]{ct300318g}
Phillips,~H.; Geva,~E.; Dunietz,~B.~D. Calculating Off-Site Excitations in
  Symmetric Donor--Acceptor Systems via Time-Dependent Density Functional
  Theory with Range-Separated Density Functionals. \emph{J. Chem. Theory
  Comput.} \textbf{2012}, \emph{8}, 2661--2668\relax
\mciteBstWouldAddEndPuncttrue
\mciteSetBstMidEndSepPunct{\mcitedefaultmidpunct}
{\mcitedefaultendpunct}{\mcitedefaultseppunct}\relax
\EndOfBibitem
\bibitem[Richard and Herbert(2011)Richard, and Herbert]{ct100607w}
Richard,~R.~M.; Herbert,~J.~M. Time-Dependent Density-Functional Description of
  the $^1L_a$ State in Polycyclic Aromatic Hydrocarbons: Charge-Transfer
  Character in Disguise? \emph{J. Chem. Theory Comput.} \textbf{2011},
  \emph{7}, 1296--1306\relax
\mciteBstWouldAddEndPuncttrue
\mciteSetBstMidEndSepPunct{\mcitedefaultmidpunct}
{\mcitedefaultendpunct}{\mcitedefaultseppunct}\relax
\EndOfBibitem
\bibitem[Wong and Hsieh(2010)Wong, and Hsieh]{ct100529s}
Wong,~B.~M.; Hsieh,~T.~H. Optoelectronic and Excitonic Properties of
  Oligoacenes: Substantial Improvements from Range-Separated Time-Dependent
  Density Functional Theory. \emph{J. Chem. Theory Comput.} \textbf{2010},
  \emph{6}, 3704--3712\relax
\mciteBstWouldAddEndPuncttrue
\mciteSetBstMidEndSepPunct{\mcitedefaultmidpunct}
{\mcitedefaultendpunct}{\mcitedefaultseppunct}\relax
\EndOfBibitem
\bibitem[Sun \latin{et~al.}(2015)Sun, Zhang, and Sun]{C4CP05470A}
Sun,~H.; Zhang,~S.; Sun,~Z. Applicability of optimal functional tuning in
  density functional calculations of ionization potentials and electron
  affinities of adenine-thymine nucleobase pairs and clusters. \emph{Phys.
  Chem. Chem. Phys.} \textbf{2015}, \emph{17}, 4337--4345\relax
\mciteBstWouldAddEndPuncttrue
\mciteSetBstMidEndSepPunct{\mcitedefaultmidpunct}
{\mcitedefaultendpunct}{\mcitedefaultseppunct}\relax
\EndOfBibitem
\bibitem[Bokareva \latin{et~al.}(2017)Bokareva, M\"{o}hle, Neubauer, Bokarev,
  Lochbrunner, and K\"{u}hn]{inorganics5020023}
Bokareva,~O.~S.; M\"{o}hle,~T.; Neubauer,~A.; Bokarev,~S.~I.; Lochbrunner,~S.;
  K\"{u}hn,~O. Chemical Tuning and Absorption Properties of Iridium
  Photosensitizers for Photocatalytic Applications. \emph{Inorganics}
  \textbf{2017}, \emph{5}, 23\relax
\mciteBstWouldAddEndPuncttrue
\mciteSetBstMidEndSepPunct{\mcitedefaultmidpunct}
{\mcitedefaultendpunct}{\mcitedefaultseppunct}\relax
\EndOfBibitem
\bibitem[Alipour and Fallahzadeh(2016)Alipour, and Fallahzadeh]{C6CP02648F}
Alipour,~M.; Fallahzadeh,~P. First principles optimally tuned range-separated
  density functional theory for prediction of phosphorus-hydrogen spin-spin
  coupling constants. \emph{Phys. Chem. Chem. Phys.} \textbf{2016}, \emph{18},
  18431--18440\relax
\mciteBstWouldAddEndPuncttrue
\mciteSetBstMidEndSepPunct{\mcitedefaultmidpunct}
{\mcitedefaultendpunct}{\mcitedefaultseppunct}\relax
\EndOfBibitem
\bibitem[Sun \latin{et~al.}(2016)Sun, Hu, Zhong, Zhang, and Sun]{sun2016jpcc}
Sun,~H.; Hu,~Z.; Zhong,~C.; Zhang,~S.; Sun,~Z. Quantitative Estimation of
  Exciton Binding Energy of Polythiophene-Derived Polymers Using Polarizable
  Continuum Model Tuned Range-Separated Density Functional. \emph{J. Phys.
  Chem. C} \textbf{2016}, \emph{120}, 8048--8055\relax
\mciteBstWouldAddEndPuncttrue
\mciteSetBstMidEndSepPunct{\mcitedefaultmidpunct}
{\mcitedefaultendpunct}{\mcitedefaultseppunct}\relax
\EndOfBibitem
\bibitem[Zheng \latin{et~al.}(2016)Zheng, Br\'{e}das, and
  Coropceanu]{zheng2016jpcl}
Zheng,~Z.; Br\'{e}das,~J.-L.; Coropceanu,~V. Description of the Charge Transfer
  States at the Pentacene/$\ce{C60}$ Interface: Combining Range-Separated
  Hybrid Functionals with the Polarizable Continuum Model. \emph{J. Phys. Chem.
  Lett.} \textbf{2016}, \emph{7}, 2616--2621\relax
\mciteBstWouldAddEndPuncttrue
\mciteSetBstMidEndSepPunct{\mcitedefaultmidpunct}
{\mcitedefaultendpunct}{\mcitedefaultseppunct}\relax
\EndOfBibitem
\bibitem[Zhuravlev \latin{et~al.}(2016)Zhuravlev, Zakharov, Shchegolev, and
  Savvateeva-Popova]{zhuravlev2016antioxidant}
Zhuravlev,~A.~V.; Zakharov,~G.~A.; Shchegolev,~B.~F.; Savvateeva-Popova,~E.~V.
  Antioxidant Properties of Kynurenines: Density Functional Theory
  Calculations. \emph{PLoS Comp. Bio.} \textbf{2016}, \emph{12}, e1005213\relax
\mciteBstWouldAddEndPuncttrue
\mciteSetBstMidEndSepPunct{\mcitedefaultmidpunct}
{\mcitedefaultendpunct}{\mcitedefaultseppunct}\relax
\EndOfBibitem
\bibitem[Bokareva \latin{et~al.}(2017)Bokareva, Shibl, Al-Marri, Pullerits, and
  K\"{u}hn]{bokareva2017jctc}
Bokareva,~O.~S.; Shibl,~M.~F.; Al-Marri,~M.~J.; Pullerits,~T.; K\"{u}hn,~O.
  Optimized Long-Range Corrected Density Functionals for Electronic and Optical
  Properties of Bare and Ligated CdSe Quantum Dots. \emph{J. Chem. Theory
  Comput.} \textbf{2017}, \emph{13}, 110--116\relax
\mciteBstWouldAddEndPuncttrue
\mciteSetBstMidEndSepPunct{\mcitedefaultmidpunct}
{\mcitedefaultendpunct}{\mcitedefaultseppunct}\relax
\EndOfBibitem
\bibitem[Garza \latin{et~al.}(2015)Garza, Osman, Asiri, and
  Scuseria]{garza2015jpcb}
Garza,~A.~J.; Osman,~O.~I.; Asiri,~A.~M.; Scuseria,~G.~E. Can Gap Tuning
  Schemes of Long-Range Corrected Hybrid Functionals Improve the Description of
  Hyperpolarizabilities? \emph{J. Phys. Chem. B} \textbf{2015}, \emph{119},
  1202--1212\relax
\mciteBstWouldAddEndPuncttrue
\mciteSetBstMidEndSepPunct{\mcitedefaultmidpunct}
{\mcitedefaultendpunct}{\mcitedefaultseppunct}\relax
\EndOfBibitem
\bibitem[Cabral~do Couto \latin{et~al.}(2015)Cabral~do Couto, Hollas, and
  Slav\'{i}\v{c}ek]{acs.jctc.5b00066}
Cabral~do Couto,~P.; Hollas,~D.; Slav\'{i}\v{c}ek,~P. On the Performance of
  Optimally Tuned Range-Separated Hybrid Functionals for X-ray Absorption
  Modeling. \emph{J. Chem. Theory Comput.} \textbf{2015}, \emph{11},
  3234--3244\relax
\mciteBstWouldAddEndPuncttrue
\mciteSetBstMidEndSepPunct{\mcitedefaultmidpunct}
{\mcitedefaultendpunct}{\mcitedefaultseppunct}\relax
\EndOfBibitem
\bibitem[Minami \latin{et~al.}(2012)Minami, Ito, and Nakano]{jz3011749}
Minami,~T.; Ito,~S.; Nakano,~M. Theoretical Study of Singlet Fission in
  Oligorylenes. \emph{J. Phys. Chem. Lett.} \textbf{2012}, \emph{3},
  2719--2723\relax
\mciteBstWouldAddEndPuncttrue
\mciteSetBstMidEndSepPunct{\mcitedefaultmidpunct}
{\mcitedefaultendpunct}{\mcitedefaultseppunct}\relax
\EndOfBibitem
\bibitem[Rangel \latin{et~al.}(2016)Rangel, Berland, Sharifzadeh,
  Brown-Altvater, Lee, Hyldgaard, Kronik, and Neaton]{PhysRevB.93.115206}
Rangel,~T.; Berland,~K.; Sharifzadeh,~S.; Brown-Altvater,~F.; Lee,~K.;
  Hyldgaard,~P.; Kronik,~L.; Neaton,~J.~B. Structural and excited-state
  properties of oligoacene crystals from first principles. \emph{Phys. Rev. B}
  \textbf{2016}, \emph{93}, 115206\relax
\mciteBstWouldAddEndPuncttrue
\mciteSetBstMidEndSepPunct{\mcitedefaultmidpunct}
{\mcitedefaultendpunct}{\mcitedefaultseppunct}\relax
\EndOfBibitem
\bibitem[Henderson \latin{et~al.}(2008)Henderson, Janesko, and
  Scuseria]{henderson2008}
Henderson,~T.~M.; Janesko,~B.~G.; Scuseria,~G.~E. Range Separation and Local
  Hybridization in Density Functional Theory. \emph{J. Phys. Chem. A}
  \textbf{2008}, \emph{112}, 12530--12542\relax
\mciteBstWouldAddEndPuncttrue
\mciteSetBstMidEndSepPunct{\mcitedefaultmidpunct}
{\mcitedefaultendpunct}{\mcitedefaultseppunct}\relax
\EndOfBibitem
\bibitem[Kohn(2018)]{kohn2018modeling}
Kohn,~A.~W. Modeling non-radiative processes in solar materials. Ph.D.\ thesis,
  Massachusetts Institute of Technology, 2018\relax
\mciteBstWouldAddEndPuncttrue
\mciteSetBstMidEndSepPunct{\mcitedefaultmidpunct}
{\mcitedefaultendpunct}{\mcitedefaultseppunct}\relax
\EndOfBibitem
\bibitem[Geva(2018)]{geva2018simulating}
Geva,~N. Simulating energy transfer between nanocrystals and organic
  semiconductors. Ph.D.\ thesis, Massachusetts Institute of Technology,
  2018\relax
\mciteBstWouldAddEndPuncttrue
\mciteSetBstMidEndSepPunct{\mcitedefaultmidpunct}
{\mcitedefaultendpunct}{\mcitedefaultseppunct}\relax
\EndOfBibitem
\end{mcitethebibliography}



\end{document}